\PassOptionsToPackage{pdfpagelabels=false}{hyperref}
\documentclass[useAMS,usenatbib]{mnras}
\usepackage{graphicx}
\usepackage{amsmath,amssymb}
\usepackage{hyperref}
\usepackage{xspace}

\newlength{\newFigurewidth}
\newcommand{\sect}[1]{Section~\ref{sec:#1}\xspace}
\newcommand{\fig}[1]{Figure~\ref{fig:#1}\xspace}
\newcommand{\msun}{\,M$_{\odot}$\xspace}
\def\Myr{{\rm\thinspace Myr}}
\def\kpc{{\rm\thinspace kpc}}

\title[Simulating cold circumgalactic gas]{Zooming in on accretion - II. Cold Circumgalactic Gas Simulated with a super-Lagrangian Refinement Scheme}

\author[Joshua Suresh et al.]{\parbox{\textwidth}{
Joshua Suresh$^{1,2}$,
Dylan Nelson$^{3}$\thanks{Corresponding author: \texttt{dnelson@mpa-garching.mpg.de}},
Shy Genel$^{4,5}$,
Kate H. R. Rubin$^{6}$, 
Lars Hernquist$^{1}$
}\vspace{0.4cm}\\
\parbox{\textwidth}{$^1$Harvard-Smithsonian Center for Astrophysics, 60 Garden Street, Cambridge, MA 02138, USA\\
$^{2}$Institute for Disease Modeling, 3150 139th Ave SE, Bellevue, WA 98005, USA \\
$^{3}$Max-Planck-Institut f\"{u}r Astrophysik, Karl-Schwarzschild-Str. 1, 85741 Garching, Germany\\ 
$^{4}$Center for Computational Astrophysics, Flatiron Institute, 162 Fifth Avenue, New York, NY 10010, USA \\
$^{5}$Columbia Astrophysics Laboratory, Columbia University, 550 West 120th Street, New York, NY 10027, USA\\
$^{6}$Department of Astronomy, San Diego State University, San Diego, CA 92182, USA \\
}}

\begin{document}

\date{}
\maketitle

\begin{abstract}
In this study we explore the complex multi-phase gas of the circumgalactic medium (CGM) surrounding galaxies. We propose and implement a novel, super-Lagrangian `CGM zoom' scheme in the moving-mesh code \textsc{arepo}, which focuses more resolution into the CGM and intentionally lowers resolution in the dense ISM. We run two cosmological simulations of the same galaxy halo, once with a simple `no feedback' model, and separately with a more comprehensive physical model including galactic-scale outflows as in the Illustris simulation. Our chosen halo has a total mass of \mbox{$\sim 10^{12}$ M$_\odot$} at $z \sim 2$, and we achieve a \textit{median} gas mass (spatial) resolution of $\simeq$ 2,200 solar masses ($\simeq$ 95 parsecs) in the CGM, six-hundred (fourteen) times better than in the Illustris-1 simulation, a higher spatial resolution than any cosmological simulation at this mass scale to date. We explore the primary channel(s) of cold-phase CGM gas production in this regime. We find that winds substantially enhance the amount of cold gas in the halo, also evidenced in the covering fractions of HI and the equivalent widths of MgII out to large radii, in better agreement with observations than the case without galactic winds. Using a tracer particle analysis to follow the thermodynamic history of gas, we demonstrate how the majority of this cold, dense gas arises due to rapid cooling of the wind material interacting with the hot halo, and how large amounts of cold, \mbox{$\sim 10^4$ K} gas can be produced and persist in galactic halos with $T_\text{vir} \sim 10^6$ K. At the resolutions presently considered, the quantitative properties of the CGM we explore are not appreciably affected by the refinement scheme.
\end{abstract}

 
 
\section{Introduction}

The collapse of dark matter and the formation of a gravitationally bound halo is followed by the accretion of baryons from the intergalactic medium. The flow of gas through the subsequently forming circumgalactic medium (CGM) regulates the growth of its central galaxy. As a result, this gaseous reservoir hosts the galactic baryon cycle, acting as the interface regime between small-scale feedback processes and large scale inflows.

Early theoretical understanding of the structure of CGM gas described a simple `single-phase' model, with dynamics dependent on the balance of the virialization timescale and the gas cooling timescale. If the cooling timescale was long, a stable virial shock could develop, and any incoming gas would be heated to the virial temperature before accreting to the galaxy (so called `hot-mode' accretion). Conversely, if the cooling timescale was short, it was argued that no stable shock could form, and the cold primordial gas could pass unperturbed through the virial sphere and accrete directly onto the central object \citep{rees77,silk77,wr78}.

This basic model is complicated by the growing wealth of evidence, both observational \citep[e.g.][]{werk13,rubin15,crighton15} and theoretical \citep{shen13,ford13,suresh15}, that the CGM may be commonly, or always, comprised of multiple phases. Indeed, even in halos which can in principle form a stable virial shock with $T_\text{vir} \gtrsim 10^6$ K, there is nevertheless a significant mass fraction of circumgalactic gas that is observed to be at $T \sim 10^4$ K. Even more surprising, this cold gas is found in abundance around not only star-forming galaxies, but around quenched galaxies at low redshift \citep{thom12}, and even massive quasar-hosting halos at high redshift \citep{prochaska13}. It is currently an open question how this phase can persist long enough to produce high covering fractions, particularly in halos with virial temperatures $T \gtrsim 10^6$ K where it should be difficult to remain cold or be long lived \citep{schaye07,mccourt15}.

The origin of this phase is unclear, and various physical channels likely contribute. Some of the gas may be primordial, unshocked gas accreted from the IGM \citep{keres05,keres09,nelson13}. The gas may be recently ejected in a galactic outflow \citep[e.g.][]{thompson16} and/or in an ongoing galactic fountain \citep{ford13}. Another possibility is that some of the gas may have cooled out of thermal instabilities in the hot halo gas \citep{maller04}. Determining the primary channel(s) of cold CGM gas production in $\sim 10^{12} M_\odot$ halos at high redshift ($z \simeq 2$) is the main scientific aim of this paper.

Because the structure and properties of the cold halo gas phase may arise from complex gas physics, hydrodynamical simulations of galaxy formation, particularly those of cosmological volumes with statistically representative samples of galaxies \citep[e.g.][]{vdv12,dubois16}, are a uniquely powerful tool to directly study the origin of this particular phase. Studying the circumgalactic regime in cosmological simulations has, however, proven difficult for a number of reasons.

First, only recently have simulations begun to achieve realistic and heterogeneous galactic populations in broad agreement with basic observables and scaling relations \citep[e.g.][]{crain15,pillepich18a}. Second, while the general properties of the IGM are now thought to be fairly well understood \citep[e.g.][]{hernquist96}, hydrodynamic solver inaccuracies have made robust conclusions about the CGM substantially more difficult \citep{torrey12,nelson13,hayward14}. Third, the CGM is difficult to study using Lagrangian-adaptive codes (such as smoothed particle hydrodynamics), as it is orders of magnitude less dense than the interstellar medium (ISM). This typically means that even in simulations where the spatial resolution in the ISM is excellent, the CGM can be resolved quite poorly. Although this should be less of an issue when modeling the volume-filling warm-hot phase \citep[e.g. the OVI ion;][]{oppenheimer09,suresh17,nelson18b}, higher resolution is likely necessary to resolve the clumpier cold phase \citep{mccourt18}. As one example, \cite{crighton15} model CGM absorbers at $z\sim 2.5$ and derive cold cloud sizes $r_\text{cloud} < 500$ pc \citep[see also][]{schaye07,rubin17,koyamada17}.

We attempt to address these issues directly. First, we use the moving-mesh code \textsc{arepo}, which is particularly accurate for modeling difficult regimes which arise in the CGM \citep{bauer12,keres12,torrey12,nelson13}. Next, we include the well-validated Illustris model for galactic-scale outflows driven by stellar feedback. Finally, in order to resolve the cold phase, we design and employ a novel mesh refinement scheme to focus resolution out of the galaxy and into the CGM (see \citealt{vdv18}, \citealt{peeples18}, \textcolor{blue}{Richardson et al. in prep} for alternate, complementary approaches). 

Since our goal is to understand the origin of the cold gas in high-redshift halos, we employ two simulations with substantially different physics implementations. The first simulation has only primordial (H, He) cooling, with no metal-line contribution. The second simulation includes both primordial and metal line-cooling, including a pre-enriched IGM, and also adds galactic winds, which allows us to investigate stirring and fountain circulation from strong outflows.

In Section~\ref{sec:methods} we describe the simulations and numerical methodology. Section~\ref{sec:results} explores the results, connecting to observational benchmarks in Section~\ref{subsec:obs} and exploring the origin of cold-phase CGM gas in Section~\ref{sec:coldphase}. We finish with a discussion in Section~\ref{sec:discussion} and summarize  in Section~\ref{sec:summary}.


\section{Methods}
\label{sec:methods}

\subsection{Simulation Details}

The simulations employed in this study were run using the moving-mesh \textsc{arepo} code \citep{springel10}. Briefly, we address the solution of the equations of continuum hydrodynamics coupled to self-gravity. The fluid dynamics are solved with a finite-volume type scheme using an unstructured, dynamic, Voronoi mesh for the spatial discretization, while gravity is treated with the standard Tree-PM approach. 

This paper presents several cosmological zoom simulations of a single galaxy halo, each starting from the same set of initial conditions (ICs), but run with increasingly complex physical models. The ICs were selected from an earlier set of eight zoom simulations carried out in \cite{nelson16}, and our variations allow us to study the individual effect of different physical mechanisms on the cold gas content in the CGM. All simulations employ a spatially uniform UV background \citep{fg09}, accounting for gas self-shielding following \cite{rahmati13}, and include a model for star formation in a dense, multi-phase interstellar medium \citep{spr03}.

The first simulation, \textbf{Primordial-Only}, includes only primordial (H, He) cooling, and neglects metal-line cooling as well as any explicit galactic feedback. This physics is identical to the model employed in \cite{nelson16}. Although we often refer to this first simulation as `cooling-only', we note that it does include the impact of unresolved supernova feedback as a source of pressurization of the ISM \citep{spr03}. The second simulation, \textbf{Galactic-Winds}, models both metal-line and primordial cooling, as well as stellar population evolution and chemical enrichment following supernovae Ia, II, as well as AGB stars. In addition, it also includes a prescription for galactic-scale outflows, or winds. Neither simulation includes any treatment of feedback from supermassive blackholes. We have intentionally omitted the Illustris BH feedback model from the present simulations to simplify interpretation of the results, and we anticipate that it will have negligible impact given our halo mass scale \citep[BH feedback becoming important only for $M_{\rm halo} > 10^{12}$\msun in Illustris;][]{vogelsberger13}.

The winds in the Galactic-Winds simulation are identical to the winds employed by the Illustris simulation. See \cite{vogelsberger13} for details of the wind implementation, and \cite{torrey14} for a comparison of their outcome versus observations. The Illustris model has been shown to produce reasonable agreement with many important baseline observations for the heterogeneous galaxy population, including the stellar mass function across redshift and the cosmic star-formation rate \citep{genel14}. This affords some confidence that the wind implementation is able to approximately regulate the buildup of the stellar mass content of dark matter halos across cosmic time \citep[see][for a comparison to the new IllustrisTNG wind model]{pillepich18a}. Nevertheless, as a subgrid prescription, the wind treatment necessarily has many simplifying assumptions, and in this study we do not focus on its details. Instead we seek to understand circumgalactic properties with the reasonable Illustris physical model. We note for reference that, after launch, winds are hydrodynamically (though not gravitationally) decoupled while still residing in the dense ISM -- they recouple at a density equal to $5\%$ of the SF threshold, namely $\log n_{\rm H}[{\rm cm}^{-3}] \sim -2.2$, or after a maximum time of 2.5\% of the concurrent Hubble time.

For the Galactic-Winds simulation we implement a spatially uniform metal background of $10^{-3}$ $Z_\odot$ throughout the box at $z=6$. This IGM pre-enrichment was chosen for two reasons: first, to approximate the effect of very early enrichment of the IGM by small galaxies, which would be unresolved in this simulation. And second, to study the possible impact of metal-line cooling on slightly enriched IGM gas, which could subsequently accrete onto, and modify the evolution of, the CGM.

For details of the initial conditions, we refer the reader to \cite{nelson16}, and here give only a brief overview. Cosmological zoom ICs were generated by first running a low-resolution cosmological dark matter only simulation of a uniform volume and identifying halos of interest (in this case, $M\sim 10^{12} M_\odot$ halos at $z \sim 2$). To re-simulate a given halo all particles within a radius of several times $r_{\rm vir}$ around the halo center were tracked back to the initial conditions, where we use their convex hull to define the `high-resolution' region of the simulation. We then increased the mass resolution in this high-resolution region of the box significantly above the rest of the simulation volume. This allows for the study of a single halo at high resolution within its proper cosmological context. The cosmology is consistent with the WMAP-9 results ($\Omega_{\Lambda,0}=0.736$, $\Omega_{m,0}=0.264$, $\Omega_{b,0}=0.0441$, $\sigma_8=0.805$, $n_s=0.967$ and $h=0.712$).

\subsection{CGM Refinement Scheme}

In the standard mode of \textsc{arepo} for galaxy formation simulations, gas cells are kept at a roughly constant mass, to within a factor of two. These gas cells `refine' (with a binary split) or `de-refine' (with a binary merge) if their mass deviates too far from a pre-defined target gas cell mass $(m_{\rm target})$. This approximately constant mass resolution implies that the spatial size of cells scales with gas density, such that the smallest cells are found in the dense centers of massive galaxies, and the largest cells in low density voids of the IGM.\footnote{By gas cell size, we always mean the spherical volume equivalent radius $r_{\rm cell} = [3V/(4\pi)]^{1/3}$, the most appropriate definition given the arbitrary shapes of the Voronoi cells, which in the simulation are regularized to be roughly spherical \protect\citep{springel10}.} The smallest and densest cells also have the smallest hydrodynamical timesteps, as required to satisfy the Courant Friedrichs Lewy \citep[CFL;][]{courant67} condition. Consequently, the dense, star-forming ISM can easily dominate the run-time of typical galaxy formation simulations, and this numerical cost arises because of the natural `Lagrangian' nature of the code, which generically results in higher resolution where it is typically desired: in the dense centers of collapsed structures, i.e. galaxies.

In this study we are, however, not interested in galaxies themselves, but rather in the gaseous reservoirs surrounding galaxies: the circumgalactic medium. In a typical galaxy simulation, the numerical resolution in the outer CGM can easily be an order of magnitude lower than in the galaxy itself \citep{nelson16}. We would like to better resolve the hydrodynamics and multi-phase gas of the CGM, but simply increasing the global resolution of the simulation would be prohibitively expensive. Therefore, we here propose and implement a novel, super-Lagrangian `CGM zoom' method, which focuses more resolution into the CGM and intentionally lowers resolution in the dense ISM. 

\begin{figure*}
\centering
  \includegraphics[width=5.4in]{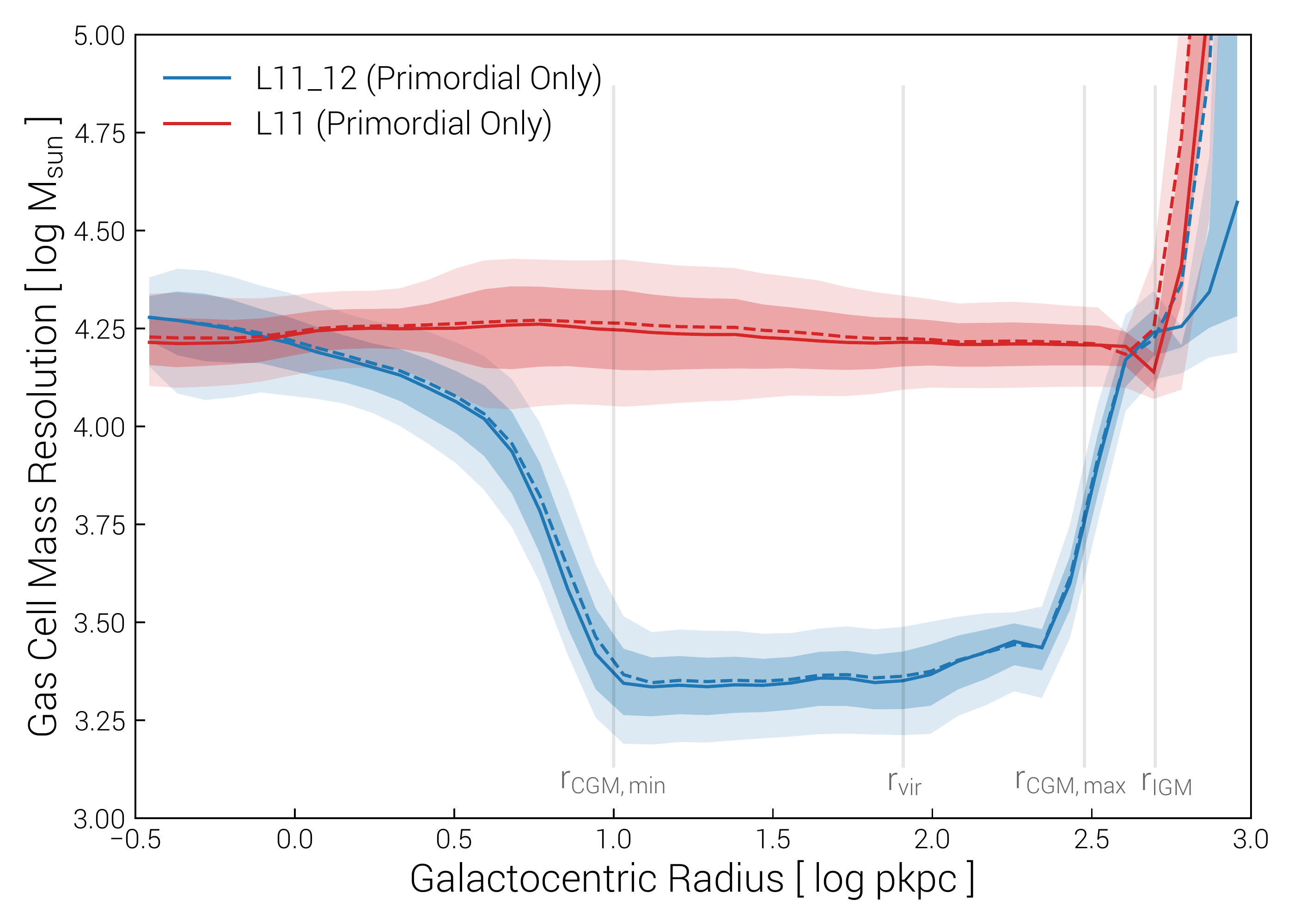}
  \includegraphics[width=3.3in]{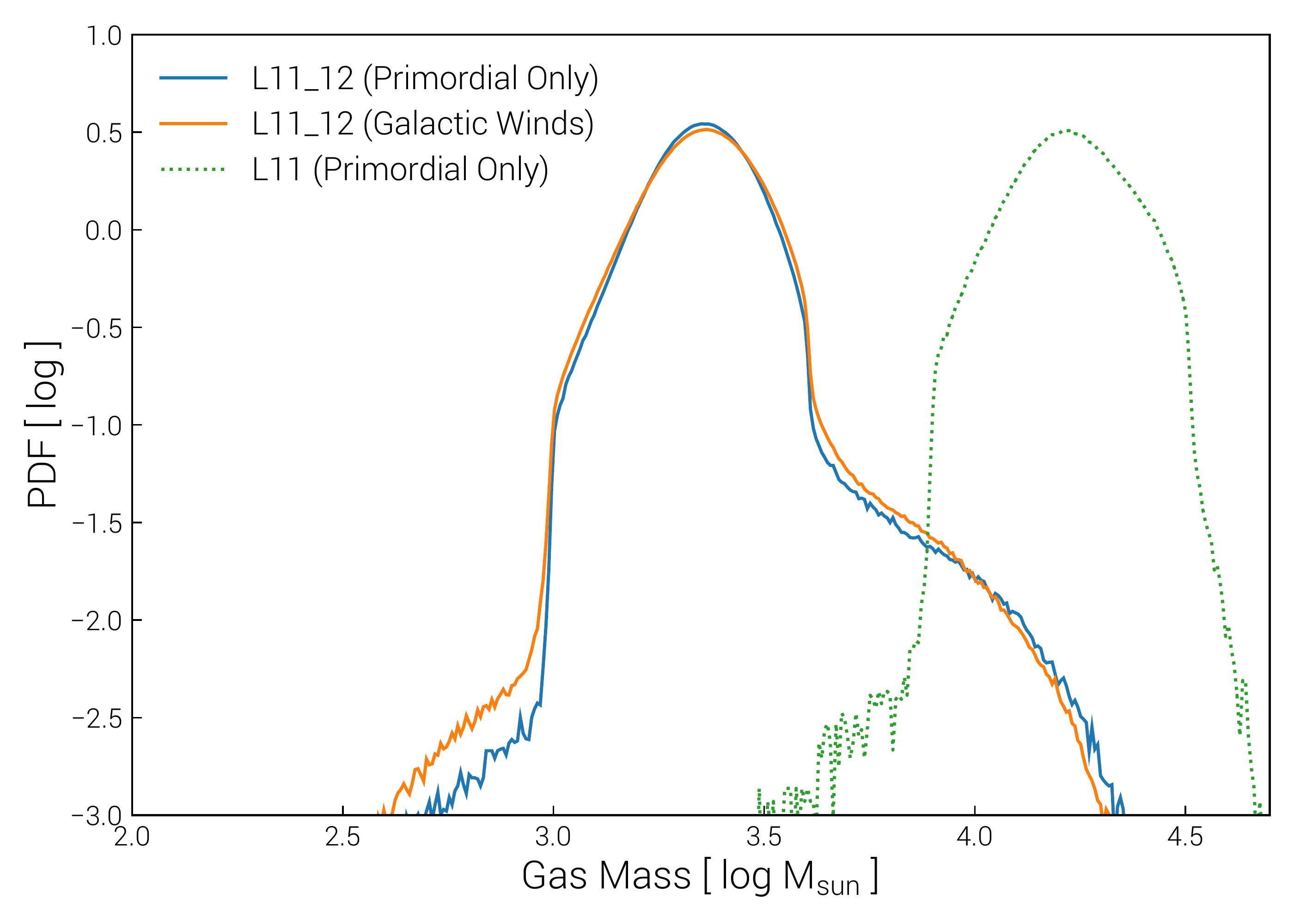}
  \includegraphics[width=3.3in]{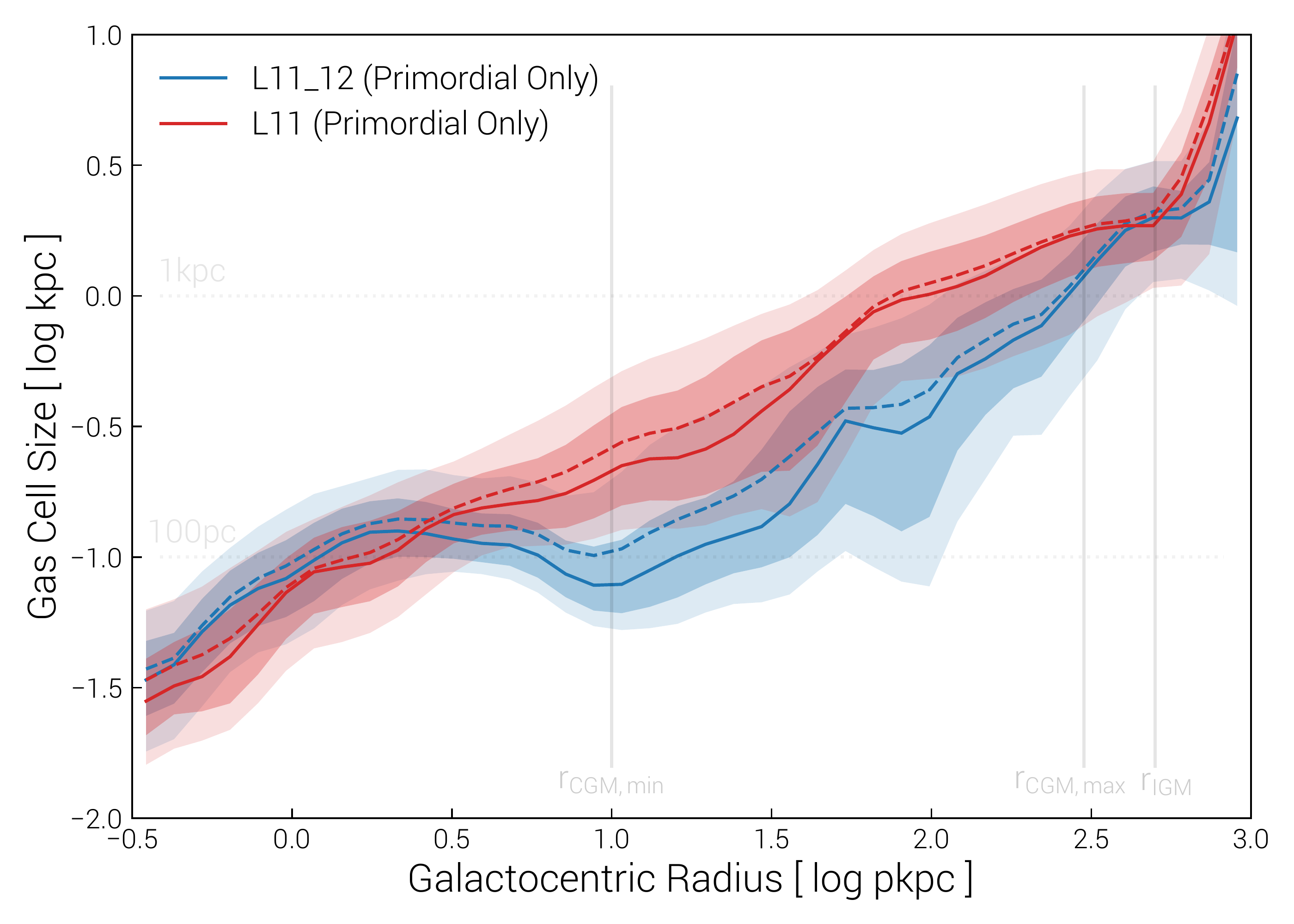}
\caption{Baryon resolution statistics and the impact of our super-Lagrangian refinement technique. The top panel shows the spherically-averaged radial profile of gas cell mass (median solid; mean dotted; 10-90 and 25-75 percentiles as shaded bands), comparing the un-modified `L11' case to the resolution boosted `L11\_12'. The baryonic mass resolution is improved by a factor of eight, as designed, specifically within $\sim$ 10 kpc to $\sim$ 300 kpc. The lower right panel shows the radial profile of the size of gas cells, which is improved to $\sim$ 100 parsecs in the inner CGM. The lower left panel compares the normalized distributions of gas cell masses: the CGM resolution has a narrow spread about an average of slightly more than 2,000 solar masses. In all cases, we include here only gas in the cool-dense phase ($T < 10^5$ K, $n_{\rm H} > 10^{-3}$ cm$^{-3}$) of the CGM ($r > 10$ kpc) at $z=2.25$. Assuming a minimum threshold of $4^3$ cells to properly resolve a CGM `cloud', we can robustly resolve features with radii as small as $\sim 400$ pc at typical densities, and even smaller at higher densities.\label{fig:cell_mass_size}}
\end{figure*}

The method proceeds as follows. First, we use the \textsc{subfind} algorithm \citep{springel01} to identify the location of the potential minimum of the target zoom halo. A single passive `ghost' particle, which has no bearing on the dynamics, is pinned to this potential minimum and moves along with it. The location of this particle is then used, internally, to define a galacto-centric radius for all gas cells in the simulation. Using this galacto-centric radius, we define a radius-dependent target gas mass, specifying a low target gas mass (high resolution) in the circumgalactic regime and a higher target gas mass (low resolution) in both the inner ISM and distant IGM, linearly interpolating in between.

We define a single mass $m_0 = 1.6 \times 10^4 M_\odot$, which in the ordinary `L11' base initial conditions is the target gas mass in the high-resolution region. Then we define three characteristic radii, $r_\text{CGM,min}$, $r_\text{CGM,max}$, and $r_\text{IGM}$, which we use to establish the spatial region where we require higher resolution cells in the resolution-boosted `L11\_12' simulation by increasing $m_{\rm target}$ as follows:

\begin{itemize}
\item $r = 0$: $m_{\rm target} = m_0$
\item $r_\text{CGM,min} < r < r_\text{CGM,max}$: $m_{\rm target} = m_0/8 = 2000 M_\odot$
\item $r > r_\text{IGM}$: $m_{\rm target} = m_0$
\end{itemize}

We linearly interpolate the target mass between the different regions, so that the effective $m_{\rm target}(r)$ is a continuous function of radius. We take \{$r_\text{CGM,min}$,$r_\text{CGM,max}$,$r_\text{IGM}$\} = \{30,900,1500\} comoving kpc, which at $z=2$ corresponds to proper distances of \{10,300,500\} kpc, respectively.

\fig{cell_mass_size} shows several resolution statistics of the cold and dense CGM gas and therefore the result of our `CGM zoom' refinement procedure. The top panel plots the radial profile of gas cell mass, from the very center of the galaxy out to the IGM. The `L11' base initial conditions (red) are contrasted against our `L11\_12' resolution-boosted realization (blue), which improves upon the mass resolution in the radial range of the CGM by a factor of eight, as designed. The lower left panel shows the distributions of gas cell masses, where our two simulations are essentially identical (blue, green), and all shifted to much lower masses than the original simulation (red). Tails of these distributions represent the inability of the (de-)refinement to gently eliminate or split cells fast enough: the least massive cells in the CGM of the Galactic-Winds simulation reach 150\msun.

The lower right panel shows the radial profile of gas cell size, which decreases towards the inner halo as a result of the increasing density. The CGM gas, taken here as gravitationally bound non-star forming cells, at the final snapshot ($z=2.25$) has a median cell mass of about 2,200 $M_\odot$ and a median cell size of $\simeq$ 95 physical parsecs. Due to the range of gas densities, these cells also occupy a distribution of sizes from 52 pc to 405 pc (5th$-$95th percentiles). For the median cell size, if we require a minimum threshold of $4^3$ cells to robustly resolve a CGM `cloud', then we capture such features with radii as small as $\sim 400$ pc at $z=2.25$. Nonetheless cold gas structures in the CGM may be yet smaller still, and mixing and hydrodynamical processes including the Kelvin-Helmholtz and Rayleigh-Taylor instabilities may still not be properly resolved \citep{agertz07,crighton15}.

\subsection{Monte Carlo Tracers} \label{sec:tracers}

\begin{figure*}
  \includegraphics[width=7.2in]{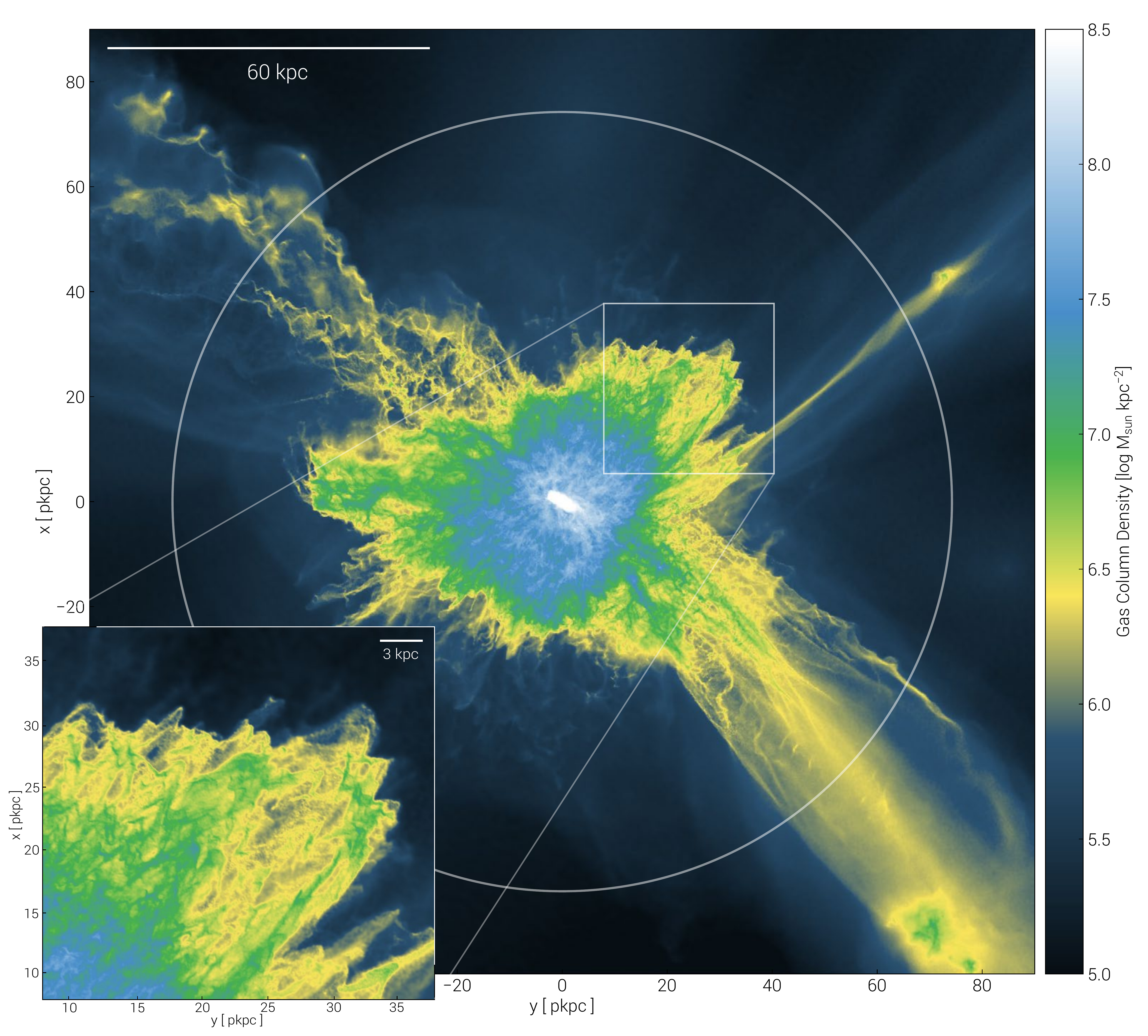}
  \caption{Large view of the Galactic-Winds halo at the final redshift of $z=2.25$, in projected gas density on the scale of the virial radius (shown as the white circle). Large scale inflow predominates from the upper left and lower right quadrants in this rotation, which is arbitrary with respect to the central galaxy. Strong galactic outflows are directed outwards towards the upper right and lower left corners. These winds are multi-phase and metal-enriched, with high radial velocity and low angular momentum; cold, over-dense structures are observed to form in the outflow with a mix of clumpy and elongated morphologies, with characteristic sizes of a few \mbox{$\sim$ 100 pc} to \mbox{$\sim$ 1 kpc}. The inset on the lower-left zooms into an outflow dominated region to emphasize this structure.\label{fig:large_image}}
\end{figure*}

\begin{figure*}
  \includegraphics[width=7.2in]{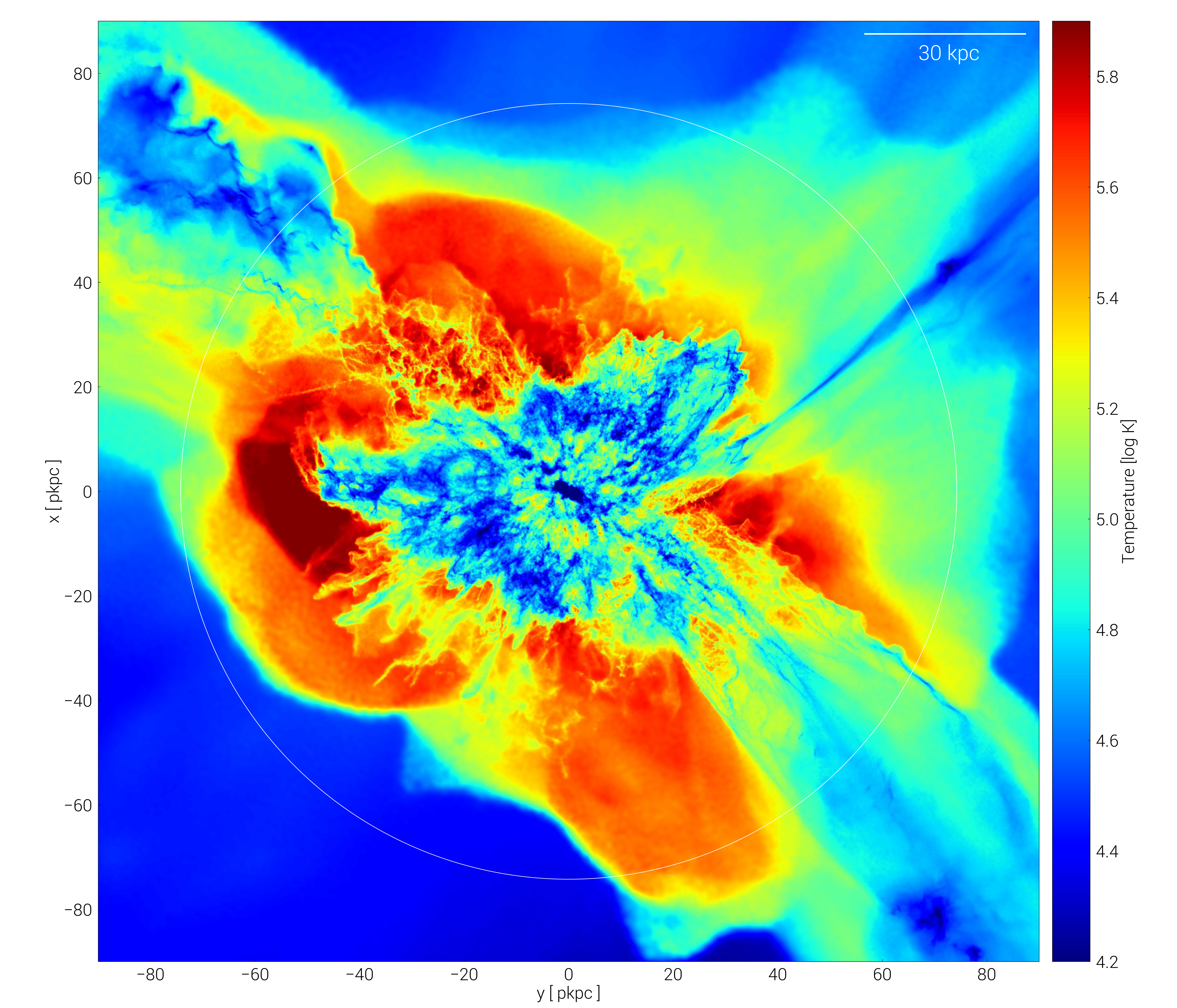}
  \caption{As in \fig{large_image}, a large view of the Galactic-Winds halo at the final redshift of $z=2.25$, except here we show the same halo in mass-weighted projected gas temperature. The scale of the virial radius (shown as the white circle) is unchanged. Ejection of cold gas from the star-forming ISM of the disk at the center results in multi-phase outflows on the scale of $\sim$ 20-30 kpc which are dominated in large part by cold/warm gas at temperatures from $\sim$ 10,000 K to $\sim$ 100,000 K.\label{fig:large_image2}}
\end{figure*}

\begin{figure*}
  \includegraphics[width=3.45in]{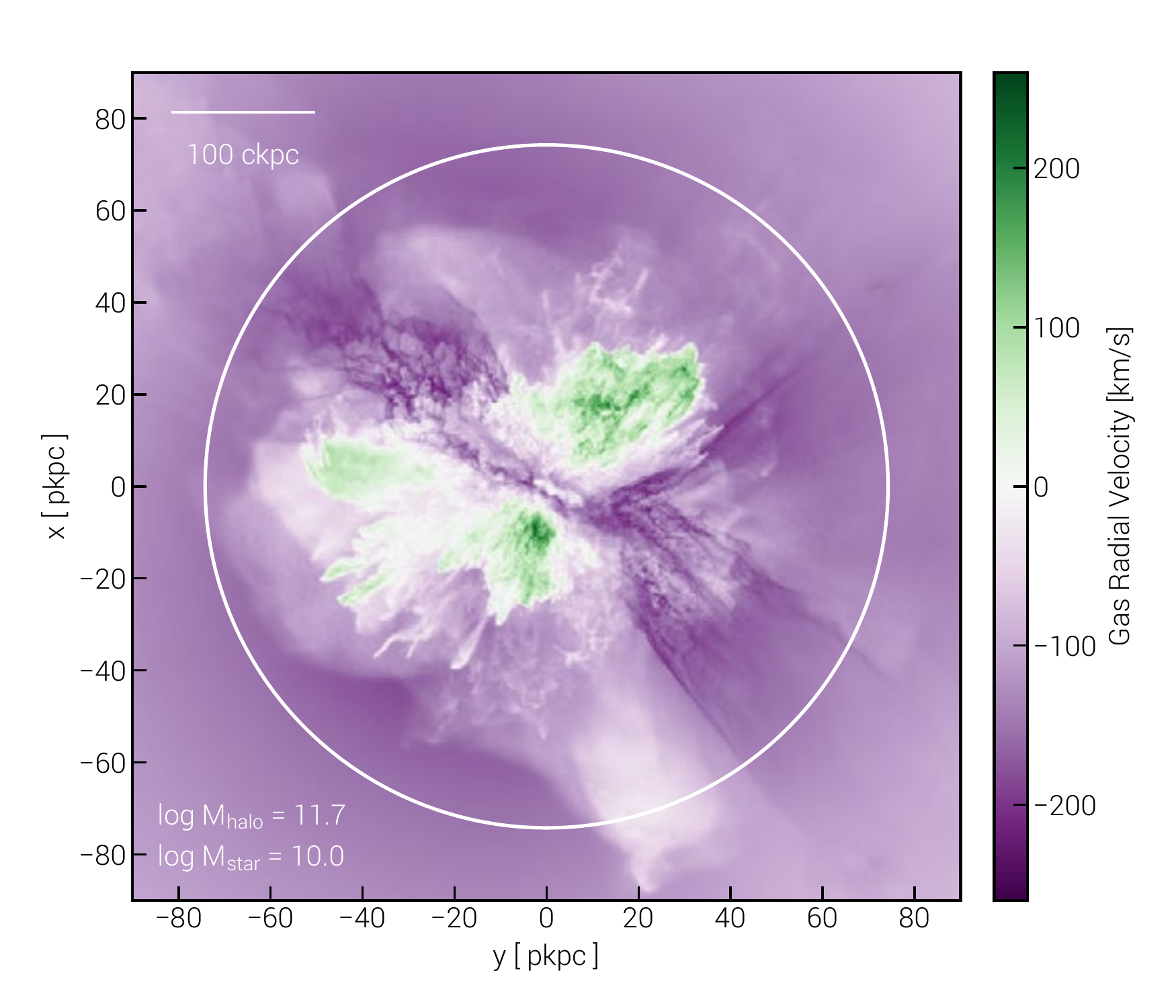}
  \includegraphics[width=3.45in]{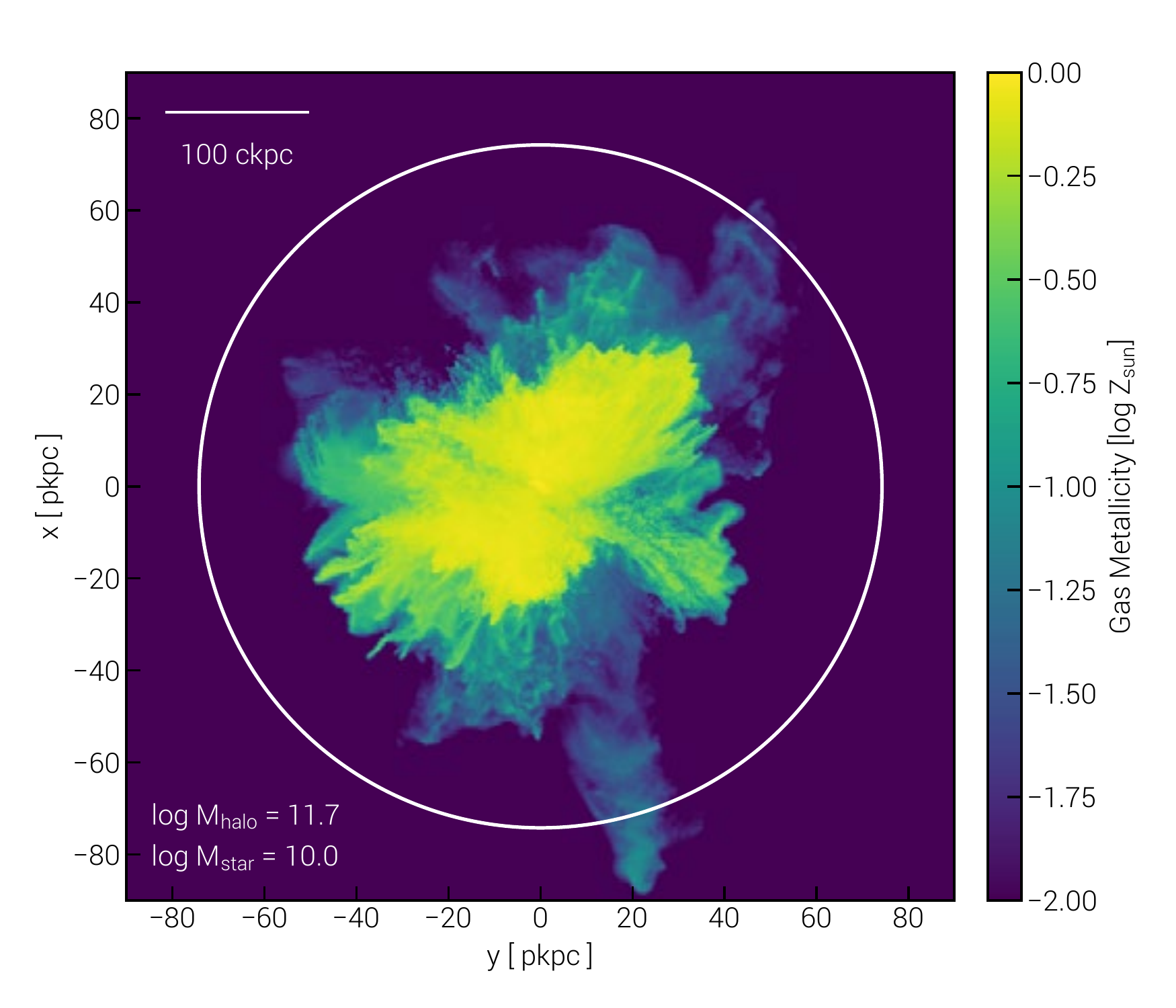}
  \caption{Two additional views of the same halo as in \fig{large_image2}, the Galactic-Winds halo at the final redshift of $z=2.25$. Here we show gas radial velocity (left), where positive denotes outflow, as well as the gas metallicity in terms of solar (right). Outflowing gas at speeds of $\sim 200$km/s and with a preferential alignment along the minor axis of the central galaxy reaches metallicity values approaching solar. In contrast, the ambient hot halo gas into which the outflow is propagating is largely at rest and relatively un-enriched with heavy elements. The virial volume is occupied by both outflows and inflows, both of which are highly anisotropic.\label{fig:large_image3}}
\end{figure*}

Each simulation includes a large number of Monte Carlo tracer particles which enable us trace the evolution of Lagrangian parcels of gas mass across time. As a result, we can quantitatively study the origin and explicitly follow the thermodynamical evolution of the gas which makes up the $z \sim 2$ CGM. A detailed description of the implementation of these tracers can be found in \cite{genel13}. Briefly, these tracers (which retain their own unique IDs throughout the simulation) are associated with individual gas elements, and are exchanged probabilistically based on the ratio of the mass flux between cells to the total mass of the parent cell. We initialize the simulations with 20 tracers per gas cell, resulting in a total number of $\simeq 6 \times 10^8$ per run. 

\subsection{Definitions of the CGM}

Unless specified otherwise, we define the circumgalactic (CGM) gas as any gas which is within the virial sphere (i.e. $r < r_{\rm vir}$) that is not star-forming and is not located within any galaxy (central or satellite). We ensure the last criterion by defining a mass-dependent ``galaxy radius" around each galaxy from which we do not draw CGM gas. In the Galactic-Winds run feedback keeps the galaxy sizes and masses roughly consistent with that of the Illustris simulation \citep[see][]{genel14,genel18}, and we use the following scheme based on those results: the galaxy radius is taken to be \{4,6,8,10\} kpc for $M_\text{gal} = \{10^{7},10^{8},10^{9},10^{10}\} M_\odot$, respectively, linearly interpolating between. While these radii are motivated by typical galaxy sizes, our results are not sensitive to the specific choices of galaxy sizes. 
We divide the CGM in two principal phases: the cool-dense CGM with $T < 10^5$ K and $n_{\rm H} > 10^{-3}$ cm$^{-3}$, and the non-cool-dense, which complements it.


\section{The structure of the CGM} \label{sec:results}

\subsection{General Trends}
\label{subsec:trends}

We begin in \fig{large_image} and \fig{large_image2} with a qualitative look at the gas distribution in the entire circumgalactic medium of the Galactic-Winds halo at its final redshift $z=2.25$. Total gas density and temperature projections are shown, respectively, including all material within a cube of side-length 180 physical kpc centered on the halo. In this orientation the galactic disk is roughly edge-on, and strong outflows are evident predominantly along its minor axes -- towards the upper right, and lower left, of the image. In general, these outflows contain a mixture of gas at different densities and temperatures. They are metal-enriched above the level of the background CGM, have high outward radial velocity (\fig{large_image3}), and low angular momentum. At small scales they show complex density structures, including a mix of clumpy and highly elongated overdensities, with typical sizes ranging from a few times $\sim 100$ parsecs up to $\sim 1$ kilo-parsec.

In contrast, cosmological inflow is preferentially arriving along the major axes of the disk -- towards the upper left, and lower right, of the image, with a lower mass flux also from the upper right. Gas accretion across the virial sphere is highly anisotropic and related to large-scale gas filaments in the IGM. Gravitationally bound structures -- i.e., future satellite galaxies -- are embedded in the accreting baryons, as for example in the lower right. Interestingly, each of the diffuse gas filaments associated with this halo have particularly distinct structure. The top left inflow is clumpy and has fragmented into a large number of small, high-density features near $r_{\rm vir}$ or just within. On the other hand, the bottom right inflow is nearly smooth crossing the virial radius; it is also much wider in cross section at high density. The top right inflow is again smooth, but much narrower.

\begin{figure}
\centering
\includegraphics[width=3.4in]{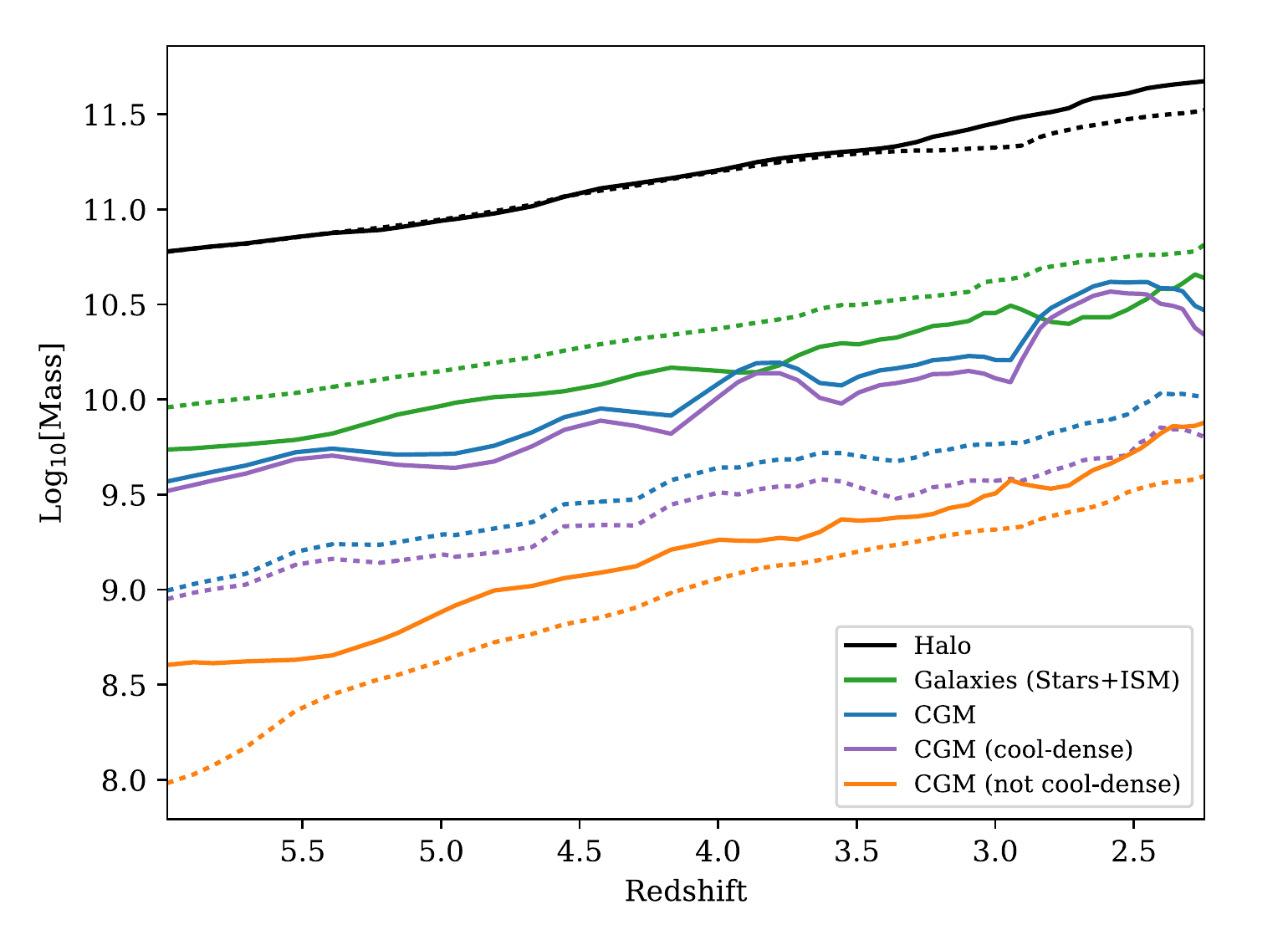}
\caption{Redshift evolution of the total mass in different halo gas components. The total halo mass (black) is decomposed into the central galaxy (green), and total CGM (blue), which is then further separated with a binary selection into cool-dense (purple; $T < 10^5$ K, $n_{\rm H} > 10^{-3}$ cm$^{-3}$) and non-cool-dense (orange) components. Different linestyles show different simulations: Primordial-Cooling Only (dotted) and Galactic-Winds (solid). When supernovae driven galactic winds are included the CGM reaches the same total mass as the galaxy itself, and is dominated in mass fraction by the cool-dense component.\label{fig:allmasses}}
\end{figure}

\begin{figure}
\includegraphics[width=3.4in]{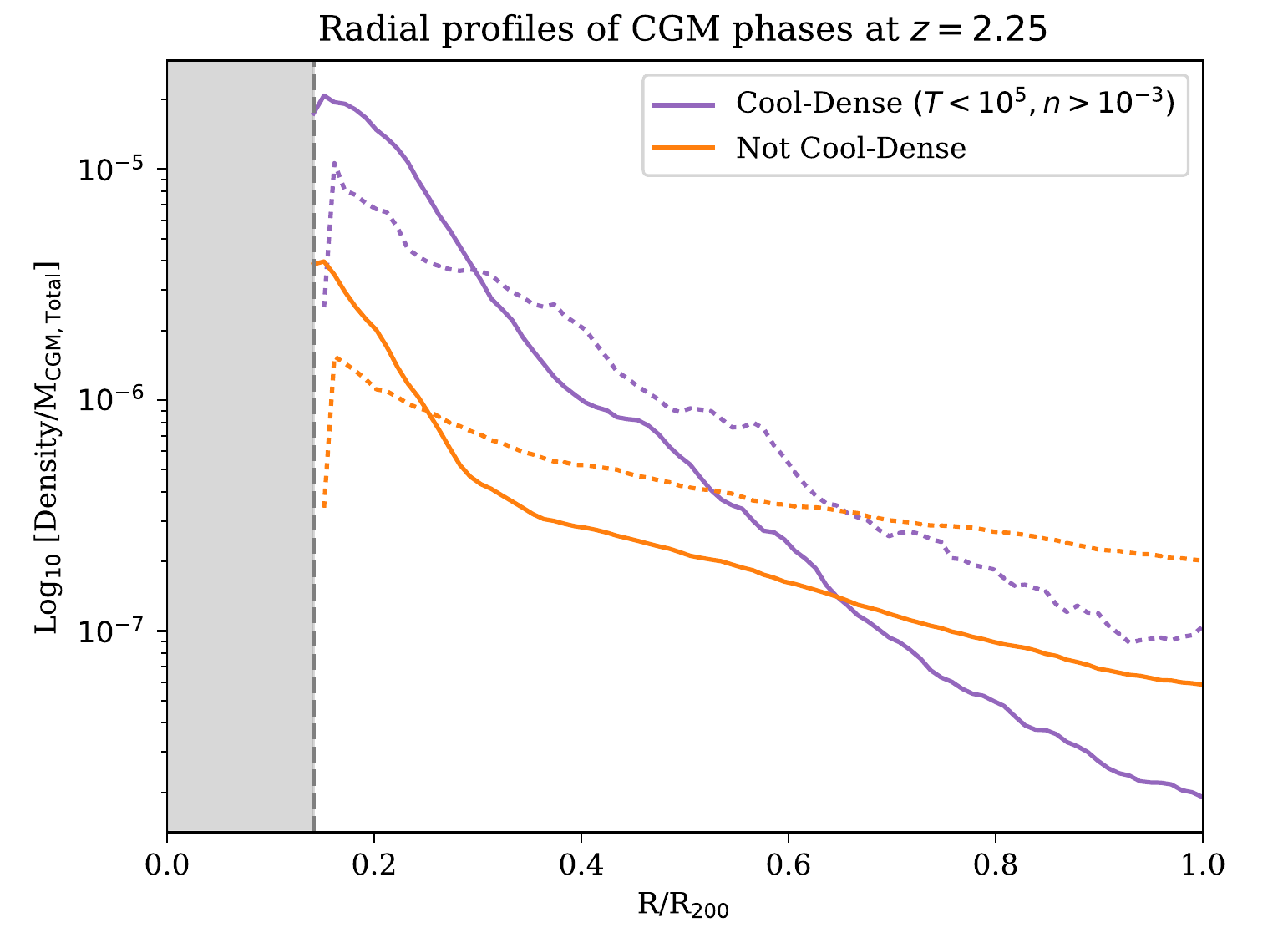}
\caption{Normalized CGM radial profiles of gas density, decomposed into the cool-dense phase (purple) and the less dense, hotter phase (orange). Different linestyles show different simulations: Primordial-Cooling Only (dotted) and Galactic-Winds (solid). In both cases, the cool-dense phase is steeper than the non-cool-dense phase, and dominates only within $r \lesssim 0.6 r_{\rm vir}$. The steep rise in both phases in the inner halo for the Galactic-Winds simulation is primarily due to the action of strong galactic outflows in this region.\label{fig:CGM_frac_R}}
\end{figure}

This cosmological accretion will, as the Universe evolves, provide the baryon reservoir of the forming dark matter halo. \fig{allmasses} shows how the halo mass of different components builds up over time, in each of the simulation variants. The cooling-only run, as expected, has more massive galaxies (including stars and ISM) than in the Galactic-Winds run, with a correspondingly less massive CGM. The cool-dense CGM phase always dominates in mass over the hotter, less dense phase. This is true by a factor of ten at $z \simeq 6$, decreasing to only a factor of two or less by $z \simeq 2$. There is little difference when adding metal-line cooling alone (dashed versus dotted lines).

Adding galactic winds substantially increases the overall mass of the CGM, by roughly a factor of three, starting already as high as $z \simeq 6$. By redshift two the total mass in the CGM and in the galaxy itself are roughly equal. The presence of winds also increases the mass (and mass fraction) in the cool-dense phase, such that by $z \simeq 2$ there is nearly an order of magnitude more total mass in this phase when compared to the hotter, less dense component.

\begin{figure*}
  \includegraphics[width=3.4in]{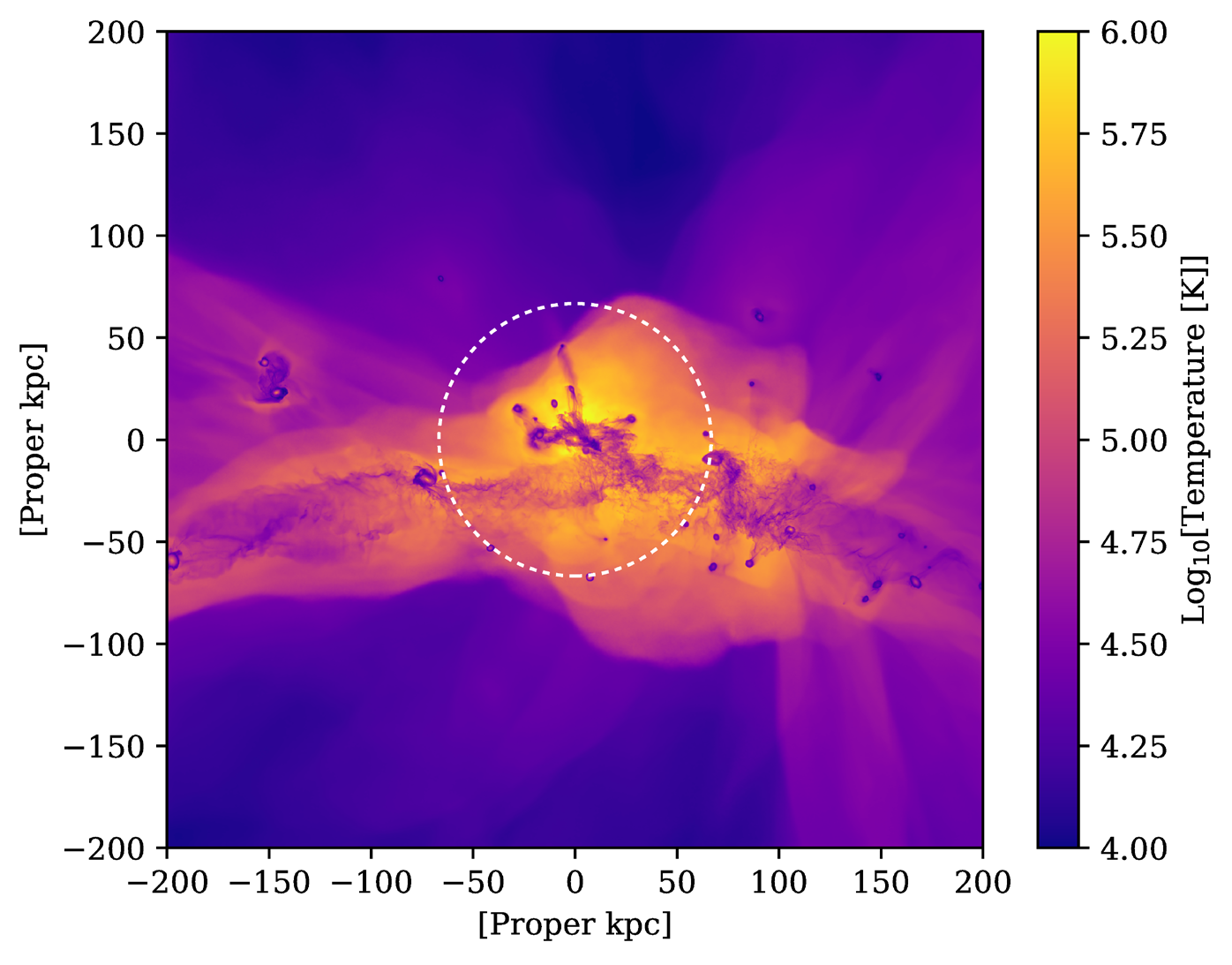}
  \includegraphics[width=3.4in]{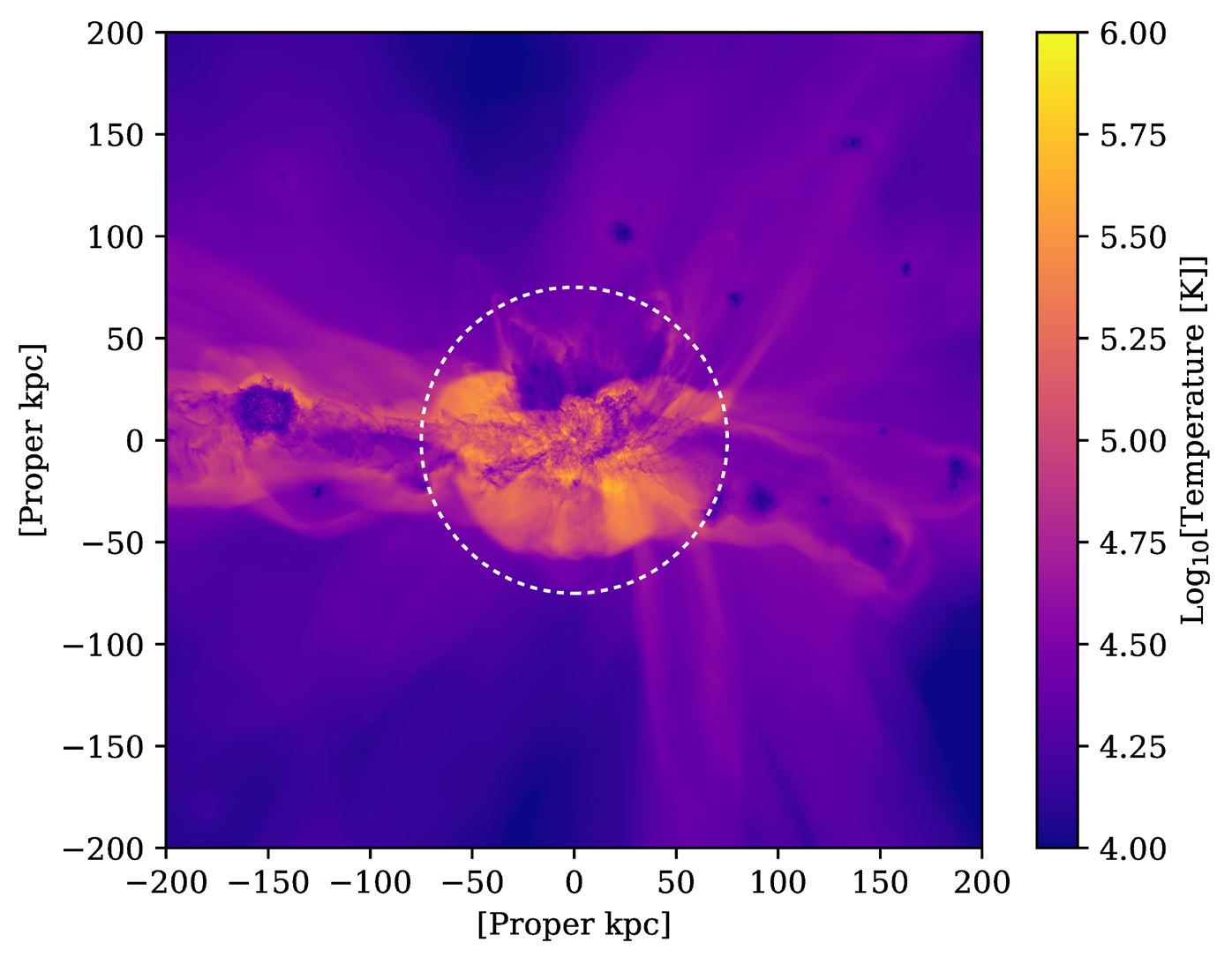}
  \includegraphics[width=3.37in]{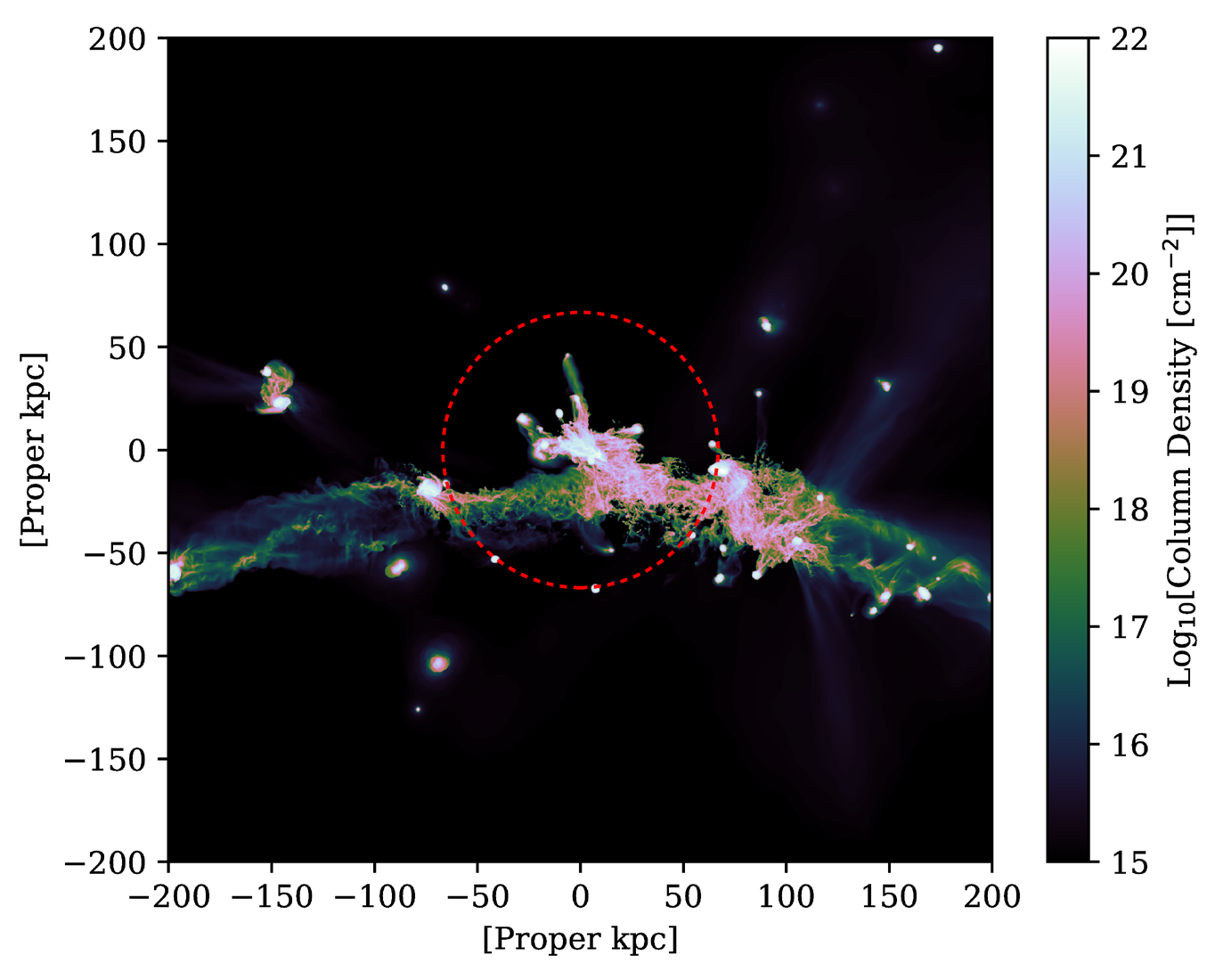}
  \includegraphics[width=3.37in]{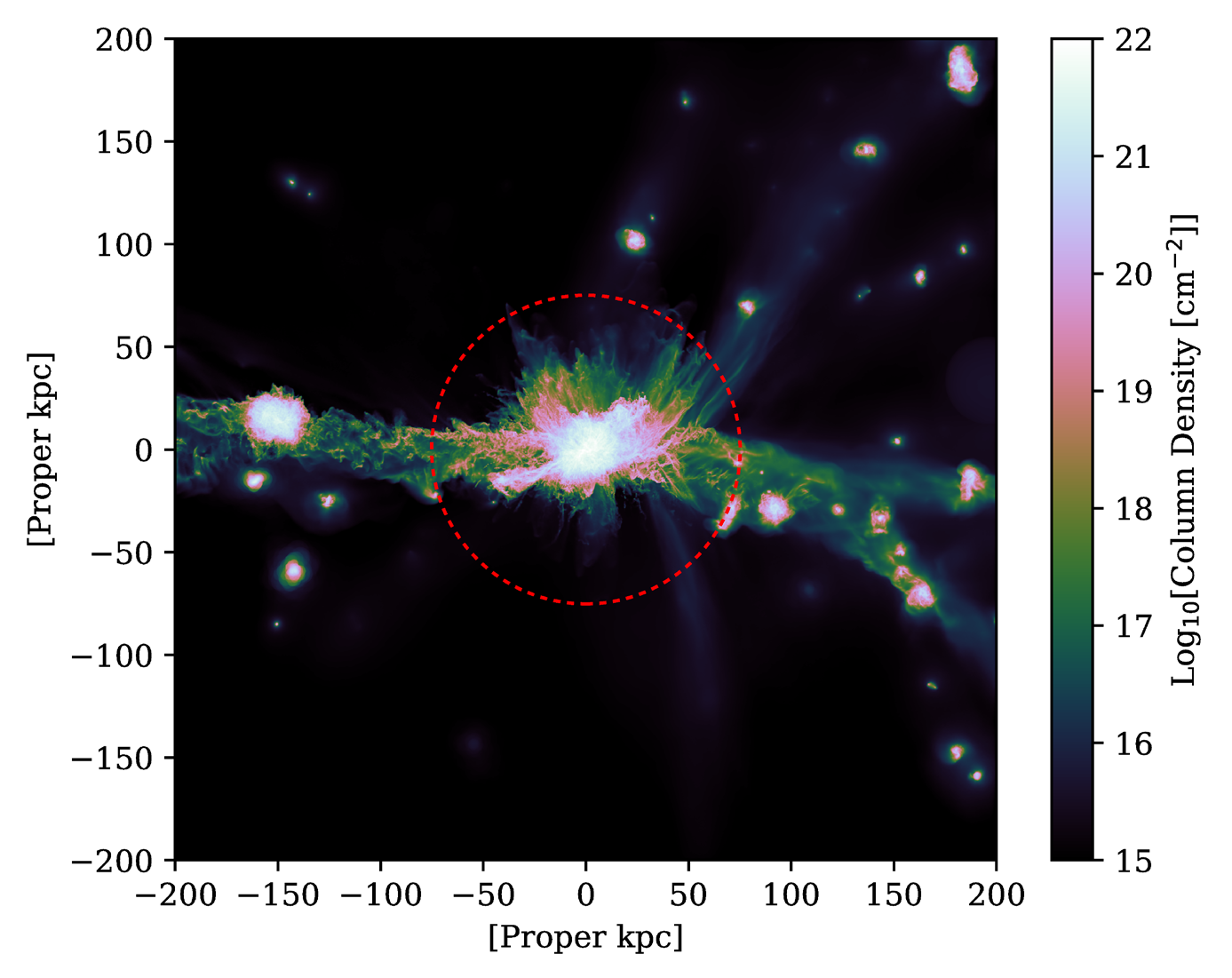}
  \includegraphics[width=3.44in]{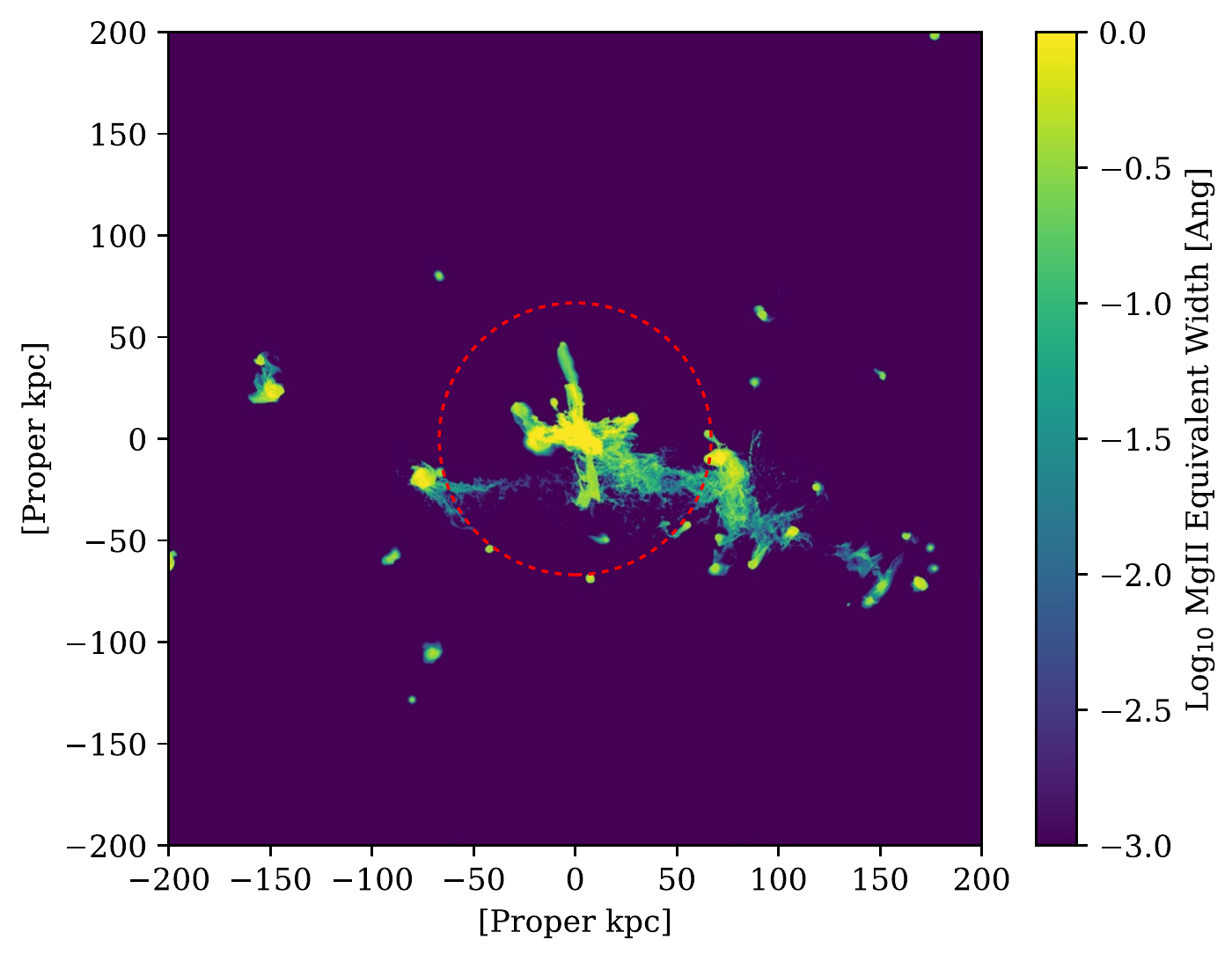}
  \includegraphics[width=3.44in]{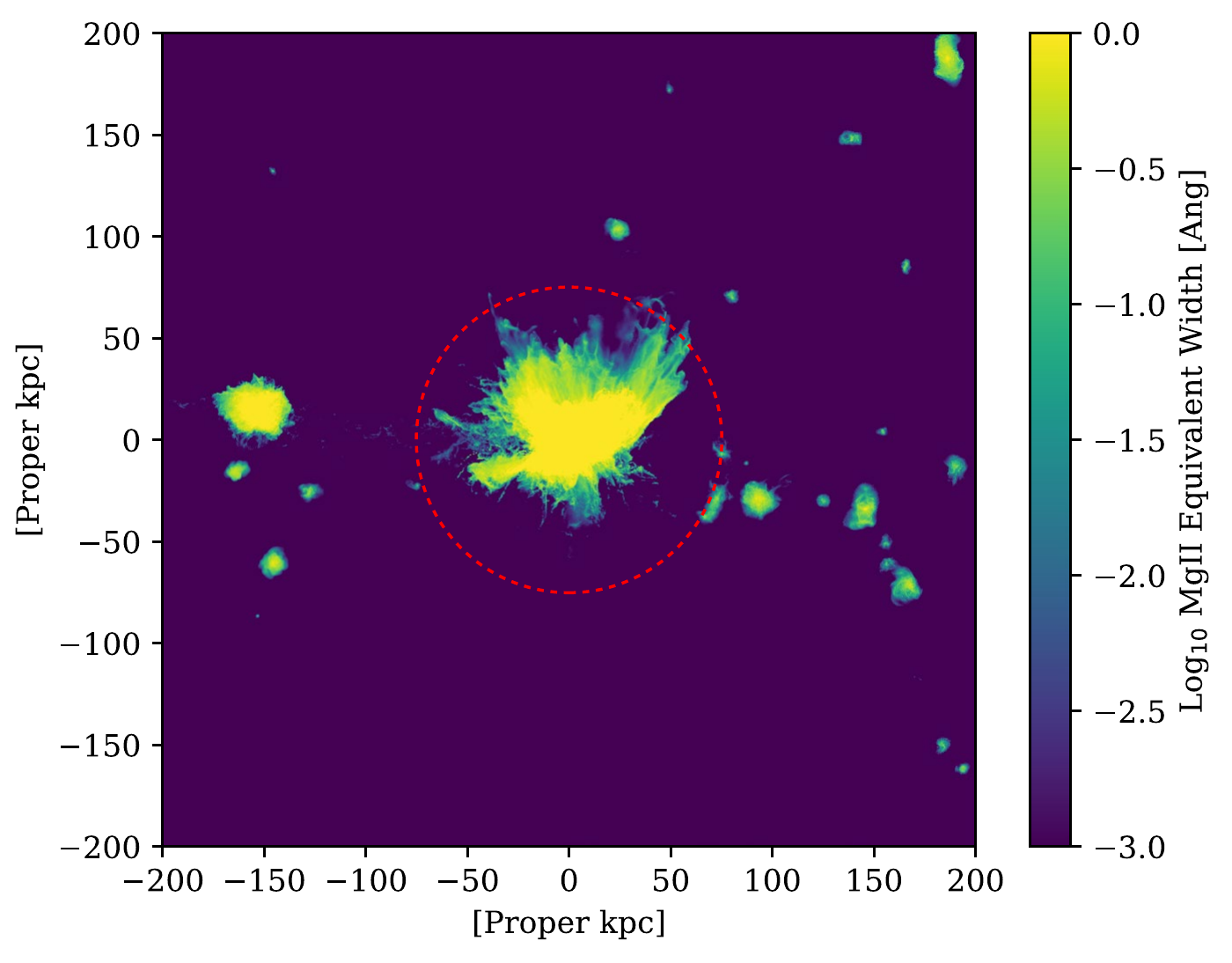}
  \caption{Visualization of the difference between the primoridal cooling only simulation (left column) and the galactic winds run (right column). We show the gas temperature projection (top), HI column density (center), and MgII equivalent width (bottom) around the central galaxy at the final redshift $z=2.25$. The latter two adopt a $\pm 1000$ km/s velocity cut along the line of sight. In the simulation with galactic winds the dense gas clearly extends to larger distances into the inner halo, and this material corresponds to high column density (or equivalent width) observational tracers, which have correspondingly larger covering fractions. The dashed circles mark the virial radii of the halos.
  \label{fig:images}}
\end{figure*}

\fig{CGM_frac_R} shows 3D radial density profiles, splitting into the cool-dense phase (purple) and the less dense, hotter phase (orange). To facilitate comparison across simulations with (slightly) varying virial radii and CGM masses, the radius is scaled to the halo virial radius, and the density is normalized by the total CGM mass. In both simulations the hot, low-density gas in the CGM has a flatter profile than the steeper cool-dense phase. There is significantly more cool-dense gas at small radii, while the rarefied hot component dominates only at $r/r_{\rm vir} \gtrsim 0.6$. In the inner halo, the cold phase has a density higher by a factor of few, while at the virial radius the opposite is true.

The Galactic-Winds simulation (solid lines) has a steeper rise towards inner radii, which results from the strong galactic outflows in this region, an effect we explore further in Section~\ref{sec:coldphase} below. Both phases are more centrally concentrated, with steeper radial profiles, in the simulation including feedback. This rising inner component is likely a mixture of hot outflow (or gas heated by the outflow) and cooling inflow (transitioning from the hot halo temperature downwards). In the hot phase a different behavior is particularly evident -- at $r/r_{\rm vir} \lesssim 0.3$ the slope of the density profile is the same as the cold, dense phase, an effect apparently caused by the winds, while beyond this radius the profile has a break and the slope transitions to a flatter profile consistent with the cooling-only simulation. The overall lower gas densities in the outer halo of the Galactic-Winds run may reflect accelerated metal-line cooling due to extended metal mixing as a result of winds and fountain flows far beyond the galactic bodies themselves.

\subsection{Comparison to HI and MgII Observations}
\label{subsec:obs}

Before moving into the space of observables, \fig{images} shows the gas temperature distributions around the primary galaxy at $z=2.25$, in the cooling only (top left) and galactic winds (top right) simulations. The morphology of the gas in the cooling-only simulation is made up of a hot gas halo mostly filling the virial sphere, while clumpy plumes of cold gas are visibly being stripped from satellites passing through this hot atmosphere. The addition of galactic winds substantially changes the amount, distribution, and state of cold gas in the halo. The CGM gas in the Galactic-Winds simulation is more well-mixed due to the stirring action of galactic winds -- individual satellites within $r_{\rm vir}$ are hardly visible, as their outflows have distributed formerly dense ISM gas over a comparatively large volume, lowering the effective density contrast. This is nowhere more clear than with the central galaxy itself, whose galactic winds fill the inner halo (i.e. $r \lesssim r_{\rm vir}/4$) with high column density ($N > 10^{22} \,\rm{cm}^{-2}$) gas. As a large fraction of this `recent wind' material is cold, it lowers the mean temperature of the inner hot halo, as evidenced in the temperature projections (right panels).

Unfortunately, neither the temperature of the gas nor its total column (or mass) density are directly accessible through absorption line based observations. We therefore decompose the total gas content into two commonly observed phases. \fig{images} also shows the corresponding atomic hydrogen (HI) column density and MgII equivalent width (EW) maps, both generated using line-of-sight velocity cuts of $\pm 1000$ km/s.  Both of these species arise in cool ($T \sim 10^4$), dense ($n > 10^{-3}$ cm$^{-3}$) gas, so these column density and EW projections especially highlight the cool-dense gas phase which we focus on in this work. The HI content of each gas cell in the simulation is derived following the method of \cite{bird13}, computing neutral hydrogen abundances using the prescription from \cite{rahmati13}. The MgII maps were generated by using \textsc{Cloudy} \citep{ferland13} to compute MgII ion fractions, then computing synthetic spectra by accumulating contributions from each gas cell intersecting every line of sight \citep{bird17}, summing up the total equivalent width in a $\pm$ 1000 km/s velocity window. Note that the cold ISM gas in the galaxy itself is modeled with the effective equation of state of \cite{spr03}, so we do not have the density/temperature information needed to model MgII in this gas and as a general rule exclude star-forming gas from such ionic calculations.

Similar to the re-distribution of the total gas density, we see that the galactic winds disperse both HI and MgII far outside of the dense ISM of galaxies themselves. These tracers of relatively cold and dense gas fill the inner halo ($r/r_{\rm vir} \lesssim 0.5$) with column densities $N_{\rm HI} \gtrsim 10^{20}$ cm$^{-2}$ and equivalent widths EW $\gtrsim 1$\,\r{A}, respectively.

\begin{figure*}
  \includegraphics[width=6.0in]{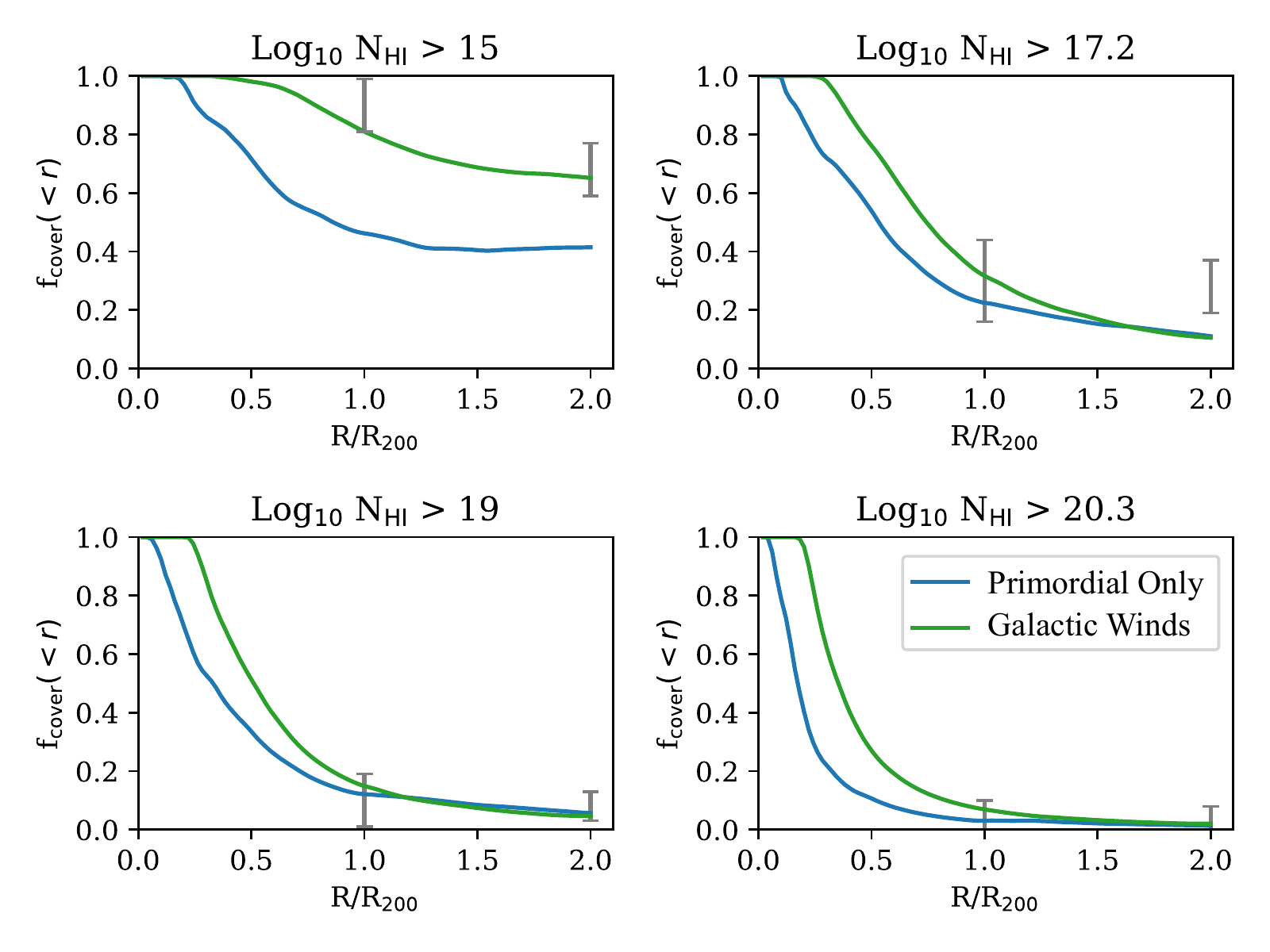}
  \caption{HI covering fractions for four different column density thresholds, compared to observational data at $z\sim 2$ for $\sim 10^{12} M_\odot$ halos from \protect\cite{rudie12}, which is shown in gray errorbars. Different colored lines show our simulations: Primordial-Cooling Only (blue) and Galactic-Winds (green). The Galactic-Winds simulation has covering fractions of neutral gas within $r_{\rm vir}$ which are appreciably enhanced above the levels seen in the cooling-only simulation, an effect which makes them roughly consistent with the observations.
 \label{fig:rudie}}
\end{figure*}

To quantify the difference between the atomic hydrogen distributions \fig{rudie} compares the HI covering fractions at $z=2.25$, plotting the radial profiles of each. The four panels show $f_{\rm cover}(r)$ for four different $N_{\rm HI}$ column density thresholds, from low 10$^{15}$ cm$^{-2}$ columns, to Lyman-limit system (LLS) equivalents with $> 10^{17.2}$ cm$^{-2}$ through damped Lyman-alpha systems (DLAs) at $> 10^{20.3}$ cm$^{-2}$. At all column density limits, the Primordial-Only simulation has similar covering fractions as a function of distance, dropping rapidly from unity at $r \lesssim [0.1-0.2] r_{\rm vir}$ to approximately 20\% (5\%) at $r_{\rm vir}$ for $\log(N_{\rm HI}) >$ 17.2 (20.3) cm$^{-2}$. In contrast, the Galactic-Winds simulation (green line) exhibits higher covering fractions regardless of $N_{\rm HI}$ threshold or radius. At the lowest column densities, this enhancement is as much as 50\% over the cooling-only simulation. For LLS columns and above this effect is closer to 10\% within the halo. This enhancement exists despite the lower overall (spherically averaged) radial number density profiles seen in \fig{CGM_frac_R}, due to the larger area covered by effectively higher column HI. At large distances ($\gtrsim$ 1.5$r_{\rm vir}$) both simulations converge to similar covering fractions, except at the lowest threshold of $N_{\rm HI} > 10^{15}$ cm$^{-2}$ where the Galactic-Winds case remains enhanced.

In \fig{rudie} we also include a comparison to observations of $\sim 10^{12} M_\odot$ halos at $z \sim 2$ from \cite{rudie12}, who measured this quantity around star-forming galaxies. Note that our halo may be slightly smaller at $z=2.25$ ($M_\text{halo} \sim 10^{11.6} M_\odot$) than the observed LBGs \citep[$M_\text{halo} \sim 10^{12} M_\odot$, although see][who determine a somewhat smaller mass]{rakic13}. The cooling-only simulation has significantly less lower-column HI gas than observed, with covering fractions low by as a much as a factor of two. On the other hand, the simulation with galactic winds alleviates this tension and is consistent with the observations, thanks to the much flatter decline of $f_{\rm cover}(r)$ for low columns \citep[see also][]{fg16}. As in \cite{suresh15}, we note a slight discrepancy between both simulations and the observational data point for the covering fraction of LLSs ($N_\text{HI} > 10^{17.2}$ cm$^{-2}$) at twice the virial radius \citep[this is likely only due to the missing contribution from other structures along the line of sight, which are not represented in our zoom simulations, e.g.][]{rahmati15,nelson18b}. Nevertheless, the HI covering fractions within the virial radius, which is our focus herein, are consistent with the available observational constraints across the full range of column density thresholds for the Galactic-Winds run. We note that at $r_{\rm vir}$ and beyond there is, at present, little discriminatory power in the covering fractions of LLS columns and above, as even the most discrepant models -- the cooling only cases, which fail to produce reasonable galaxy stellar masses for example -- cannot be distinguished from the simulation with the more realistic Illustris galactic winds model.

\begin{figure}
  \includegraphics[width=3.4in]{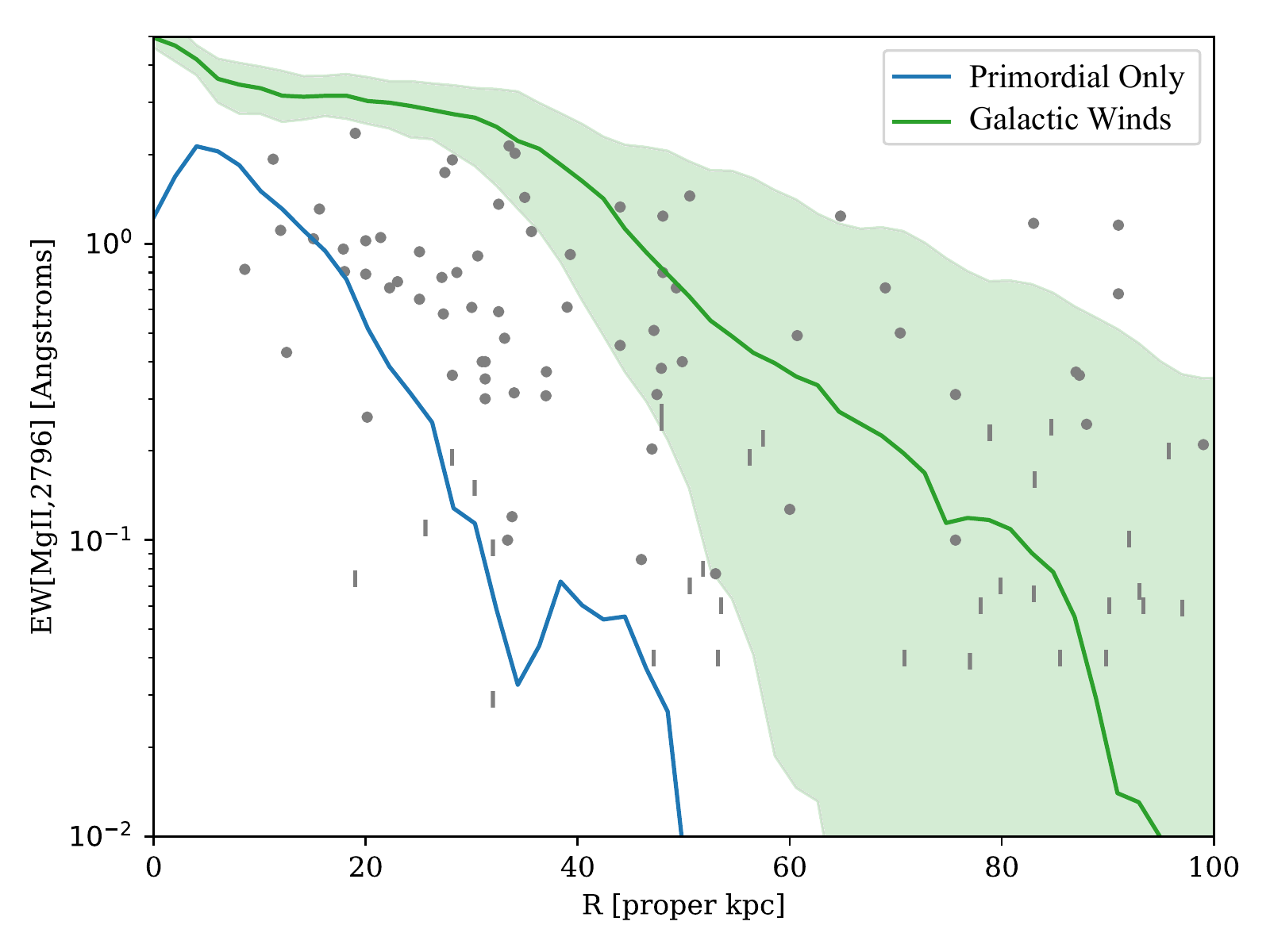}
  \caption{The MgII (2796\r{A}) equivalent width as a function of impact parameter, compared to observational data (different symbols) from lower redshifts taken from \protect\cite{barton09,chen10,werk13}. Different colored lines show medians for our simulations: Primordial Cooling Only (blue) and Galactic Winds (green). The colored band gives the 10-90th percentile scatter. Enhancement of the MgII EW in the Galactic Winds simulation is significant -- an order of magnitude effect at $r_{\rm vir}/2$ which persists all the way to the virial radius and beyond.
  \label{fig:MgIIprof}}
\end{figure}

Going beyond hydrogen and considering a relatively low ionization energy metal ion, \fig{MgIIprof} compares the MgII equivalent width profiles at $z=2.25$ with observations taken at lower redshifts ($z<1$). Note that this plot is meant to be \textit{suggestive only}, since the difference in redshift is considerable, and furthermore, our simulated galaxy does not necessarily represent the typical system in these observations. Nonetheless, the differences between the simulations are quite instructive, and the Galactic-Winds equivalent width results intriguing even at high redshift. 

As with HI, \fig{MgIIprof} shows that the MgII profiles for the cooling-only simulation are significantly less extended than the Galactic-Winds run, as expected. The extent of MgII outside of galaxies in particular may represent an interesting, observationally accessible constraint on feedback models \citep[e.g.][]{kauffmann17}. While the cooling-only simulation has equivalent widths exceeding $\sim$\,1\,\r{A} only within 20 kpc, this is true for the simulation with winds out to nearly 50 kpc. The relative enhancement in EW is an order of magnitude at half the virial radius, and this increase continues even beyond $r_{\rm vir}$. Interestingly the MgII profile at $R \gtrsim 40$ kpc in the Galactic-Winds simulation would be broadly consistent with the $z<1$ data. Conversely, the simulated profile is significantly higher than the low-redshift observations for $R \lesssim 40$ kpc, which is where the galactic winds begin to dominate the cool-dense gas mass profile, as we explore below in Section~\ref{sec:coldphase}. Galactic outflows are significantly stronger at $z\gtrsim2$ than at $z<1$, both in simulations and observations, so it is unsurprising that the MgII EW is stronger than observed at low redshift. 

Taken together, \fig{rudie} and \fig{MgIIprof} demonstrate that the Galactic-Winds simulation may plausibly produce a reasonable distribution of cold, dense gas in the CGM. This tentative and somewhat speculative agreement with observational expectations lends at least some confidence to our further exploration into the origin of this gas.

\begin{figure}
  \centering
  \includegraphics[width=3.35in]{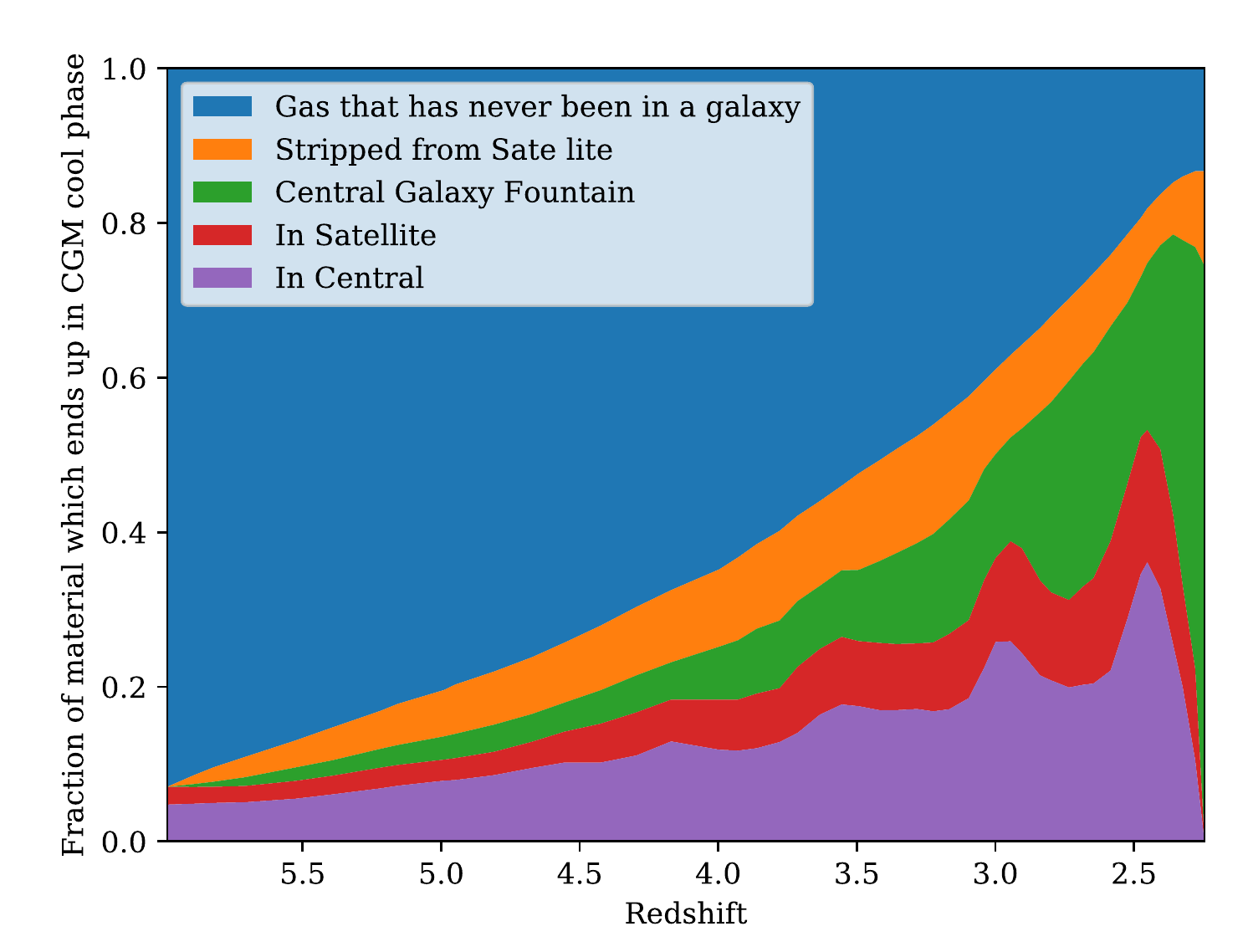}
  \caption{Breakdown of gas which will end up in the cold CGM phase at $z=z_f=2.25$ as a function of redshift. By the final redshift $z_f$, when including galactic winds, approximately 3/4 of the cool-dense CGM phase is gas which has, at one point, been in the central galaxy and is now expelled through winds. The remaining quarter is split equally between primordial, 'smooth' cosmological accretion and gas stripped from satellite galaxies.
  \label{fig:cg_tracer_fractions}}
\end{figure}


\section{Tracing the Origins of the Cool-Dense CGM Phase}
\label{sec:coldphase}

\subsection{Accretion History}
\label{sec:acc}

Using the Monte Carlo tracers (see description in \sect{tracers}; there are on average 20 tracers per gas cell or star particle) we directly identify the mass which will give rise to the cold CGM phase, and track its thermal and dynamical history back in time. Noting that the final redshift is $z_f = 2.25$, we define the following tracer categories:

\begin{itemize}
\item `Primordial' is material that has never been inside a galaxy by $z_f$. This is intergalactic baryonic mass.
\item `Fountain' is material that was once part of the central galaxy, and is in the CGM at $z_f$, having been ejected at some point either dynamically or directly by our wind model.
\item `Stripped' is material that was once part of a satellite (but never the central), and is in the CGM at $z_f$, having been stripped from its original system.
\end{itemize}

The second category could also be called `central galactic fountain', and includes gas ejected in the distant past as well as gas just entering the CGM from the wind at $z_f$. In addition to studying if and when material has been inside of a galaxy, we can also quantify the contribution of the hot halo cooling into the cold CGM by further decomposing the `Primordial' category into two sub-categories, depending on previous heating. We use a threshold on the ratio of the max past temperature of the tracer particle, $T_\text{max}$, and the virial temperature $T_\text{vir}$ of the halo at that time. If $T_\text{max}/T_\text{vir} < 0.5$ we refer to this gas as `unvirialized', otherwise it is `virialized' \citep{nelson13}.

Given these definitions, \fig{cg_tracer_fractions} shows a mass budget of the material which ends up in the cold phase of the CGM. We focus here, and below, on the Galactic-Winds simulation. We select all the gas in the cool-dense phase at $z_f$ and show the evolution of that material at earlier times. For instance, roughly 15\% of the cool-dense CGM has just originated from cosmological accretion, `gas that has never been in a galaxy', while about as much ($\sim$\,15\%) has been likewise accreted after previously having belonged to a satellite. The majority, $\sim$\,75\%, has joined this phase of the CGM as a direct result of the central galactic fountain, about a third out of which has been deposited since the previous snapshot, within the last $\sim$ 50 Myr.\footnote{Since $z=2.25$ is our reference snapshot, all of the cold CGM gas at this snapshot is, by construction, either primordial, fountain, or stripped. If any gas has been in \textit{both} a satellite and a central galaxy, and is now not associated with any galaxy, it is defined as being part of the galactic fountain.}

Of the $\sim$\,15\% component arising from `smooth' cosmological accretion, roughly half has been shock heated and half has remained unvirialized. Of the total, the vast majority originates from the central galactic fountain, which accounts for the origin of approximately 3/4 of the cool-dense CGM gas. We note that the cooling-only run exhibits behaviors that are distinctly different from the Galactic-Winds simulation. In particular, in the cooling-only simulation $\sim 60 \%$ of the cold CGM phase arises from gas that is `primordial' (never been part of a galaxy), and about $25\%$ has been stripped from a satellite (at $z_f=2.25$). The remaining $15\%$ is gas which was at one point associated with the central galaxy. This fits the general picture that, in the absence of galactic feedback, it is difficult for gas to escape once it has entered any galaxy. Gas is more easily stripped from the orbiting satellites than the central galaxy at the center of the potential well, which helps the small satellites contribute to the cold phase. Conversely, when winds are turned on, much of the material which passes through the central is now able to cycle back out. 

\begin{figure}
  \centering
  \includegraphics[width=\newFigurewidth]{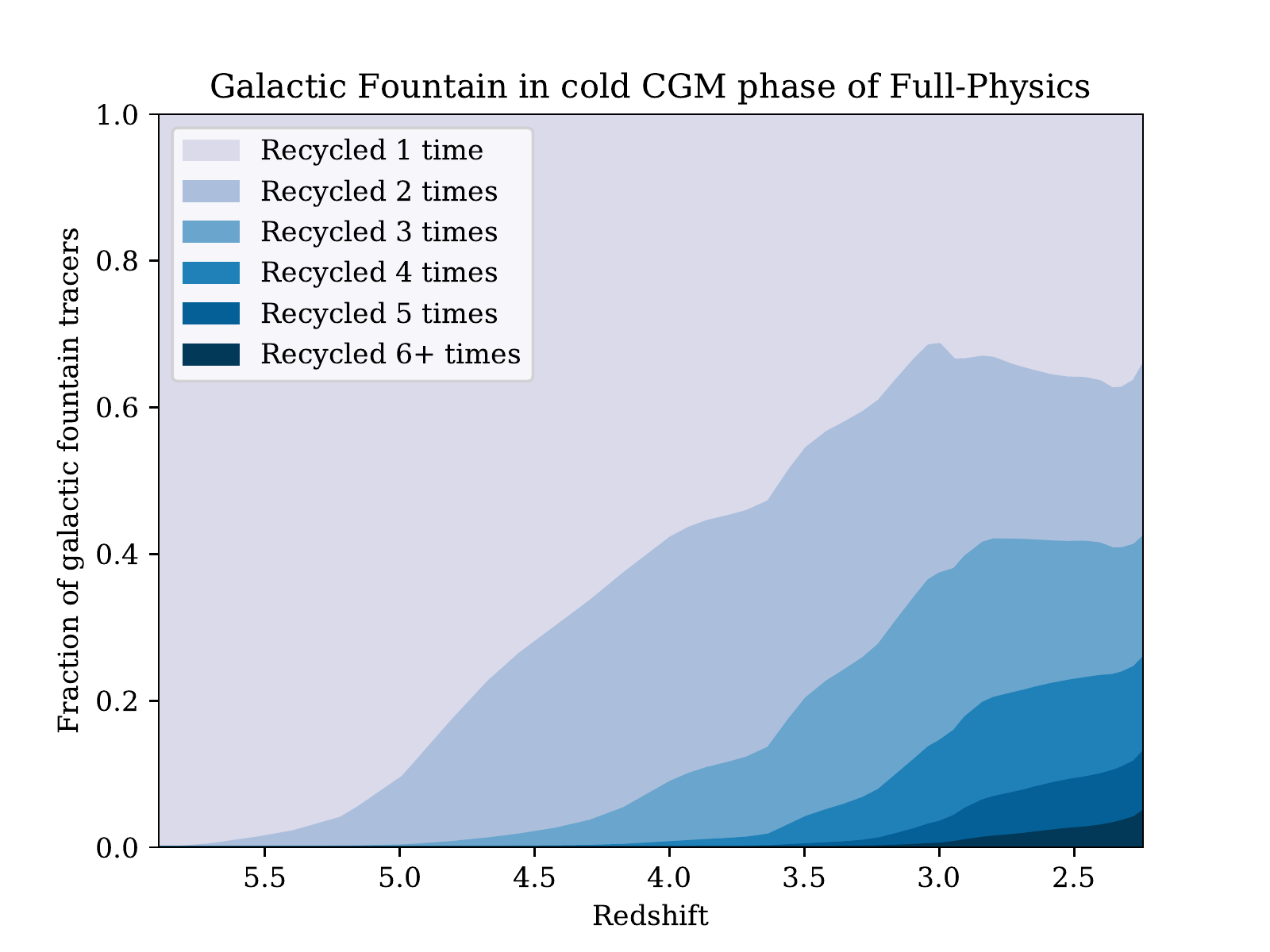}
  \caption{The number of times each tracer of `galactic fountain' material has cycled through the central galaxy, as a function of redshift. By $z=2.25$, $\sim 60\%$ of the galactic fountain has been through the galaxy more than once, and $\sim 25\%$ has cycled through the central galaxy four or more times. Only $\sim$\,1/3 of gas in this phase has been in a galactic wind just once.\label{fig:fp_fountain}}
\end{figure}

The vigor of the galactic fountain flow is related to a recycling timescale -- that is, what fraction of ejected gas returns to the central galaxy and on what timescale. Relatedly, we measure the multiplicity of outflow events for individual parcels of gas. \fig{fp_fountain} shows how many times the gas which is in this galactic fountain which contributes to the cold CGM phase at $z=2.25$ has been recycled. At the end of the simulation, only $\sim$\,1/3 of gas in this phase has participated in a galactic outflow just once, while $\sim 60\%$ has cycled through the galaxy two or more times. The profusion of multiple-outflow tracers builds up with redshift; at $z \gtrsim 5$ there hasn't yet been enough time for any baryons to participate in the cycle more than once. This strong recycling is a signpost of the `ejective' nature of the galactic wind model, which acts primarily by removing cold, dense gas from the ISM, rendering it unable to form stars for some period of time. The non-negligible wind mass-loading at injection ($\eta = \dot{M}_{\rm wind} / \dot{M}_\star \simeq 2-3$ at this mass and redshift scale) implies that a significant amount of gas mass is shifted from the ISM to the inner CGM. \cite{anglesalcazar17} present similar recycling event distributions for the FIRE model (although at $z=0$), and in rough comparison it appears that we have somewhat more frequent recycling (i.e. more recycling events implying shorter recycling timescales).

\begin{figure}
  \centering
  \includegraphics[width=3.35in]{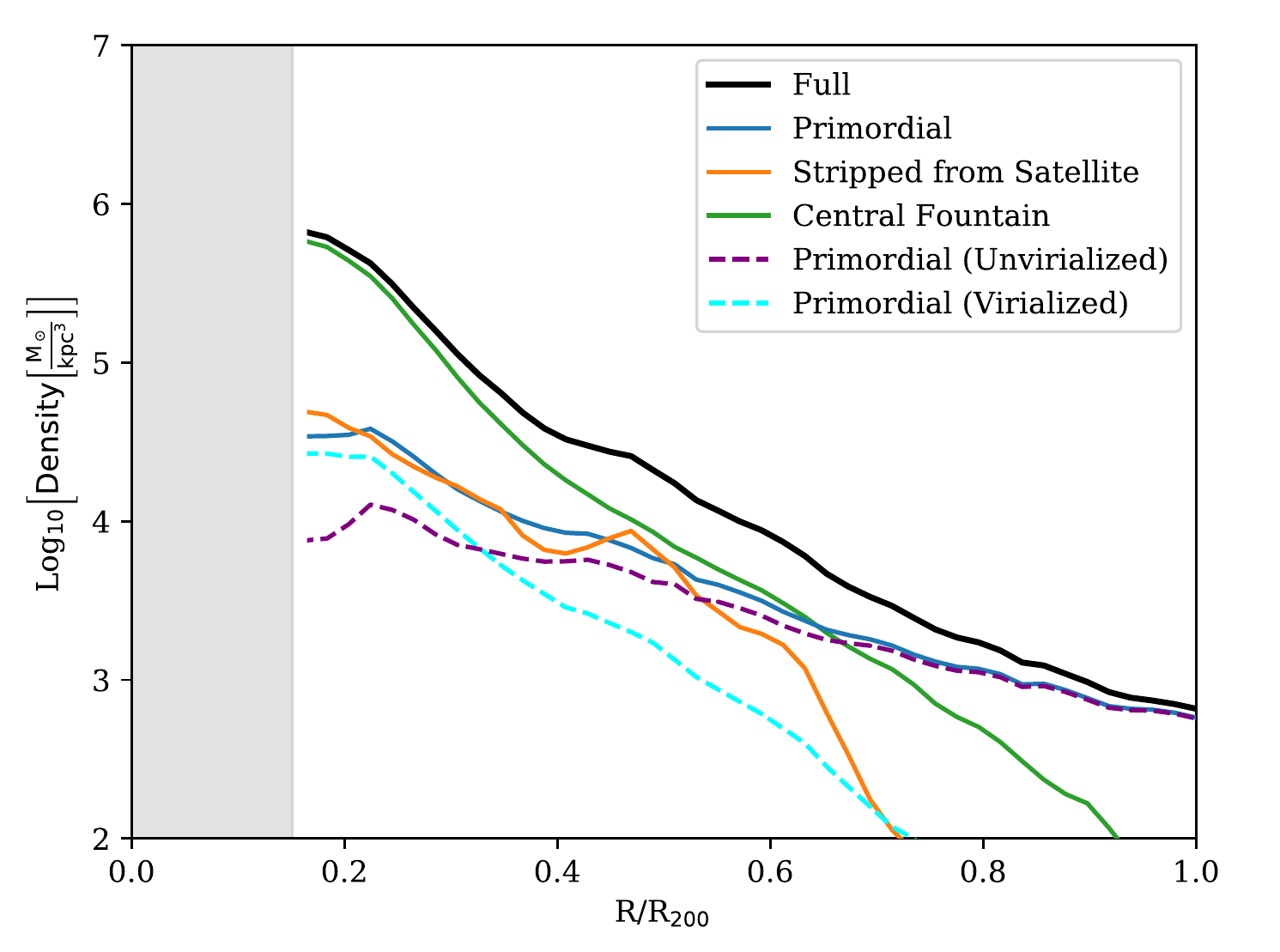}
  \caption{Mass profiles of the cool-dense CGM at $z=2.25$ for the Galactic-Winds run, broken down by the various production channels. The strong galactic fountain material is the most significant component out to $\sim 0.65 \,r_{\rm vir}$ ($\sim 45$ pkpc), beyond which primordial gas begins to dominate. Cool-dense gas which has cooled out of the hot halo and gas which has been stripped from satellites are, at every radius, sub-dominant to cool-dense gas from the galactic fountain.
  \label{fig:mprof_cg}}
\end{figure}


\subsection{Density and Kinematics}
\label{sec:denskin}

In \fig{mprof_cg} we therefore show the mass profile of the cold CGM phase at $z=2.25$, broken down by the production channels defined above. In the Galactic-Winds simulation a strong galactic fountain dominates out to $\sim 0.65 \,r_{\rm vir}$ ($\sim 45$ proper kpc), beyond which primordial gas becomes the largest component. While the total density and slope of the inner halo are set by fountain gas, they are likewise set by unvirialized primordial gas in the outer halo. Cool-dense gas which has cooled out of the hot halo -- the `primordial (virialized)' channel (light blue line) -- and gas which has been stripped from satellites (orange line) are both, at every radius, sub-dominant to cool-dense gas originating in the galactic fountain (green line). Interestingly, the shape of the fountain gas and the virialized primordial (hot halo cooled) gas profiles are similar, the latter consistent with a simple scaling down by roughly one order of magnitude in density. This suggests a direct relationship between the two, whereby wind material interacting with the dense gas of the inner halo produces a rapidly cooling hot halo phase. We explore this idea of `wind stimulated cooling' and its implications for the presence of cold gas in the CGM below.

\begin{figure}
  \centering
  \includegraphics[width=\newFigurewidth]{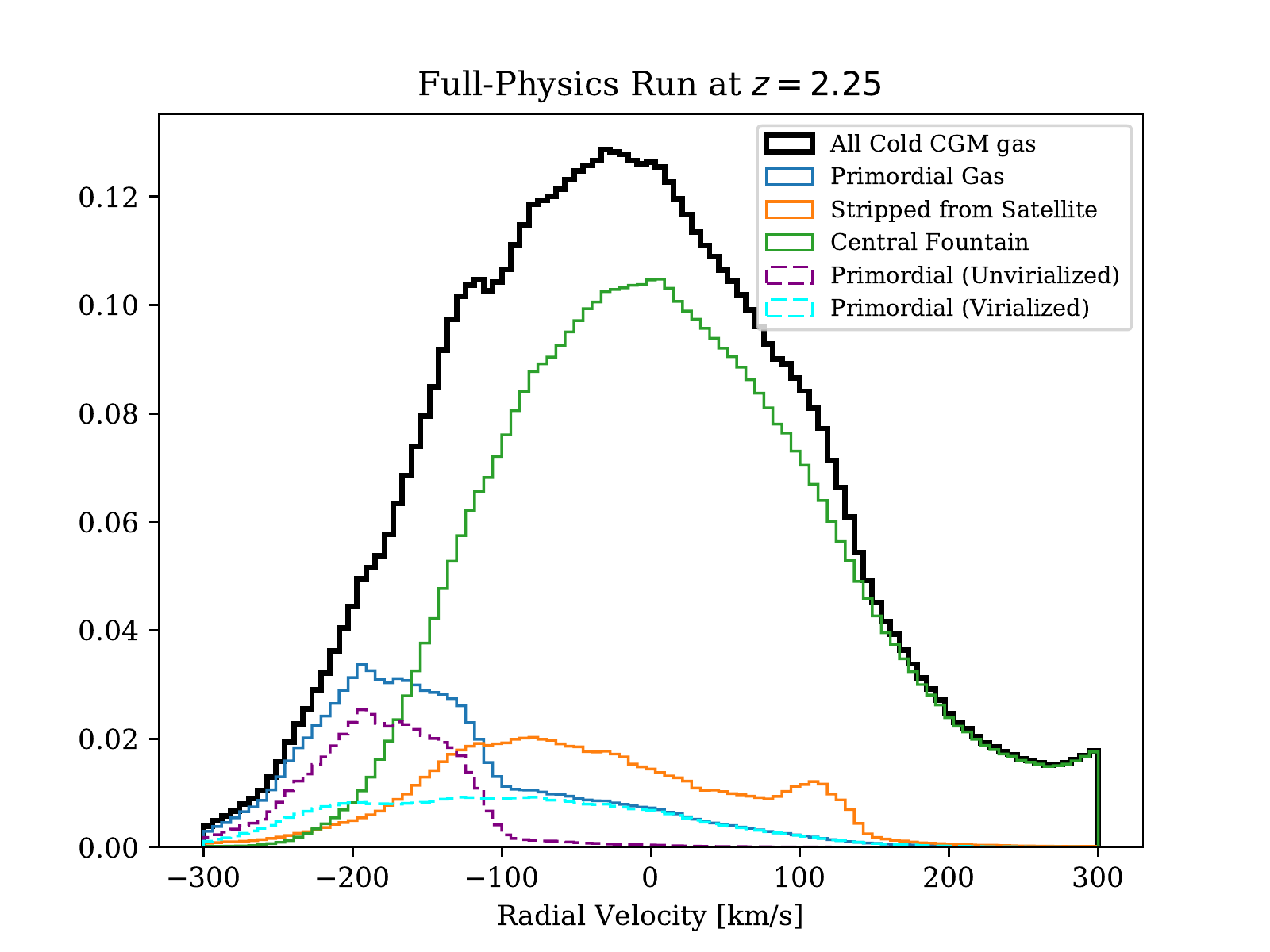}
  \caption{Kinematics: radial velocity distributions of the cold CGM gas at $z=2.25$ in the Galactic-Winds simulation, decomposed into the different production channels. Negative velocities denote gas which is infalling towards the central galaxy, while positive velocities are outflowing material. As expected, the vast majority of gas which is outflowing was produced by the galactic fountain, especially for $v_{\rm rad} > 100$ km/s. All other components are net inflowing, although each has distinct $v_{\rm rad}$ distributions, including the notably different behaviors of virialized and unvirialized primordial gas.
  \label{fig:vrhist}}
\end{figure}

Beyond density and temperature, we next consider the kinematics of the cool-dense phase. \fig{vrhist} shows the distribution of cold CGM gas radial velocities at $z=2.25$ for the Galactic-Winds simulation, binned from -300 km/s (inflow) to +300 km/s (outflow). We again decompose this phase into its different origins, according to the same definitions. The dominant galactic-fountain component demonstrates the existence of strong (significant mass) outflows with typical velocities of $\sim$ 100-200 km/s. Indeed, essentially all gas which is outflowing has been produced by the central fountain, especially for $v_{\rm rad} > 100$ km/s. The fountain origin material is the dominant component of the cool-dense phase by mass. It also has a far more symmetric radial velocity profile than any of the other components, indicative of a near equilibrium balance between inflow and outflow through the halo \citep[i.e. leading to nearly zero net inflow available to the galaxy;][]{nelson15}.  

Interestingly, the kinematics of primordial-origin cool-dense gas depends strongly on past heating history. Unvirialized primordial gas (purple line) is unambiguously inflowing, with an average velocity between -150 and -200 km/s, i.e. approximately equal to the circular velocity of the halo. In contrast, cool-dense gas with a virialized primordial origin (light blue line) is both inflowing and outflowing, with a broad and nearly flat $v_{\rm rad}$ distribution between -250 km/s and 100 km/s. This tail towards less negative radial velocities hints that the `primordial (virialized)' gas is cosmic accretion mixed with, or influenced by, $v_{\rm rad} > 0$ material -- i.e. the outflowing winds. Accretion shocks could likewise decrease the inflow speed and so make $v_{\rm rad}$ less negative.


\subsection{Metallicity}

\begin{figure}
  \centering
  \includegraphics[width=\newFigurewidth]{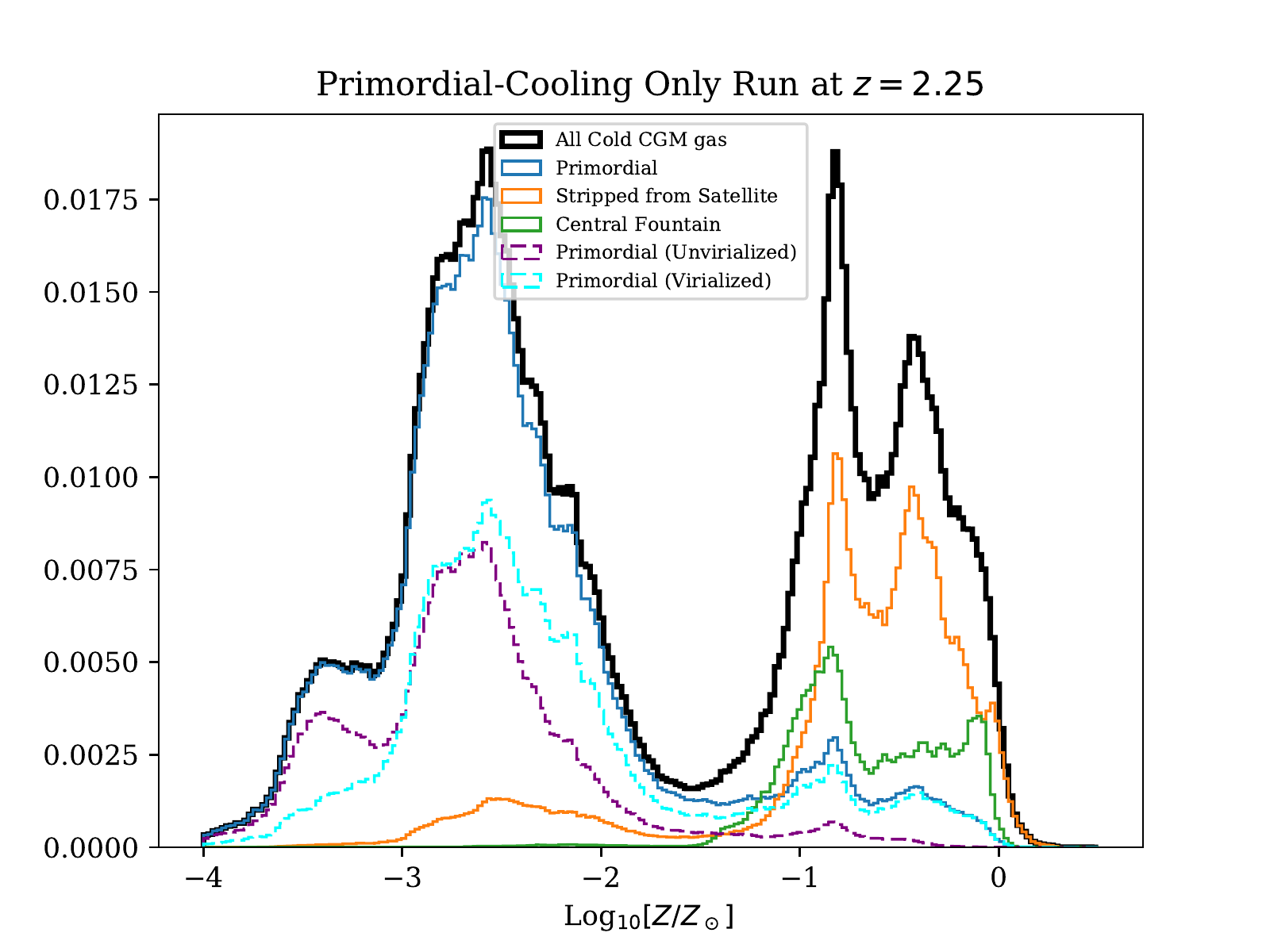}
  \includegraphics[width=\newFigurewidth]{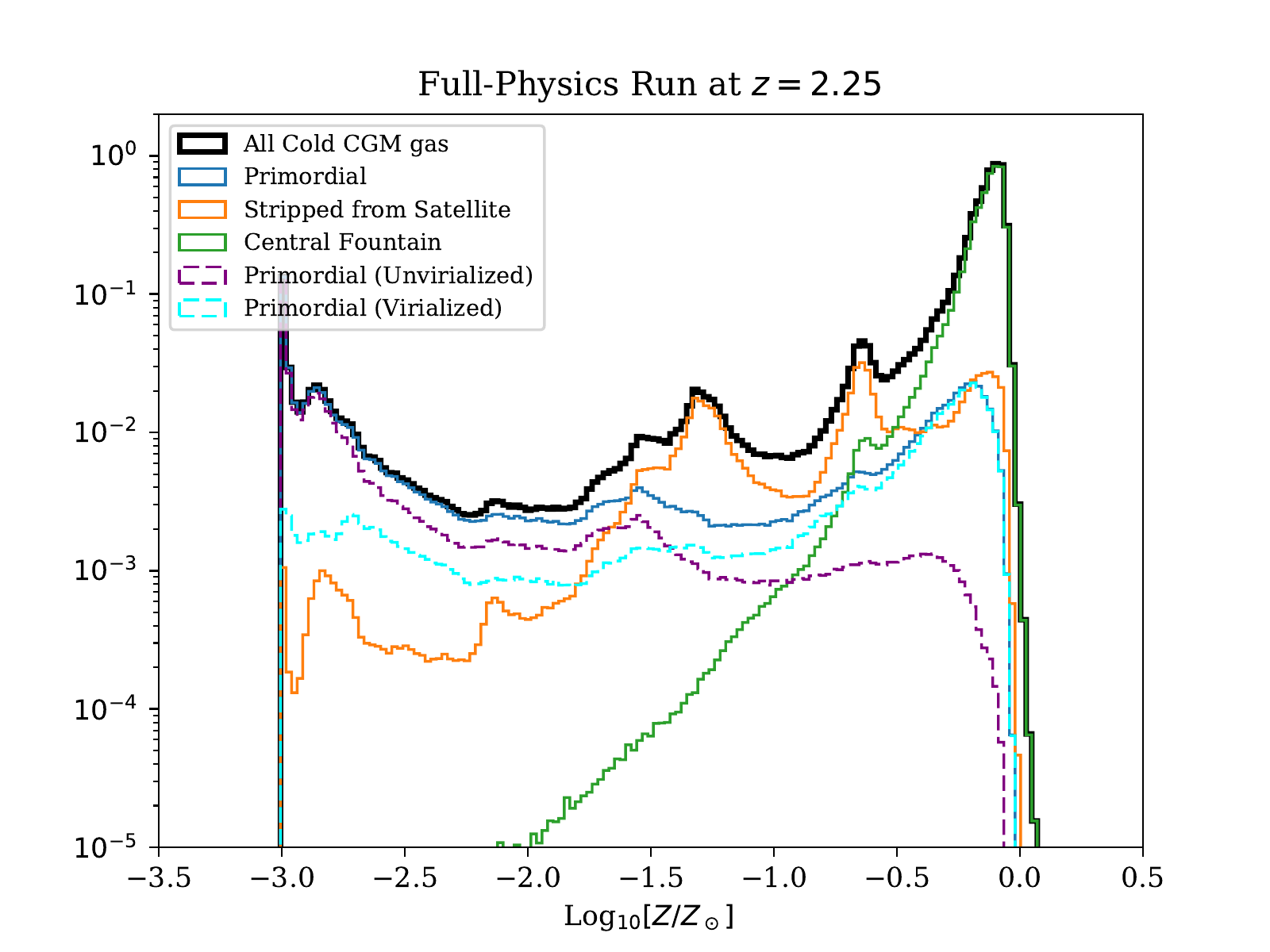}
  \caption{The metallicity distribution of the cold CGM phase, split by production channel. Both of our simulations show a metallicity bimodality between gas that has been inside a galaxy versus primordial gas, particularly the cooling-only run. The former, arising from stripped gas or the central fountain, has $Z/Z_\odot \gtrsim 1\%$. Primordial inflow is largely restricted to lower metallicity in the cooling-only run, flattening considerably in the Galactic-Winds case due to enhanced enrichment prior to infall.
  \label{fig:zhist}}
\end{figure}

Traditionally, in either simulations or observations, the enrichment level of gas provides a tempting target for the characterization of inflow versus outflow. \fig{zhist} therefore addresses the metallicity of cool-dense CGM gas at $z=2.25$, again split by production channel -- the top panel shows the primordial cooling-only run, while the bottom shows the Galactic-Winds run. The metallicity distribution shows a bimodality in all simulation variants, and is particularly strong in the cooling only cases. Specifically, gas which has at some point been inside a galaxy typically has metallicities $Z/Z_\odot > 10\%$ (orange and green lines), while gas which has never entered a galaxy typically has metallicities $Z/Z_\odot \lesssim 1\%$ (purple and blue lines). These two classes of gas are nearly disjoint in metallicity space in the cooling-only run, although we see that a small mass fraction of `primordial' gas is enriched to 0.1$Z_\odot$ while an even smaller fraction of stripped infall remains below 0.01$Z_\odot$. This strong separation in $Z$ is much less pronounced in the Galactic-Winds case, where more enrichment of even `primordial' gas can occur prior to infall due to mixing and pre-enrichment. Note that the quantitative values of these dividing lines largely separating inflows and outflows are only appropriate in the regime of $M_{\rm halo} \sim 10^{12}$\msun at $z \sim 2$. This bimodal distribution is suggestively similar to the bimodality in LLS metallicities observed by \cite{lehner13,wotta16} at $z \sim 1$, who found two peaks centered at $\sim 2.5\%$ and $\sim 50\%$ solar metallicity. At our higher redshift these values should naturally shift lower.

As an important caveat, the role of mixing in CGM gas including the impact on e.g. metallicity distributions of LLS absorbers, is an issue which requires further study. In particular, due to the finite resolution of numerical simulations, some level of numerical diffusion (mixing) is unavoidable, and it is unclear if, at present resolutions, this is comparable or not to the expected amount of physical mixing. Understanding the role of mixing in setting CGM properties, and the influence of CGM refinement on mixing processes, will be an important topic for future work.

\begin{figure}
  \centering
  \includegraphics[width=3.3in]{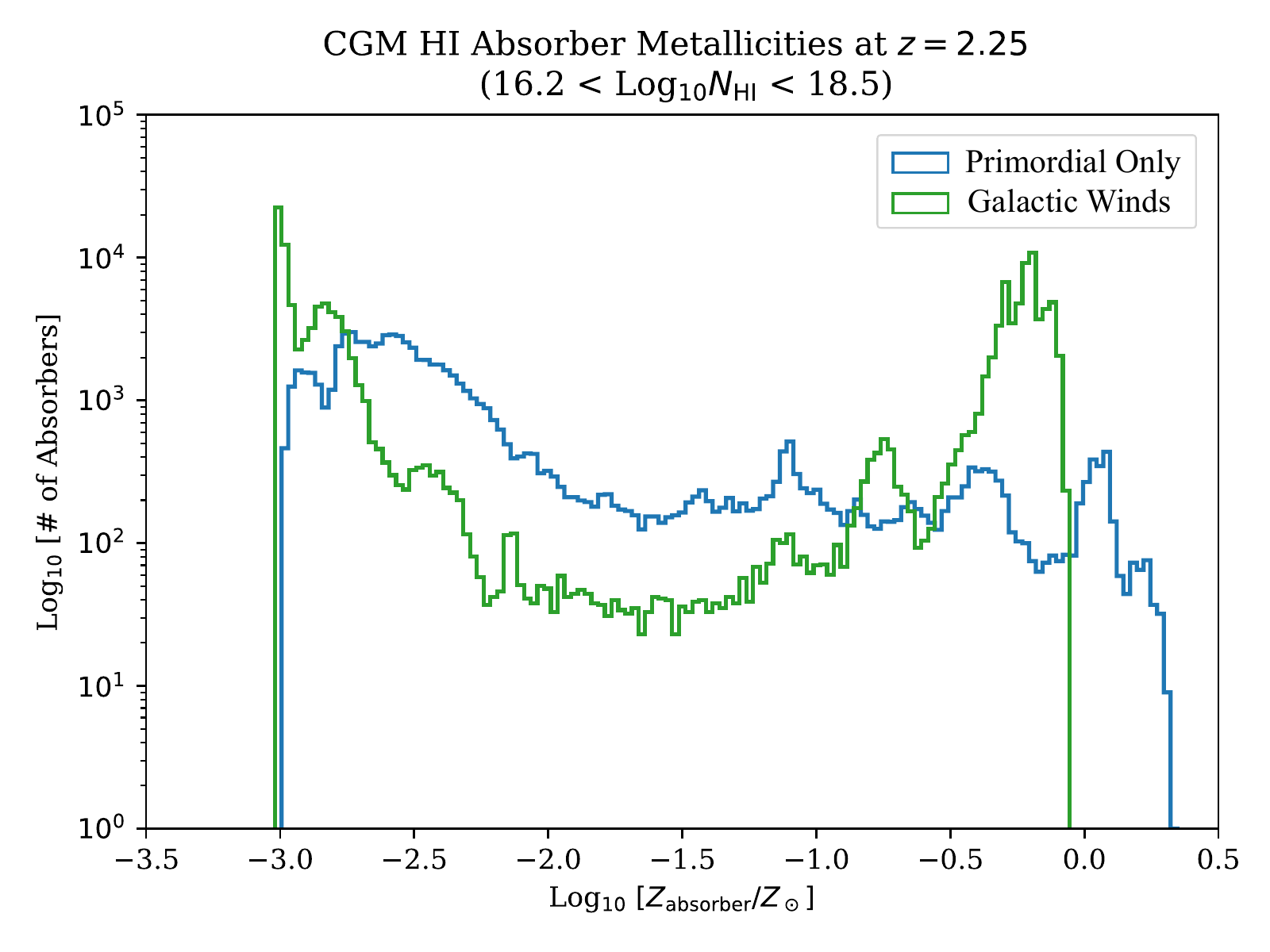}
  \caption{Metallicity distribution of HI absorbers with impact parameter $r_\text{2D} < r_{\rm vir}$, restricted to $10^{16.2} \text{cm}^{-2} < N_\text{HI} < 10^{18.5} \text{cm}^{-2}$. This column density range is chosen for comparison with the bimodality identified in $z \lesssim 1$ HI absorbers \protect\citep{lehner13,wotta16}. Metallicities span a large range, from one thousandth solar to super-solar values. The Galactic-Winds simulation shows the strongest bimodality, with two distinct peaks: one at low, `primordial' (IGM) metallicities of $\lesssim 1\%$ solar, and the second at high, `enriched' (outflow) metallicities of $\sim$\,$Z_\odot$/2.
  \label{fig:LLS_z}}
\end{figure}

\fig{LLS_z} shows a slightly more direct comparison to the observations, considering the metallicity of HI absorbers in the column density range $10^{16.2} \,\text{cm}^{-2} < N_\text{HI} < 10^{18.5} \,\text{cm}^{-2}$ and impact parameters of $r_{\rm 2D} < r_{\rm vir}$, using the same line of sight velocity cut of $\pm 1000$ km/s. A strong bimodality is still visible in the HI absorbing sightlines in the Galactic-Winds simulation, where absorbers at the two extrema are about two orders of magnitude more common than at intermediate metallicities of $Z \sim 10^{-1.5} Z_\odot$. This bimodality is present in the cooling only run, although less pronounced, presumably because no outflows (other than pure dynamical motions) are present to redistribute metals out of the ISM and into these sightlines.

We note there are some strong caveats in this qualitative comparison with the observational data: first, we have not attempted to compute the metallicity directly from the spectrum as would be required observationally. Second, we have taken only sightlines which are within the virial circle of the central galaxy, whereas the observations are derived from an HI-selected sample. This may be minor, as the column density cut employed suggests that these absorbers are likely to reside within the CGM of galaxies (albeit with uncertain halo masses). Third, we have simulated here only one halo, and at higher redshift. Nevertheless, our finding tentatively suggests that the observed metallicity bimodality is revealing two largely distinct populations of gas: a high metallicity peak of gas which has been cycled through the central galaxy and ejected through a wind, and a low metallicity peak of cold halo gas which has never accreted into the ISM of a galaxy \citep[see also][who analyze LLS metallicities in EAGLE and do not find any clear bimodality]{rahmati17b}. We would anticipate this $z \sim 1$ observed dichotomy should be present and even more pronounced at $z \sim 2$.


\subsection{Thermodynamical Evolution}

\begin{figure}
  \centering
  \includegraphics[width=3.3in]{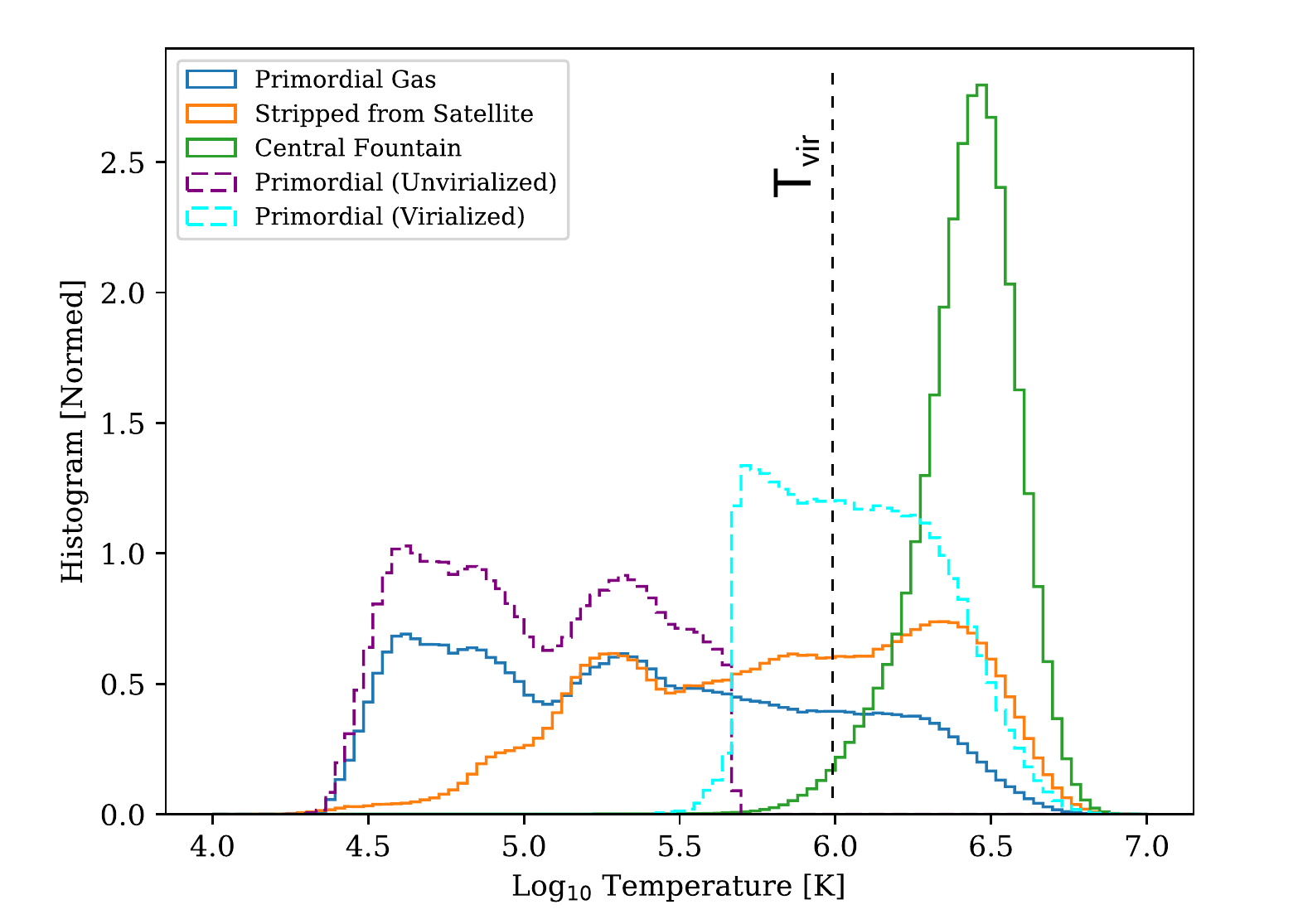}
  \caption{Maximum past temperature of all gas which is in the cold CGM phase at $z=2.25$, in the Galactic-Winds simulation. Note that the histograms are independently normalized to highlight the $T_{\rm max}$ distribution within each origin category. Essentially all of the gas which is in the central fountain reaches high maximum temperatures, symptomatic of the outflowing winds shocking on the existing quasi-static hot halo gas at relatively small radii.
  \label{fig:Tmax}}
\end{figure}

\begin{figure}
  \centering
  \includegraphics[width=\newFigurewidth]{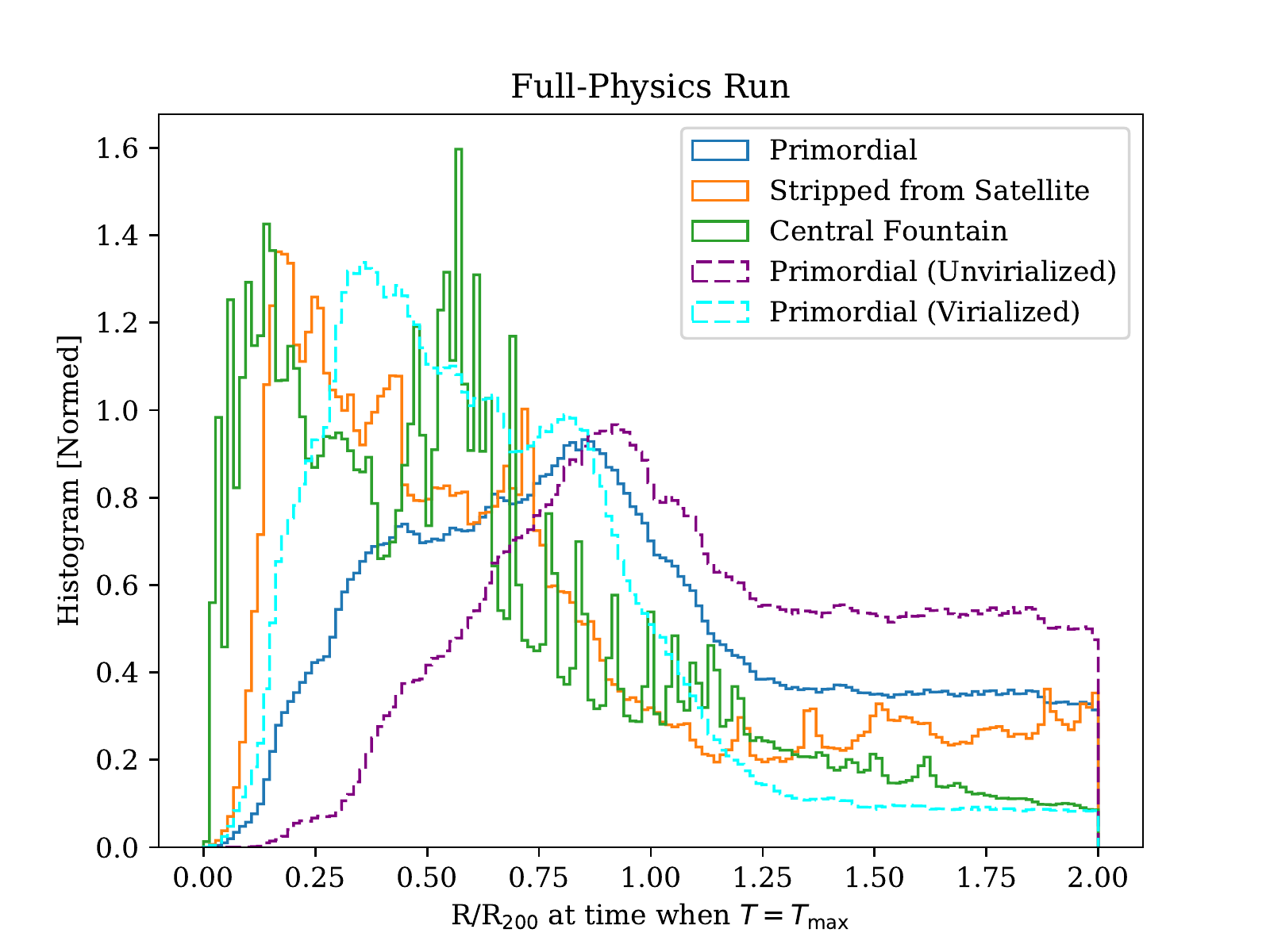}
  \caption{Radial distance between gas element and central galaxy at the time that the gas cell reaches its maximum temperature, normalized by $r_{\rm vir}$ at that time. All gas in the cold CGM phase at $z=2.25$ in the Galactic-Winds simulation is included. The histograms are independently normalized to highlight the distribution of each origin channel. Much of the galactic fountain reaches its maximum temperature very close to the galaxy, as winds shock immediately upon interacting with the hot halo gas, though there is a sizable tail to larger radii and evidence for a `double-shock' structure within the halo.
  \label{fig:r_at_Tmax}}
\end{figure}

We conclude with a look at the temperature history of gas, decomposed into each of the usual production channels, focusing as above on the $z=2.25$ cool-dense phase of the CGM. \fig{Tmax} shows the distributions of maximum past temperature $T_\text{max}$, while \fig{r_at_Tmax} gives the distribution of radii at which these $T_\text{max}$ events occurred, scaled to the virial radius of the halo at that time\footnote{Our knowledge of the maximum temperature is not limited by the snapshot time separation, as the tracers continuously, namely in every simulation time step, keep track of this quantity. In contrast, the location of any tracer at the time it sees its maximum temperature is not known to us exactly. Hence, its approximate value here is taken to be the location of the tracer at the first available snapshot after the maximum temperature has been recorded.}.

The maximum past temperature distributions of stripped (orange) and primordial (dark blue) gas are broad; the former is skewed towards higher temperatures, and the latter towards lower temperatures. Gas from these two origins also experiences heating at different locations within the halo. In particular, stripped gas appears biased towards heating at smaller radii, possibly due to its distinctive orbits and/or phase structure. Unvirialized primordial gas achieves its $T_{\rm max}$ in a narrow peak between $0.8-1.0$ times $r_{\rm vir}$. This clear virial shock signature implies that our $T_\text{max}/T_\text{vir} < 0.5$ criterion to distinguish unvirialized gas is an imperfect separation.

Gas above the $T_\text{max}=0.5T_\text{vir}$ threshold (light blue line) has either heated to higher temperatures at this same radius, or, more commonly, experienced additional heating within the halo at an inner shock. The presence of galactic winds modifies the radius at which this `primordial (virialized)' gas reaches its $T_\text{max}$. In the absence of winds, the primordial gas which does virialize typically reaches its $T_\text{max}$ at $r \sim 0.7 r_{\rm vir}$ (not shown), but when winds are turned on, much of this gas reaches its maximum temperature much closer to the galaxy, as seen in \fig{r_at_Tmax}. This feature may be expected based on the finding of \cite{schaal16} that high redshift galaxies in Illustris commonly have a double-shock structure. Similarly, \cite{nelson16} discuss how the radius of the virial shock preferentially moves towards smaller radii in directions experiencing rapid accretion across the virial sphere, albeit in simulations without winds. Here the outflows may also partially disrupt the stability of the outer virial shock and hence allow more of the primordial gas to enter the inner halo before being virialized.

Most interestingly, gas originating in the central fountain (green line), which constitutes the majority of the cool-dense CGM in the Galactic-Winds simulation, has high $T_\text{max}$ values, peaking at $\simeq 10^{6.5}$K, and even exceeding the virial temperature in many cases. There are two typical halocentric distances where this occurs: in the inner halo, at radii of order ten kiloparsecs ($\sim 0.1-0.25 r_{\rm vir}$) and just beyond half the virial radius at $\sim 0.5-0.7 r_{\rm vir}$. Here we measure $T_\text{max}$ over the full history of the tracers, hence these statistics represent a mix of maximum temperatures of individual tracers that could have been achieved either before or after they are in the wind phase.

\begin{figure*}
  \centering
  \includegraphics[width=6.9in]{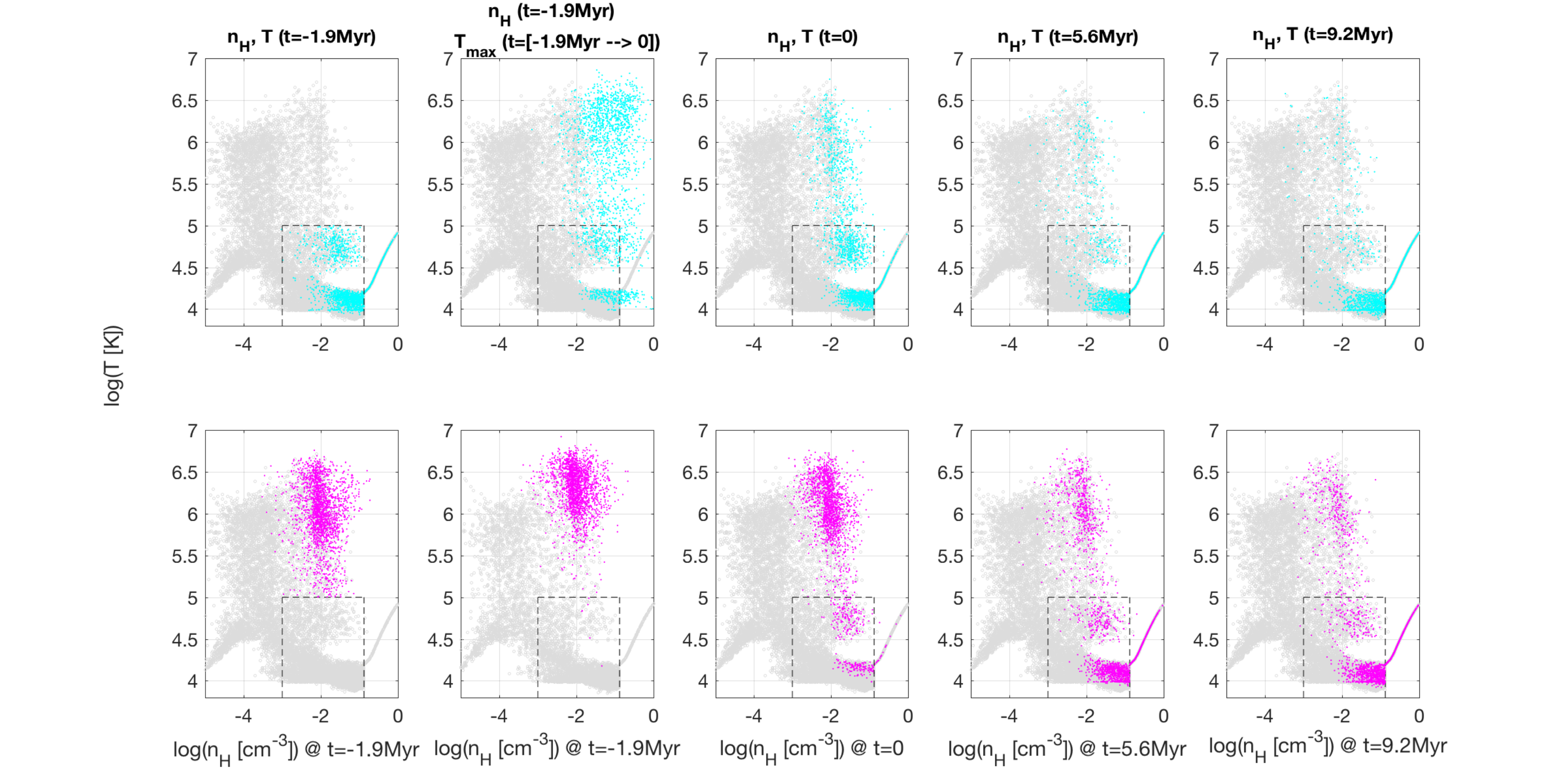}
  \caption{Thermodynamical evolution of gas due to wind recoupling. In all panels, gray shows the general distribution of gas cells on the phase diagram, and the dashed lines highlight our definition of the cool-dense phase. The colored points in each row represent a certain fixed selection of tracer particles, shown at different times in the different columns. The selection is for tracers that at $z=2.25$, defined here as $t=0$, reside in gas cells \textit{into which} wind-phase tracers have recoupled between $t=-1.9\Myr$ and $t=0$. The two rows separately show the evolution of tracers according to their temperature prior to recoupling: either \textbf{initially cold (top row)} defined as $T_{\rm t=-1.9\Myr}<10^5K$, or \textbf{initially hot (bottom row)}, with $T_{\rm t=-1.9\Myr}>10^5K$. The time evolution of the columns is: {\it Leftmost (first) column:} Pre-recoupling. {\it Second column:} Maximum temperature between $t=-1.9\Myr$ and $t=0$ versus pre-recoupling density. {\it Third column:} Post-recoupling. {\it Fourth column:} At $\sim6-7\Myr$ after recoupling. {\it Fifth column:} At $\sim9-11\Myr$ after recoupling.
  \label{fig:tracerevolution}}
\end{figure*}

\begin{figure}
  \centering
  \includegraphics[width=3.25in]{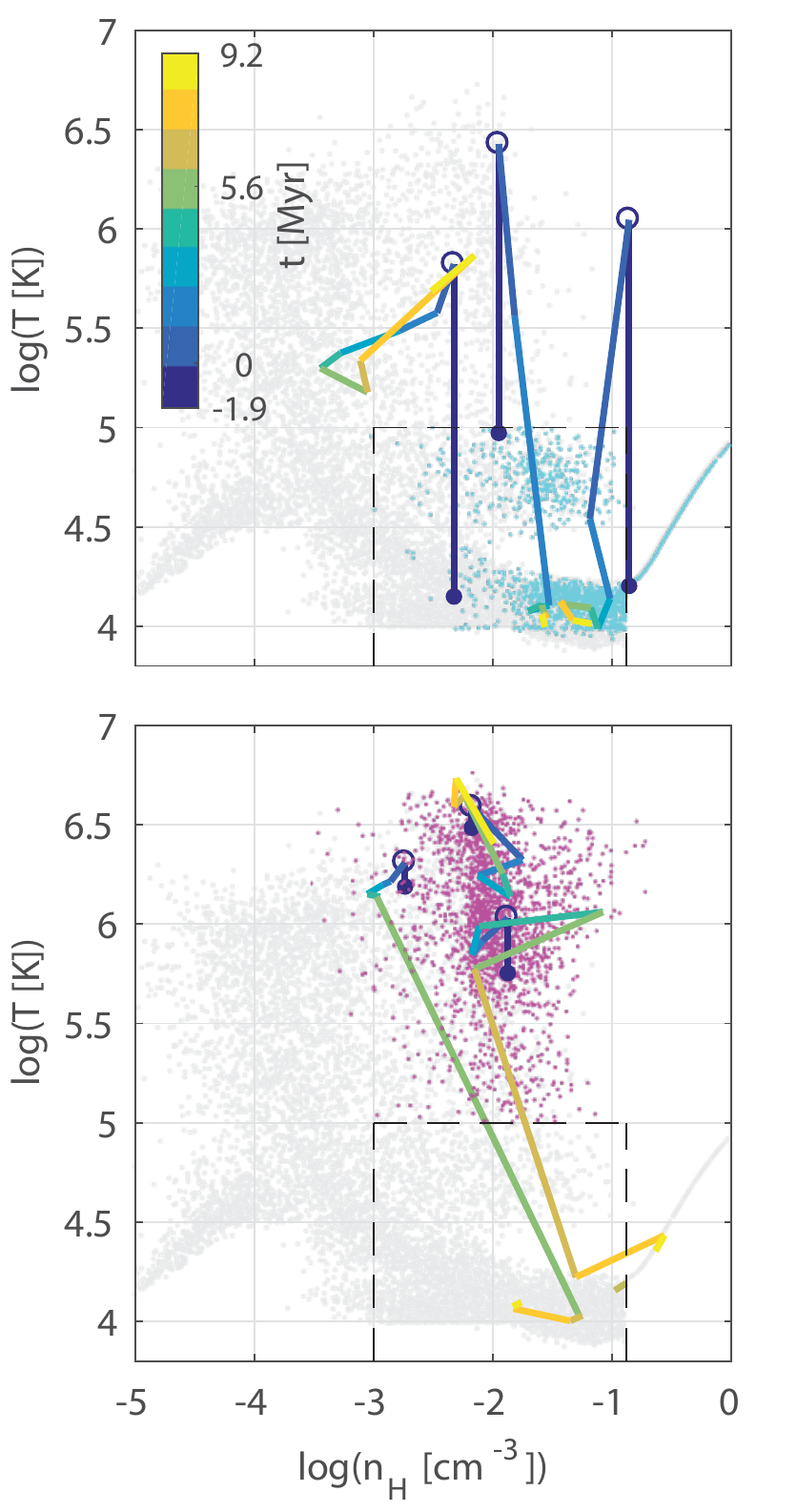}
  \caption{Evolution examples of individual tracer particles on the phase diagram due to recoupling. The background is identical to the left column in \fig{tracerevolution}, representing the overall gas distribution (gray) and tracers that reside in cold (top) or hot (bottom) gas cells that are about to experience a recoupling by a wind particle. Each track represents the evolution of a single `typical' tracer that starts from the colored point distribution prior to recoupling (t=-1.9\Myr, filled circle). The maximum temperature between $t=0$ and $t=-1.9\Myr$ is shown by the empty circle on each track. The color on the tracks and the colorbar indicate the time, in Myr, relative to the post-recoupling snapshot ($t=0$).
  \label{fig:tracertracks}}
\end{figure}

To understand the nature of these very high maximum temperatures of the central fountain gas, we perform an analysis of the thermodynamical evolution of gas around wind recoupling events, which are the defining feature of central fountain gas. To do so, we generated simulation snapshots separated by short time intervals of $\sim1-2\Myr$, for a total period of $11\Myr$ between $z=2.2466$ and $z=2.2383$. For convenience, in the following discussion we define the second snapshot in this series, $z=2.2452$, as $t=0$. We make a selection of tracer particles based on information from two adjacent snapshots, as follows. We select all tracers that i) belong to normal gas cells throughout $t=-1.9\Myr$ to $t=9.2\Myr$, and ii) at $t=0$ reside in gas cells that also contain tracers which at $t=-1.9\Myr$ belonged to a wind particle, rather to a normal gas cell. That is, we select tracers that at $t=0$ belong to cells that have `just' (namely within the last 1.9\,Myr, which is roughly the cooling time) experienced a wind particle recoupling into them. Since all tracers in any given cell are fully mixed, at $t\geq0$ this is equivalent to following the evolution of tracers that come directly from the wind itself. We instead follow the tracers that have been `recoupled into', as this allows us to make a distinction according to pre-recoupling properties. In particular, we split this tracer sample by their temperature prior to recoupling: cold ($T_{\rm t=-1.9\Myr} < 10^5$K) or hot ($T_{\rm t=-1.9\Myr} > 10^5$K).

In \fig{tracerevolution} we present the evolution of these two tracer populations as colored points on the $(\rho,\rm{T})$ phase diagram. The first column from the left shows the $t=-1.9\Myr$ properties of gas that will soon (by $t=0$) be recoupled into by one or more wind-phase particles. At this point, cold gas (top row) has densities from $\log n_{\rm H}[{\rm cm}^{-3}] \sim -2$ up to the density threshold for star-formation $\log n_{\rm H}[{\rm cm}^{-3}]\sim-0.9$ and even beyond, while hot gas (bottom row) has a narrower pre-recoupling density of $\log n_{\rm H}[{\rm cm}^{-3}] \sim -2$. Recall that the recoupling density is a parameter of our model, and is set to $5\%$ of the star-formation threshold, namely $\log n_{\rm H}[{\rm cm}^{-3}] \sim -2.2$. This explains the lower limit evident in the pre-recoupling density. The distribution extends towards larger values, however, because wind particles can also recouple based on a maximum time criterion, equal to 2.5\% times the current Hubble time. Overall, wind particles recouple into gas with a broad distribution of temperature and density, subject to a lower limit of $\log n_{\rm H}[{\rm cm}^{-3}] \sim -2.2$.

The second column from the left in \fig{tracerevolution} shows the same quantity on the horizontal axis, namely the pre-recoupling density, while the vertical axis now indicates the maximum temperature that each of the selected tracers has experienced {\it between} these two adjacent snapshots, which bracket its `recoupling event'. The majority of the initially cold ($T<10^5$K, top row) tracers have been heated during this short time window to temperatures above $10^{5.5}$K, and the initially hot ($T>10^5K$, bottom row) tracers have largely heated up to above $10^6K$. We conclude that this heating occurs primarily due to a rapid conversion of the kinetic energy of the wind into thermal energy through a local shock \citep[e.g.][]{schaal16,fielding17}.\footnote{It is worth noting that some tracers in the upper row of the second column of \fig{tracerevolution} show a maximum temperature between $t=-1.9\Myr$ and $t=0$ that is cold, at $\sim10^{4.2}$K. These are tracers that have in fact not experienced the recoupling event, but have joined the cells that contain previously-wind tracers {\it after} an already-completed recoupling-heating-cooling sequence. In this sense, our selection is not `pure' due to the finite time window of $1.9\Myr$ (of order $t_{\rm cool}$) between the two snapshots we used for the selection.}

The third column from the left in \fig{tracerevolution} shows the properties of these same tracers, now at $t=0$, the snapshot in which they are identified to be residing in the same gas cells as tracers that have just been converted from wind particle. It is evident that the densities in the top row have typically dropped by roughly an order of magnitude, representing the expansion of over-pressurized cells that have started off at $t=-1.9\Myr$ with densities $\log n_{\rm H}[{\rm cm}^{-3}]\sim-1$ and subsequently been heated to $\sim10^6$K by the recoupling. A significant drop in temperatures is also evident. Within less than two Myr after recoupling, a significant population of cool-dense gas accumulates, mostly from tracers that were cool at $t=-1.9\Myr$, but also from those that were hot at that time. By $\sim6\Myr$ after the recoupling event (fourth column), the majority of initially hot gas is already part of the cool-dense phase, and this transition is almost fully complete by $t=9.2\Myr$, as seen in the fifth column.

In \fig{tracertracks} we show six typical examples for evolutionary tracks of individual tracers through time. The gray and colored cyan/magenta points are identical to the $t=-1.9\Myr$ distributions of \fig{tracerevolution} and represent the starting point of the evolution. Each tracer track begins with a filled dark blue circle at that time, and its maximum temperature between $t=-1.9\Myr$ and $t=0$ is shown with an empty circle. The tracks then follow their evolution until $t=9.2\Myr$, with color indicating time in Myr according to the color bar. These tracks confirm our previous analysis of the evolutionary progression of the wind-heated CGM. Namely, gas in the top panel, with pre-recoupling temperature $T<10^5$K, typically heats up to $T\gtrsim10^6$K but cools extremely rapidly back down into the cool-dense phase within $\sim1\Myr$.\footnote{One example is also shown of a track where this does not happen. This example is, as is typical of these cases, a tracer starting out with relatively low density.} Similarly, gas in the bottom panel, with pre-recoupling temperature $T>10^5$K, typically heats up to $T\sim10^{6-6.5}$ and then cools back down into the cool-dense phase, however over a somewhat longer timescale of a few million years, which is still much shorter than the dynamical timescale of the halo.

It is clear that the high maximum temperatures of the central fountain gas, previously seen in \fig{Tmax}, are short-lived. These rapid evolutionary timescales are expected, given the densities of the inner halo gas. At $\log n_{\rm H}[{\rm cm}^{-3}]=-2$, the cooling time of the gas (in photo-ionization equilibrium with the $z=2.25$ metagalactic background) is $\simeq 3\Myr$ at $T=10^{5.5}$\,K and $\simeq 10\Myr$ at $T=10^6$\,K. We also examined the distribution of halocentric distances of tracer particles at the time they undergo recoupling (not shown), and found that the typical distances are $\sim10-20\kpc$, with a sharp cutoff at $\approx22\kpc$. The distribution of wind-phase particles occupies a similar volume of the inner halo. This indicates that, in our previous analysis of the radii where $T_{\rm max}$ is achieved by `central fountain' gas (\fig{r_at_Tmax}), the $\sim 0.15-0.25 r_{\rm vir}$ peak corresponds to maximum temperatures achieved as a result of the wind shock heating into the inner CGM. It is worth noting that the particular location of this peak is a rather strong function of the recoupling criteria, encapsulated in free parameters of the sub-grid model (see also Section \ref{sec:discussion}). On the other hand, the $\sim 0.5-0.7 r_{\rm vir}$ peak corresponds to tracers whose temperature during this process did not exceed the maximum temperature achieved earlier during their initial accretion into the halo and encounter with the halo virial shock. As these two peaks have roughly the same integrals, about half of galactic fountain material, as it recycles in the inner halo, shocks to a higher temperature than as a result of the virial shock itself.


\subsection{Impact of the CGM refinement scheme}

\begin{figure*}
  \includegraphics[width=7.0in]{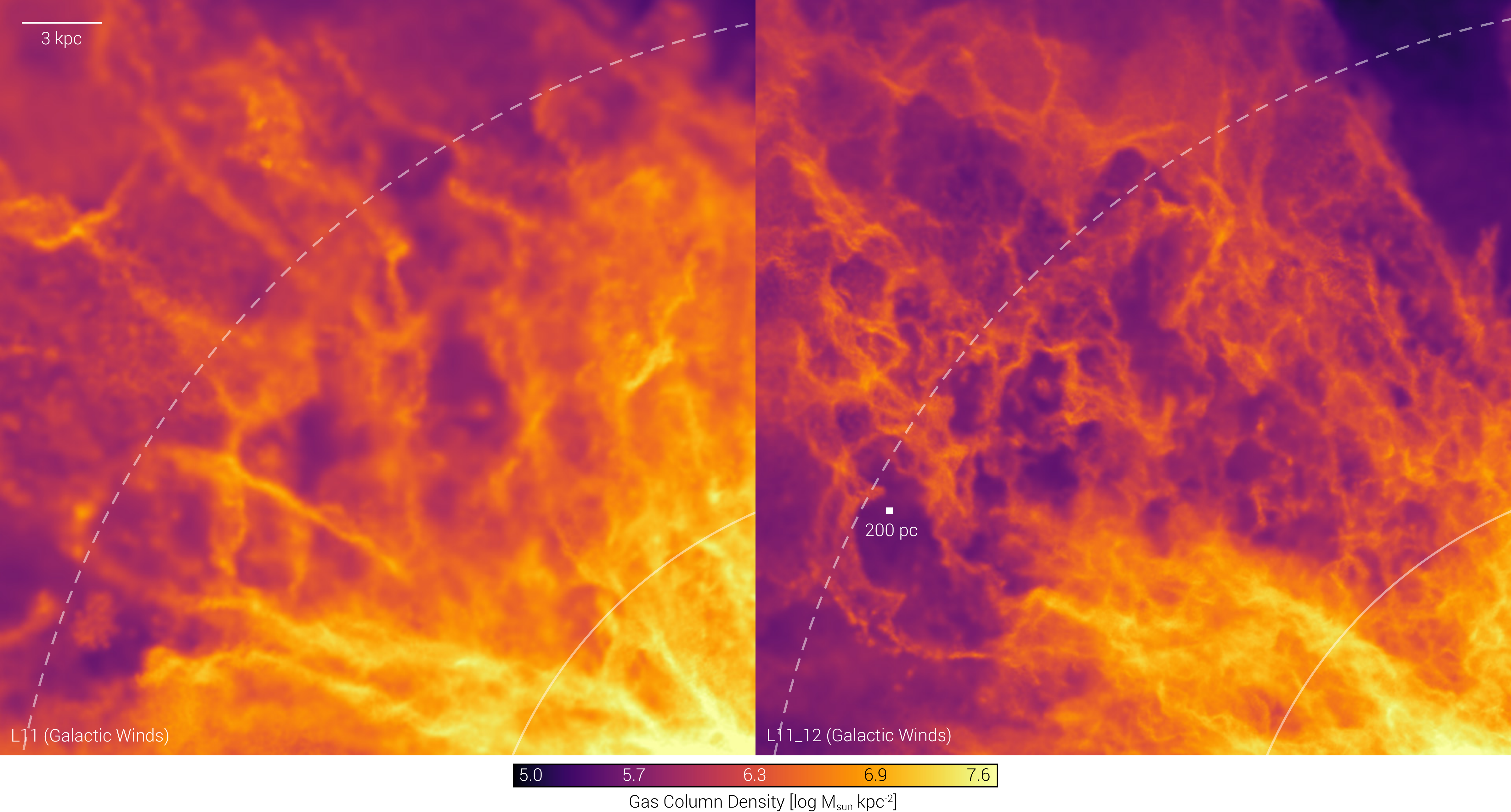}
  \caption{Comparison of gas structure in the Galactic Winds halo at the final redshift of $z=2.25$, contrasting a `normal' L11 resolution run (left panel) to our fiducial, CGM refinement boosted run (right panel). We show projected gas density, focusing on the upper left quadrant of images previously shown in this paper. The solid circles show $0.25 r_{\rm vir}$ while the dotted circles show $0.5 r_{\rm vir}$. The enhanced resolution with the CGM refinement scheme leads to a striking ability to resolve halo gas structures, in both outflows and inflows. For reference, the small white square is 200 physical parsecs on a side -- the fiducial simulation analyzed in this work \textit{well resolves} gas structures with $100 - 200$ pc sizes. Nonetheless, we find little quantitative difference between these two simulations in the metrics, such as HI covering fractions or the radial profiles of cold gas mass, considered herein. \label{fig:comp_image}}
\end{figure*}

Until now we have primarily emphasized the differences between the Primordial Cooling and Galactic Winds runs, both executed using our new CGM refinement scheme. Here we discuss briefly the impact of the additional refinement itself, i.e. we compare the fiducial L11\_12 Galactic Winds simulation with an otherwise equivalent, `unboosted' L11 run. Note that this unboosted run with galactic winds has not been considered above, nor in previous work. The resolution properties of the halo gas, for instance the gas cell mass and spatial resolution as a function of galactocentric distance, are the same as previously presented in \fig{cell_mass_size}. In the majority of the halo volume, the unboosted run has gas cells which are a factor of two larger in size, and a factor of eight larger in mass. We note that the galaxy properties themselves appear, reassuringly, resilient to the CGM refinement procedure. The final stellar mass changes by less than 0.1 dex, and the gas-phase metallicity of the ISM is also unchanged ($\lesssim\,$20\%).

In \fig{comp_image} we show a visual comparison of these two simulations, in projected gas density within a small subset of the halo -- each panel is 40\% of the virial radius across. In comparison to the un-modified, unboosted run (left), the CGM refinement scheme (right) enables us to resolve much finer halo gas structures, as designed. In particular, we show with the small white square a reference size-scale of 200 physical parsecs, concluding that the resolution-boosted simulation is able to resolve gas structures with $100 - 200$ pc sizes. Note that this requires gas cell sizes well below the $\sim 95$ pc \textit{median} resolution quoted previously, however these occur naturally in denser flows thanks to the natural spatial adaptivity of the code, reaching 12 pc at their smallest.

Despite the impressive qualitative differences in the morphology and structure of halo outflows as well as cosmological inflows, we find, as a general conclusion, that the majority of our quantitative results are more or less robust to the enhanced resolution. In particular, the redshift evolution of the masses of different halo component (\fig{allmasses}) is unchanged, at the $\sim$ 0.1 dex level. The radial profiles of cool-dense and non cool-dense gas (\fig{CGM_frac_R}) are also consistent between the two runs, at all radii $0.15 < r/r_{\rm vir} < 1.2$. The HI covering fractions as a function of distance out to twice the virial radius (\fig{rudie}), for the four different column density thresholds, are converged at the level of 5\%. The column density profiles of MgII versus impact parameter (\fig{MgIIprof}) show a hint of an increase in the resolution-boosted run, but the signal is marginal and $\lesssim$ 0.2 dex in log cm$^{-2}$. Overall, the metrics (and observables) explored herein are not strongly modified by the enhanced CGM resolution. 

In a recent work considering a different, AMR-based CGM refinement procedure, \cite{peeples18} came to similar conclusions, finding that integrated quantities such as mass surface densities and ionic covering fractions changed only at the $\lesssim$ 30\% level (HI covering fractions by less than 10\%). In contrast, \cite{vdv18} find larger changes in HI column densities and covering fractions of $\sim 1$ dex and up to 20\%, respectively (although at $z=0$ instead of $z \simeq 2$). As all of these studies, including ours, are essentially resolution convergence tests, we speculate that the different conclusions as to the impact of an additional CGM refinement may result from the different `base' resolutions of the simulations, i.e. where on the resolution ladder the refinement step is taken. In particular, our `normal' L11 run has already quite a high resolution -- the halo spatial resolution is already below 1 kpc everywhere within 100 kpc, which is the boosted resolution of \cite{vdv18} at which point they note a difference (with respect to a much lower resolution simulation). Our results imply, therefore, that our base run is already converged in e.g. HI covering fraction, explaining the lack of change when moving to the super-Lagrangian CGM refinement case. In the future, careful comparisons will be able to determine the spatial and/or mass resolution required to properly resolve particular processes and structures of circumgalactic gas. For example, it is important to keep in mind the likely possibility that there exists a relevant physical scale that is still much smaller than the scales that any of the aforementioned CGM resolution-boosted simulations, including ours, is able to resolve \citep[e.g.][]{mccourt18}. In other words, the apparent convergence we see may be broken in future simulations that resolve even much smaller scales, for example by utilizing an even more aggressive or targeted super-Lagrangian CGM refinement scheme (\textcolor{blue}{Nelson et al. in prep}).

Given the lack of substantiative changes in the structural properties of the halo gas and its evolution, we also expect little change in the subsequent tracer based analysis of accretion and cold gas production in the halo. The notable exceptions are \fig{tracerevolution} and \fig{tracertracks}, where we appreciate an important difference in the functioning of our galactic winds model in the `normal' versus boosted runs. Namely, wind-phase particles are launched from the dense ISM gas, which has the same base resolution in both simulations. This is by design, whereby the resolution of the central galaxy is unchanged by our CGM refinement scheme. However, these wind particles recouple into halo gas which has a factor of eight different mass in the two cases -- i.e., the mass ratio of the wind and CGM gas cell is unity in the normal run, but eight in the CGM refinement run. As a result, the impulsive energy deposition is much higher in the latter case. This results in a more efficient thermalization of the kinetic energy and hence stronger heating, as discussed in \fig{tracerevolution}, but also a factor of a few larger density increase, which enhances the cooling rate, allowing this newly heated cell to rapidly cool. In contrast, recoupling into cold CGM gas the un-modified L11 simulation less frequently results in heating much beyond $\sim 10^5$\,K. The post recoupling cooling is however similarly rapid in both cases. As a result, the overall discussion of \fig{tracertracks} is largely unchanged and this channel for cool halo gas production is qualitatively similar in both simulations.

Finally, we note that the upcoming TNG50 cosmological volume simulation (\textcolor{blue}{Nelson et al. in prep}, \textcolor{blue}{Pillepich et al. in prep}) lies just shy of our `normal' L11 resolution, and will hydrodynamically resolve the circumgalactic medium of halos with cells only $\sim 60$\% larger in spatial extent. For the mass scale considered herein, $M_{\rm halo} \sim 10^{12}$\msun and $z=2$, TNG50 will resolve the CGM at the 1 kpc level within $r \lesssim 60$ kpc, and for a full cosmological sample of halos. It is reasonable to expect the upcoming generation of cosmological hydrodynamical simulations to reach such resolutions, which have previously only been accessible in zoom scenarios.


\section{Discussion}
\label{sec:discussion}

Our principal goal was to explore the primary channel(s) by which cold-phase CGM gas is produced and maintained in the regime of $\sim 10^{12}$\msun halos at $z \sim 2$. We find that multiple channels contribute to the cool-dense CGM phase; galactic outflows, satellite stripping, primordial unshocked gas, and cooling out of the hot halo all play a role. Galactic winds are the primary contributor to cold phase production, with roughly 3/4 of the cold phase mass originating in a large-scale galactic fountain. Gas stripping from satellites and primordial accretion both contribute roughly equally to the remaining $\simeq$ 25\% of the cold phase. Compared to cooling-only simulation, the Galactic-Winds run significantly enhanced the amount of cold gas in the halo, which is reflected in the larger covering fractions of neutral hydrogen \citep[e.g.][]{fg16} and in the equivalent widths of MgII out to large radii \citep[e.g.][]{zhu14}. A large fraction of this cool gas can then reaccrete and provide the fuel for subsequent star formation \citep{oppenheimer10,fg11}; we expect at least half of cold gas accretion onto galaxies at $z=2$ to originate from fountain recycling \citep{nelson15}.

We emphasize that since our sample size consists of only one galaxy, the specific percentage breakdown of cool-dense gas production is only suggestive, and also undoubtedly varies as a function of halo mass as well as redshift. 

We have seen that primordial inflows and fountain outflows occupy largely distinct regimes in terms of radial distribution, kinematics, and metallicity. In the cooling-only run, the metallicity distribution in particular shows a clear dichotomy, whereby gas which has at some point been inside a galaxy typically has metallicities $Z/Z_\odot > 10\%$, while gas which has never entered a galaxy typically has metallicities $Z/Z_\odot \lesssim 1\%$. This could qualitatively explain similar bimodal distributions seen for $z \lesssim 1$ HI absorbers as found by \cite{lehner13} and \cite{wotta16} \citep[see also][at $z \sim 3$]{lehner16}. However, this signal is much weaker in the Galactic-Winds run due to non-negligible pre-enrichment of `primordial gas' prior to accretion. Further, these distinctions are by no means absolute, and there is overlap. For instance, we observed that a non-negligible fraction of cool-dense halo gas whose origin was shock-heated primordial accretion was enriched and even outflowing. Therefore not all outflow signatures necessarily arise from a wind driven from the central galaxy \citep{steidel10,rubin14,turner15}. It may simply be the case that no particular observation can uniquely be identified as inflow versus outflow, nor e.g. its origin be determined on the basis of its metallicity, although we anticipate that simulated ensembles could provide guidelines in the statistical sense and endeavor to provide more observational diagnostics in future work.

We caution that our model for galactic-scale winds driven by stellar feedback is necessarily limited in physical fidelity. Namely, as we do not attempt to resolve the cold neutral or molecular phases of the interstellar medium, nor the explosions of individual supernovae, we cannot capture the interaction between the two. Thus we employ the effective model of \cite{spr03} which is designed to function in this regime by hydrodynamically decoupling wind-phase gas for some period of time, until they leave the star-forming disk and then are allowed to `recouple' into the inner halo gas. While the decoupling and recoupling events themselves are clearly not physical, the overall effect of the model is to uplift gas from the galaxy to the inner halo, where it can then cool rapidly at high density while still outflowing, in resemblance to some more detailed numerical models, as discussed below.

This effective behaviour of our model, which is critical for the thermodynamical evolution of the CGM and the origin of cool-dense CGM in our simulations, can be expected to be sensitive to the numerical recoupling parameters that are employed. For a preliminary exploration of this possibility, we have continued the Galactic-Winds simulation from $z=2.25$ further down to $z=2$ with a recoupling density that is ten times lower, implying that wind particles interact with the CGM at somewhat larger radii, and where the cooling times are longer. We examine the outcome close to $z=2$ and find that this results in tracer tracks that are much more likely to not join the cool-dense phase within $10\Myr$ after recoupling. This is in accordance with the typical cooling times of hundreds of $\Myr$ for gas at $T\sim10^6$K and the lower recoupling density of $\log n_{\rm H}[{\rm cm}^{-3}]\sim-3$. What follows are also a lower overall cool-dense CGM mass, as well as a lower star-formation rate in the galaxy itself. We anticipate that further research along these lines, together with more detailed simulations of wind interaction with the CGM \citep{zhang17,fielding17,schneider18}, will pave the way towards a more robust treatment of galactic winds in large-scale cosmological simulations.

It should be emphasized that the primary action of the `kinetic' wind model we employ is to generate wind-phase particles with direct momentum input. Despite being launched cold, virtually all of the gas which has been in the galactic fountain has had a previous maximum temperature of $T > 10^6$ K after interacting with the CGM. This is the case due to the kinetic energy impulse and the subsequent shocking on the surrounding CGM gas. Our tracer particle analysis has demonstrated that despite high post-recoupling temperatures, cool-dense gas then arises due to rapid cooling \citep[see also][]{fraternali13}. The implied rapid loss of the energy wind particles carry makes our wind feedback scheme `momentum-driven' in nature, despite often being described as `energy-driven' in the sense that the overall energy at launch is kept proportional to the star-formation rate of a galaxy regardless of its other properties. The scenario whereby outflowing gas can rapidly cool as it expands has qualitative similarities to that suggested by \cite{thompson16}, where more direct modeling of the wind-CGM interaction phase and its subsequent cooling behavior has also shown mechanisms for cold gas production in outflows \citep{scannapieco17,schneider18}. 

In our model, cool \mbox{$\sim 10^4$ K} gas is produced in the CGM largely due to uplifting of gas from the galaxy that is subsequently deposited under conditions that leads to rapidly cooling. Nonetheless, large amounts of this cool gas can persist in this galactic halo within a background atmosphere with temperatures of order $T_\text{vir} \sim 10^6$ K. This may provide a theoretical picture to interpret observations of high covering factors of `cool ions' in quasar-host and higher mass halos \citep{prochaska13,battaia15,johnson15}. For instance, the MgII equivalent width is elevated above $\sim$ 0.1 \r{A} all the way out to the virial radius \citep[at $z \sim 2$, see also][]{lau16}. Although we have not considered the kinematics of the different CGM gas phases, we anticipate that cool gas produced as a result of the cooling outflow will be traveling outward at (relatively) high velocity. It is unclear if this will result in a significant relative motion with respect to a slower moving hot background, or if the local hot gas will be effectively comoving in the outflow. Regardless, it is unclear that at the current resolutions we would properly capture instabilities such as Kelvin-Helmholtz which could possibly disrupt the cold component \citep[e.g.][]{hess10,schneider17}.

We note that, in addition to the exploration of MgII herein, we also find intriguing differences in the nature of absorption by higher ionization energy species -- OVI, in particular. At the $\simeq$ 2,200 solar mass resolution afforded by our new simulations, OVI absorption appears more clumpy or spatially inhomogeneous \citep[e.g.][]{simcoe06} and predominantly arises from higher density gas (by $\sim$1 dex) in comparison to analogous halos of IllustrisTNG \citep{nelson18b}. However, we cannot yet isolate the impacts of differing resolution as well as physical model, and postpone a detailed comparison of observable absorption signatures for future simulations which can leverage the `CGM zoom' technique for a larger halo sample simulated to $z\sim0$ in the context of the new TNG model for galaxy formation \citep{weinberger17,pillepich18a}.

\section{Summary} \label{sec:summary}

In this paper we present cosmological simulations of the same galaxy halo with increasingly comprehensive physical models, ultimately including a treatment of galactic-scale outflows driven by stellar feedback. Our chosen halo has a total mass of $\sim 10^{12}$ M$_\odot$ at $z \sim 2$, and we use these simulations to explore the complex multi-phase gas of the circumgalactic medium (CGM). With a particular emphasis on the origin of cold, dense gas in the halo, our main results are as follows:

\begin{itemize}
\item First, we propose and implement a novel, super-Lagrangian `CGM zoom' scheme, which focuses more resolution into the CGM and intentionally lowers resolution in the dense ISM. In the present work, this enables us to achieve a median gas mass resolution of $\simeq$ 2,200 solar masses in the CGM, or equivalently a spatial resolution of $\simeq$ 95 parsecs, the latter of which is, for instance, ten times better than in Illustris, TNG100, or similar large-volume simulations.

\item We find that winds substantially enhance the amount of cold gas in the halo. By $z \simeq 2$, the total mass in the cool-dense phase is nearly an order of magnitude larger than in the hotter, less dense component. This enhancement also manifests in the covering fractions of HI and the equivalent widths of MgII out to large radii. Specifically, the simulation with winds has higher covering fractions of HI regardless of $N_{\rm HI}$ threshold or radius, by as much as 50\% over the cooling-only simulation (at low columns). Within the halo, our simulated halo is consistent with the HI covering fraction measurements of \cite{rudie12} at all column densities. For MgII, while the cooling-only simulation has equivalent widths exceeding $\sim$\,1\,\r{A} only within 20 kpc, this is true for the simulation with winds out to nearly 50 kpc. The relative enhancement in EW is an order of magnitude at half the virial radius. Interestingly, the MgII profile at $R \gtrsim 40$ kpc in the Galactic-Winds simulation are suggestively consistent with the $z<1$ data of e.g. \cite{chen10} and \cite{werk13}, although we do not continue the simulation down to such low redshift.

\item The metallicity distribution of Lyman-limit system (LLS) sightlines shows a strong bimodality in the cooling-only run. Specifically, gas which has at some point been inside a galaxy typically has metallicities $Z/Z_\odot > 10\%$, while gas which has never entered a galaxy typically has metallicities $Z/Z_\odot \lesssim 1\%$ (for $M_{\rm halo} \sim 10^{12}$\msun at $z \sim 2$). This dichotomy is suggestively similar to the bimodality in LLS metallicities observed by \cite{lehner13} and \cite{wotta16} at $z \sim 1$, although again we do not simulate this regime directly. The inclusion of galactic winds in the simulation, however, makes this signal noticeably weaker, suggestive of the pre-enrichment of `primordial' accretion. Winds also increase the relative importance of metal-rich, non-primordial accretion components arising from both satellite stripping and the central fountain.

\item Using a tracer particle analysis we decompose the cool-dense CGM gas at $z \sim 2$ into its different production channels. Of the total, the vast majority originates from the central galactic fountain, which accounts for the origin of approximately 3/4 of cool-dense gas in the halo. To the remaining $\sim$ 25\% stripped material and cosmological accretion contribute roughly equally. This second, primordial accretion component is also split $\sim$ 50/50 between virialized (virial shock heated) inflow and unvirialized cold inflow.

\item We demonstrate that the majority of this cool-dense gas arises due to rapid cooling of the wind material interacting with the CGM. As a result, large amounts of cold, \mbox{$\sim 10^4$ K} gas are effectively deposited in this galactic halo that contains also gas with $T_\text{vir} \sim 10^6$ K. We anticipate that the details therein will depend on the details of our wind model, and that more physical treatments of the wind-halo interaction may be crucial.
\end{itemize}

\noindent We expect that our super-Lagrangian refinement scheme can be extended and even more aggressively applied, possibly allowing fully cosmological simulations to reach parsec level resolution in the circumgalactic medium and thereby resolve additional physical phenomena of interest \citep[e.g.][]{field65}.


\section*{Acknowledgements}

JS and DN thank X. Prochaska and the Esalen Institute for hosting the IMPS where many early discussions for this paper took place.
JS and the authors thank Federico Marinacci and Mark Vogelsberger for assistance in the early stages of this project, and Simeon Bird for tool development and for reading an earlier draft.
SG is supported by the Simons Foundation through the Flatiron Institute. Computations in this paper were simulation on the Odyssey cluster supported by the FAS Division of Science, Research Computing Group at Harvard University, and on the Freya cluster of the Max Planck Institute for Astrophysics (MPA).

\bibliographystyle{mnras}
\bibliography{refs}

\begin{thebibliography}{}
\makeatletter
\relax
\def\mn@urlcharsother{\let\do\@makeother \do\$\do\&\do\#\do\^\do\_\do\%\do\~}
\def\mn@doi{\begingroup\mn@urlcharsother \@ifnextchar [ {\mn@doi@}
  {\mn@doi@[]}}
\def\mn@doi@[#1]#2{\def\@tempa{#1}\ifx\@tempa\@empty \href
  {http://dx.doi.org/#2} {doi:#2}\else \href {http://dx.doi.org/#2} {#1}\fi
  \endgroup}
\def\mn@eprint#1#2{\mn@eprint@#1:#2::\@nil}
\def\mn@eprint@arXiv#1{\href {http://arxiv.org/abs/#1} {{\tt arXiv:#1}}}
\def\mn@eprint@dblp#1{\href {http://dblp.uni-trier.de/rec/bibtex/#1.xml}
  {dblp:#1}}
\def\mn@eprint@#1:#2:#3:#4\@nil{\def\@tempa {#1}\def\@tempb {#2}\def\@tempc
  {#3}\ifx \@tempc \@empty \let \@tempc \@tempb \let \@tempb \@tempa \fi \ifx
  \@tempb \@empty \def\@tempb {arXiv}\fi \@ifundefined
  {mn@eprint@\@tempb}{\@tempb:\@tempc}{\expandafter \expandafter \csname
  mn@eprint@\@tempb\endcsname \expandafter{\@tempc}}}

\bibitem[\protect\citeauthoryear{{Agertz} et~al.,}{{Agertz}
  et~al.}{2007}]{agertz07}
{Agertz} O.,  et~al., 2007, \mn@doi [\mnras]
  {10.1111/j.1365-2966.2007.12183.x}, \href
  {http://adsabs.harvard.edu/abs/2007MNRAS.380..963A} {380, 963}

\bibitem[\protect\citeauthoryear{{Angl{\'e}s-Alc{\'a}zar},
  {Faucher-Gigu{\`e}re}, {Kere{\v s}}, {Hopkins}, {Quataert}  \&
  {Murray}}{{Angl{\'e}s-Alc{\'a}zar} et~al.}{2017}]{anglesalcazar17}
{Angl{\'e}s-Alc{\'a}zar} D.,  {Faucher-Gigu{\`e}re} C.-A.,  {Kere{\v s}} D.,
  {Hopkins} P.~F.,  {Quataert} E.,   {Murray} N.,  2017, \mn@doi [\mnras]
  {10.1093/mnras/stx1517}, \href
  {http://adsabs.harvard.edu/abs/2017MNRAS.470.4698A} {470, 4698}

\bibitem[\protect\citeauthoryear{{Arrigoni Battaia}, {Hennawi}, {Prochaska}  \&
  {Cantalupo}}{{Arrigoni Battaia} et~al.}{2015}]{battaia15}
{Arrigoni Battaia} F.,  {Hennawi} J.~F.,  {Prochaska} J.~X.,   {Cantalupo} S.,
  2015, \mn@doi [\apj] {10.1088/0004-637X/809/2/163}, \href
  {http://adsabs.harvard.edu/abs/2015ApJ...809..163A} {809, 163}

\bibitem[\protect\citeauthoryear{{Barton} \& {Cooke}}{{Barton} \&
  {Cooke}}{2009}]{barton09}
{Barton} E.~J.,  {Cooke} J.,  2009, \mn@doi [\aj]
  {10.1088/0004-6256/138/6/1817}, \href
  {http://adsabs.harvard.edu/abs/2009AJ....138.1817B} {138, 1817}

\bibitem[\protect\citeauthoryear{{Bauer} \& {Springel}}{{Bauer} \&
  {Springel}}{2012}]{bauer12}
{Bauer} A.,  {Springel} V.,  2012, \mn@doi [\mnras]
  {10.1111/j.1365-2966.2012.21058.x}, \href
  {http://adsabs.harvard.edu/abs/2012MNRAS.423.2558B} {423, 2558}

\bibitem[\protect\citeauthoryear{{Bird}}{{Bird}}{2017}]{bird17}
{Bird} S.,  2017, {FSFE: Fake Spectra Flux Extractor}, Astrophysics Source Code
  Library (\mn@eprint {ascl} {1710.012})

\bibitem[\protect\citeauthoryear{{Bird}, {Vogelsberger}, {Sijacki},
  {Zaldarriaga}, {Springel}  \& {Hernquist}}{{Bird} et~al.}{2013}]{bird13}
{Bird} S.,  {Vogelsberger} M.,  {Sijacki} D.,  {Zaldarriaga} M.,  {Springel}
  V.,   {Hernquist} L.,  2013, \mn@doi [\mnras] {10.1093/mnras/sts590}, \href
  {http://adsabs.harvard.edu/abs/2013MNRAS.429.3341B} {429, 3341}

\bibitem[\protect\citeauthoryear{{Chen}, {Helsby}, {Gauthier}, {Shectman},
  {Thompson}  \& {Tinker}}{{Chen} et~al.}{2010}]{chen10}
{Chen} H.-W.,  {Helsby} J.~E.,  {Gauthier} J.-R.,  {Shectman} S.~A.,
  {Thompson} I.~B.,   {Tinker} J.~L.,  2010, \mn@doi [\apj]
  {10.1088/0004-637X/714/2/1521}, \href
  {http://adsabs.harvard.edu/abs/2010ApJ...714.1521C} {714, 1521}

\bibitem[\protect\citeauthoryear{Courant, Friedrichs  \& Lewy}{Courant
  et~al.}{1967}]{courant67}
Courant R.,  Friedrichs K.,   Lewy H.,  1967, \mn@doi [IBM J. Res. Dev.]
  {10.1147/rd.112.0215}, 11, 215

\bibitem[\protect\citeauthoryear{{Crain} et~al.,}{{Crain}
  et~al.}{2015}]{crain15}
{Crain} R.~A.,  et~al., 2015, \mn@doi [\mnras] {10.1093/mnras/stv725}, \href
  {http://adsabs.harvard.edu/abs/2015MNRAS.450.1937C} {450, 1937}

\bibitem[\protect\citeauthoryear{{Crighton}, {Hennawi}, {Simcoe}, {Cooksey},
  {Murphy}, {Fumagalli}, {Prochaska}  \& {Shanks}}{{Crighton}
  et~al.}{2015}]{crighton15}
{Crighton} N.~H.~M.,  {Hennawi} J.~F.,  {Simcoe} R.~A.,  {Cooksey} K.~L.,
  {Murphy} M.~T.,  {Fumagalli} M.,  {Prochaska} J.~X.,   {Shanks} T.,  2015,
  \mn@doi [\mnras] {10.1093/mnras/stu2088}, 446, 18

\bibitem[\protect\citeauthoryear{{Dubois}, {Peirani}, {Pichon}, {Devriendt},
  {Gavazzi}, {Welker}  \& {Volonteri}}{{Dubois} et~al.}{2016}]{dubois16}
{Dubois} Y.,  {Peirani} S.,  {Pichon} C.,  {Devriendt} J.,  {Gavazzi} R.,
  {Welker} C.,   {Volonteri} M.,  2016, \mn@doi [\mnras]
  {10.1093/mnras/stw2265}, \href
  {http://adsabs.harvard.edu/abs/2016MNRAS.463.3948D} {463, 3948}

\bibitem[\protect\citeauthoryear{Faucher-Gigu{\`e}re, Lidz, Zaldarriaga  \&
  Hernquist}{Faucher-Gigu{\`e}re et~al.}{2009}]{fg09}
Faucher-Gigu{\`e}re C.-A.,  Lidz A.,  Zaldarriaga M.,   Hernquist L.,  2009,
  ApJ, 703, 1416

\bibitem[\protect\citeauthoryear{{Faucher-Gigu{\`e}re}, {Kere{\v s}}  \&
  {Ma}}{{Faucher-Gigu{\`e}re} et~al.}{2011}]{fg11}
{Faucher-Gigu{\`e}re} C.-A.,  {Kere{\v s}} D.,   {Ma} C.-P.,  2011, \mn@doi
  [\mnras] {10.1111/j.1365-2966.2011.19457.x}, \href
  {http://adsabs.harvard.edu/abs/2011MNRAS.417.2982F} {417, 2982}

\bibitem[\protect\citeauthoryear{{Faucher-Gigu{\`e}re}, {Feldmann}, {Quataert},
  {Kere{\v s}}, {Hopkins}  \& {Murray}}{{Faucher-Gigu{\`e}re}
  et~al.}{2016}]{fg16}
{Faucher-Gigu{\`e}re} C.-A.,  {Feldmann} R.,  {Quataert} E.,  {Kere{\v s}} D.,
  {Hopkins} P.~F.,   {Murray} N.,  2016, \mn@doi [\mnras]
  {10.1093/mnrasl/slw091}, \href
  {http://adsabs.harvard.edu/abs/2016MNRAS.461L..32F} {461, L32}

\bibitem[\protect\citeauthoryear{{Ferland} et~al.,}{{Ferland}
  et~al.}{2013}]{ferland13}
{Ferland} G.~J.,  et~al., 2013, \rmxaa, \href
  {http://adsabs.harvard.edu/abs/2013RMxAA..49..137F} {49, 137}

\bibitem[\protect\citeauthoryear{{Field}}{{Field}}{1965}]{field65}
{Field} G.~B.,  1965, \mn@doi [\apj] {10.1086/148317}, \href
  {http://adsabs.harvard.edu/abs/1965ApJ...142..531F} {142, 531}

\bibitem[\protect\citeauthoryear{{Fielding}, {Quataert}, {McCourt}  \&
  {Thompson}}{{Fielding} et~al.}{2017}]{fielding17}
{Fielding} D.,  {Quataert} E.,  {McCourt} M.,   {Thompson} T.~A.,  2017,
  \mn@doi [\mnras] {10.1093/mnras/stw3326}, \href
  {http://adsabs.harvard.edu/abs/2017MNRAS.466.3810F} {466, 3810}

\bibitem[\protect\citeauthoryear{{Ford}, {Oppenheimer}, {Dav{\'e}}, {Katz},
  {Kollmeier}  \& {Weinberg}}{{Ford} et~al.}{2013}]{ford13}
{Ford} A.~B.,  {Oppenheimer} B.~D.,  {Dav{\'e}} R.,  {Katz} N.,  {Kollmeier}
  J.~A.,   {Weinberg} D.~H.,  2013, \mn@doi [\mnras] {10.1093/mnras/stt393},
  \href {http://adsabs.harvard.edu/abs/2013MNRAS.432...89F} {432, 89}

\bibitem[\protect\citeauthoryear{{Fraternali}, {Marasco}, {Marinacci}  \&
  {Binney}}{{Fraternali} et~al.}{2013}]{fraternali13}
{Fraternali} F.,  {Marasco} A.,  {Marinacci} F.,   {Binney} J.,  2013, \mn@doi
  [\apjl] {10.1088/2041-8205/764/2/L21}, \href
  {http://adsabs.harvard.edu/abs/2013ApJ...764L..21F} {764, L21}

\bibitem[\protect\citeauthoryear{{Genel}, {Vogelsberger}, {Nelson}, {Sijacki},
  {Springel}  \& {Hernquist}}{{Genel} et~al.}{2013}]{genel13}
{Genel} S.,  {Vogelsberger} M.,  {Nelson} D.,  {Sijacki} D.,  {Springel} V.,
  {Hernquist} L.,  2013, \mn@doi [\mnras] {10.1093/mnras/stt1383}, \href
  {http://adsabs.harvard.edu/abs/2013MNRAS.435.1426G} {435, 1426}

\bibitem[\protect\citeauthoryear{{Genel} et~al.,}{{Genel}
  et~al.}{2014}]{genel14}
{Genel} S.,  et~al., 2014, \mn@doi [\mnras] {10.1093/mnras/stu1654}, \href
  {http://adsabs.harvard.edu/abs/2014MNRAS.445..175G} {445, 175}

\bibitem[\protect\citeauthoryear{{Genel} et~al.,}{{Genel}
  et~al.}{2018}]{genel18}
{Genel} S.,  et~al., 2018, \mn@doi [\mnras] {10.1093/mnras/stx3078}, \href
  {http://adsabs.harvard.edu/abs/2018MNRAS.474.3976G} {474, 3976}

\bibitem[\protect\citeauthoryear{Hayward, Torrey, Springel, Hernquist  \&
  Vogelsberger}{Hayward et~al.}{2014}]{hayward14}
Hayward C.~C.,  Torrey P.,  Springel V.,  Hernquist L.,   Vogelsberger M.,
  2014, MNRAS, 442, 1992

\bibitem[\protect\citeauthoryear{{Hernquist}, {Katz}, {Weinberg}  \&
  {Miralda-Escud{\'e}}}{{Hernquist} et~al.}{1996}]{hernquist96}
{Hernquist} L.,  {Katz} N.,  {Weinberg} D.~H.,   {Miralda-Escud{\'e}} J.,
  1996, \mn@doi [\apjl] {10.1086/309899}, \href
  {http://adsabs.harvard.edu/abs/1996ApJ...457L..51H} {457, L51}

\bibitem[\protect\citeauthoryear{{He{\ss}} \& {Springel}}{{He{\ss}} \&
  {Springel}}{2010}]{hess10}
{He{\ss}} S.,  {Springel} V.,  2010, \mn@doi [\mnras]
  {10.1111/j.1365-2966.2010.16892.x}, \href
  {http://adsabs.harvard.edu/abs/2010MNRAS.406.2289H} {406, 2289}

\bibitem[\protect\citeauthoryear{{Johnson}, {Chen}  \& {Mulchaey}}{{Johnson}
  et~al.}{2015}]{johnson15}
{Johnson} S.~D.,  {Chen} H.-W.,   {Mulchaey} J.~S.,  2015, \mn@doi [\mnras]
  {10.1093/mnras/stv1481}, \href
  {http://adsabs.harvard.edu/abs/2015MNRAS.452.2553J} {452, 2553}

\bibitem[\protect\citeauthoryear{{Kauffmann}, {Nelson}, {M{\'e}nard}  \&
  {Zhu}}{{Kauffmann} et~al.}{2017}]{kauffmann17}
{Kauffmann} G.,  {Nelson} D.,  {M{\'e}nard} B.,   {Zhu} G.,  2017, \mn@doi
  [\mnras] {10.1093/mnras/stx639}, \href
  {http://adsabs.harvard.edu/abs/2017MNRAS.468.3737K} {468, 3737}

\bibitem[\protect\citeauthoryear{{Kere{\v s}} \& {Hernquist}}{{Kere{\v s}} \&
  {Hernquist}}{2009}]{keres09}
{Kere{\v s}} D.,  {Hernquist} L.,  2009, \mn@doi [\apjl]
  {10.1088/0004-637X/700/1/L1}, \href
  {http://adsabs.harvard.edu/abs/2009ApJ...700L...1K} {700, L1}

\bibitem[\protect\citeauthoryear{{Kere{\v s}}, {Katz}, {Weinberg}  \&
  {Dav{\'e}}}{{Kere{\v s}} et~al.}{2005}]{keres05}
{Kere{\v s}} D.,  {Katz} N.,  {Weinberg} D.~H.,   {Dav{\'e}} R.,  2005, \mn@doi
  [\mnras] {10.1111/j.1365-2966.2005.09451.x}, \href
  {http://adsabs.harvard.edu/abs/2005MNRAS.363....2K} {363, 2}

\bibitem[\protect\citeauthoryear{{Kere{\v s}}, {Vogelsberger}, {Sijacki},
  {Springel}  \& {Hernquist}}{{Kere{\v s}} et~al.}{2012}]{keres12}
{Kere{\v s}} D.,  {Vogelsberger} M.,  {Sijacki} D.,  {Springel} V.,
  {Hernquist} L.,  2012, \mn@doi [\mnras] {10.1111/j.1365-2966.2012.21548.x},
  \href {http://adsabs.harvard.edu/abs/2012MNRAS.425.2027K} {425, 2027}

\bibitem[\protect\citeauthoryear{{Koyamada}, {Misawa}, {Inada}, {Oguri},
  {Kashikawa}  \& {Okoshi}}{{Koyamada} et~al.}{2017}]{koyamada17}
{Koyamada} S.,  {Misawa} T.,  {Inada} N.,  {Oguri} M.,  {Kashikawa} N.,
  {Okoshi} K.,  2017, \mn@doi [\apj] {10.3847/1538-4357/aa9a3a}, \href
  {http://adsabs.harvard.edu/abs/2017ApJ...851...88K} {851, 88}

\bibitem[\protect\citeauthoryear{{Lau}, {Prochaska}  \& {Hennawi}}{{Lau}
  et~al.}{2016}]{lau16}
{Lau} M.~W.,  {Prochaska} J.~X.,   {Hennawi} J.~F.,  2016, \mn@doi [\apjs]
  {10.3847/0067-0049/226/2/25}, \href
  {http://adsabs.harvard.edu/abs/2016ApJS..226...25L} {226, 25}

\bibitem[\protect\citeauthoryear{{Lehner} et~al.,}{{Lehner}
  et~al.}{2013}]{lehner13}
{Lehner} N.,  et~al., 2013, \mn@doi [\apj] {10.1088/0004-637X/770/2/138}, \href
  {http://adsabs.harvard.edu/abs/2013ApJ...770..138L} {770, 138}

\bibitem[\protect\citeauthoryear{{Lehner}, {O'Meara}, {Howk}, {Prochaska}  \&
  {Fumagalli}}{{Lehner} et~al.}{2016}]{lehner16}
{Lehner} N.,  {O'Meara} J.~M.,  {Howk} J.~C.,  {Prochaska} J.~X.,   {Fumagalli}
  M.,  2016, \mn@doi [\apj] {10.3847/1538-4357/833/2/283}, \href
  {http://adsabs.harvard.edu/abs/2016ApJ...833..283L} {833, 283}

\bibitem[\protect\citeauthoryear{Maller \& Bullock}{Maller \&
  Bullock}{2004}]{maller04}
Maller A.~H.,  Bullock J.~S.,  2004, Monthly Notices of the Royal Astronomical
  Society, 355, 694

\bibitem[\protect\citeauthoryear{{McCourt}, {O'Leary}, {Madigan}  \&
  {Quataert}}{{McCourt} et~al.}{2015}]{mccourt15}
{McCourt} M.,  {O'Leary} R.~M.,  {Madigan} A.-M.,   {Quataert} E.,  2015,
  \mn@doi [\mnras] {10.1093/mnras/stv355}, \href
  {http://adsabs.harvard.edu/abs/2015MNRAS.449....2M} {449, 2}

\bibitem[\protect\citeauthoryear{{McCourt}, {Oh}, {O'Leary}  \&
  {Madigan}}{{McCourt} et~al.}{2018}]{mccourt18}
{McCourt} M.,  {Oh} S.~P.,  {O'Leary} R.,   {Madigan} A.-M.,  2018, \mn@doi
  [\mnras] {10.1093/mnras/stx2687}, \href
  {http://adsabs.harvard.edu/abs/2018MNRAS.473.5407M} {473, 5407}

\bibitem[\protect\citeauthoryear{{Nelson}, {Vogelsberger}, {Genel}, {Sijacki},
  {Kere{\v s}}, {Springel}  \& {Hernquist}}{{Nelson} et~al.}{2013}]{nelson13}
{Nelson} D.,  {Vogelsberger} M.,  {Genel} S.,  {Sijacki} D.,  {Kere{\v s}} D.,
  {Springel} V.,   {Hernquist} L.,  2013, \mn@doi [\mnras]
  {10.1093/mnras/sts595}, \href
  {http://adsabs.harvard.edu/abs/2013MNRAS.429.3353N} {429, 3353}

\bibitem[\protect\citeauthoryear{{Nelson}, {Genel}, {Vogelsberger}, {Springel},
  {Sijacki}, {Torrey}  \& {Hernquist}}{{Nelson} et~al.}{2015}]{nelson15}
{Nelson} D.,  {Genel} S.,  {Vogelsberger} M.,  {Springel} V.,  {Sijacki} D.,
  {Torrey} P.,   {Hernquist} L.,  2015, \mn@doi [\mnras]
  {10.1093/mnras/stv017}, \href
  {http://adsabs.harvard.edu/abs/2015MNRAS.448...59N} {448, 59}

\bibitem[\protect\citeauthoryear{{Nelson}, {Genel}, {Pillepich},
  {Vogelsberger}, {Springel}  \& {Hernquist}}{{Nelson} et~al.}{2016}]{nelson16}
{Nelson} D.,  {Genel} S.,  {Pillepich} A.,  {Vogelsberger} M.,  {Springel} V.,
   {Hernquist} L.,  2016, \mn@doi [\mnras] {10.1093/mnras/stw1191}, 460, 2881

\bibitem[\protect\citeauthoryear{{Nelson} et~al.,}{{Nelson}
  et~al.}{2017}]{nelson18b}
{Nelson} D.,  et~al., 2017, preprint, \href
  {http://adsabs.harvard.edu/abs/2017arXiv171200016N} {} (\mn@eprint {arXiv}
  {1712.00016})

\bibitem[\protect\citeauthoryear{{Oppenheimer} \& {Dav{\'e}}}{{Oppenheimer} \&
  {Dav{\'e}}}{2009}]{oppenheimer09}
{Oppenheimer} B.~D.,  {Dav{\'e}} R.,  2009, \mn@doi [\mnras]
  {10.1111/j.1365-2966.2009.14676.x}, \href
  {http://adsabs.harvard.edu/abs/2009MNRAS.395.1875O} {395, 1875}

\bibitem[\protect\citeauthoryear{{Oppenheimer}, {Dav{\'e}}, {Kere{\v s}},
  {Fardal}, {Katz}, {Kollmeier}  \& {Weinberg}}{{Oppenheimer}
  et~al.}{2010}]{oppenheimer10}
{Oppenheimer} B.~D.,  {Dav{\'e}} R.,  {Kere{\v s}} D.,  {Fardal} M.,  {Katz}
  N.,  {Kollmeier} J.~A.,   {Weinberg} D.~H.,  2010, \mn@doi [\mnras]
  {10.1111/j.1365-2966.2010.16872.x}, \href
  {http://adsabs.harvard.edu/abs/2010MNRAS.406.2325O} {406, 2325}

\bibitem[\protect\citeauthoryear{{Peeples} et~al.,}{{Peeples}
  et~al.}{2018}]{peeples18}
{Peeples} M.~S.,  et~al., 2018, preprint, \href
  {http://adsabs.harvard.edu/abs/2018arXiv181006566P} {} (\mn@eprint {arXiv}
  {1810.06566})

\bibitem[\protect\citeauthoryear{{Pillepich} et~al.,}{{Pillepich}
  et~al.}{2018}]{pillepich18a}
{Pillepich} A.,  et~al., 2018, \mn@doi [\mnras] {10.1093/mnras/stx2656}, \href
  {http://adsabs.harvard.edu/abs/2018MNRAS.473.4077P} {473, 4077}

\bibitem[\protect\citeauthoryear{{Prochaska}, {Hennawi}  \&
  {Simcoe}}{{Prochaska} et~al.}{2013}]{prochaska13}
{Prochaska} J.~X.,  {Hennawi} J.~F.,   {Simcoe} R.~A.,  2013, \mn@doi [\apjl]
  {10.1088/2041-8205/762/2/L19}, 762, L19

\bibitem[\protect\citeauthoryear{{Rahmati} \& {Oppenheimer}}{{Rahmati} \&
  {Oppenheimer}}{2017}]{rahmati17b}
{Rahmati} A.,  {Oppenheimer} B.~D.,  2017, preprint, \href
  {http://adsabs.harvard.edu/abs/2017arXiv171203988R} {} (\mn@eprint {arXiv}
  {1712.03988})

\bibitem[\protect\citeauthoryear{Rahmati, Pawlik, Rai{\v c}evi{\'c}  \&
  Schaye}{Rahmati et~al.}{2013}]{rahmati13}
Rahmati A.,  Pawlik A.~H.,  Rai{\v c}evi{\'c} M.,   Schaye J.,  2013, MNRAS,
  430, 2427

\bibitem[\protect\citeauthoryear{{Rahmati}, {Schaye}, {Bower}, {Crain},
  {Furlong}, {Schaller}  \& {Theuns}}{{Rahmati} et~al.}{2015}]{rahmati15}
{Rahmati} A.,  {Schaye} J.,  {Bower} R.~G.,  {Crain} R.~A.,  {Furlong} M.,
  {Schaller} M.,   {Theuns} T.,  2015, \mn@doi [\mnras]
  {10.1093/mnras/stv1414}, \href
  {http://adsabs.harvard.edu/abs/2015MNRAS.452.2034R} {452, 2034}

\bibitem[\protect\citeauthoryear{{Rakic}, {Schaye}, {Steidel}, {Booth}, {Dalla
  Vecchia}  \& {Rudie}}{{Rakic} et~al.}{2013}]{rakic13}
{Rakic} O.,  {Schaye} J.,  {Steidel} C.~C.,  {Booth} C.~M.,  {Dalla Vecchia}
  C.,   {Rudie} G.~C.,  2013, \mn@doi [\mnras] {10.1093/mnras/stt950}, \href
  {http://adsabs.harvard.edu/abs/2013MNRAS.433.3103R} {433, 3103}

\bibitem[\protect\citeauthoryear{{Rees} \& {Ostriker}}{{Rees} \&
  {Ostriker}}{1977}]{rees77}
{Rees} M.~J.,  {Ostriker} J.~P.,  1977, \mnras, \href
  {http://adsabs.harvard.edu/abs/1977MNRAS.179..541R} {179, 541}

\bibitem[\protect\citeauthoryear{{Rubin}, {Prochaska}, {Koo}, {Phillips},
  {Martin}  \& {Winstrom}}{{Rubin} et~al.}{2014}]{rubin14}
{Rubin} K.~H.~R.,  {Prochaska} J.~X.,  {Koo} D.~C.,  {Phillips} A.~C.,
  {Martin} C.~L.,   {Winstrom} L.~O.,  2014, \mn@doi [\apj]
  {10.1088/0004-637X/794/2/156}, \href
  {http://adsabs.harvard.edu/abs/2014ApJ...794..156R} {794, 156}

\bibitem[\protect\citeauthoryear{{Rubin}, {Hennawi}, {Prochaska}, {Simcoe},
  {Myers}  \& {Lau}}{{Rubin} et~al.}{2015}]{rubin15}
{Rubin} K.~H.~R.,  {Hennawi} J.~F.,  {Prochaska} J.~X.,  {Simcoe} R.~A.,
  {Myers} A.,   {Lau} M.~W.,  2015, \mn@doi [\apj]
  {10.1088/0004-637X/808/1/38}, \href
  {http://adsabs.harvard.edu/abs/2015ApJ...808...38R} {808, 38}

\bibitem[\protect\citeauthoryear{{Rubin} et~al.,}{{Rubin}
  et~al.}{2017}]{rubin17}
{Rubin} K.~H.~R.,  et~al., 2017, preprint, \href
  {http://adsabs.harvard.edu/abs/2017arXiv170705873R} {} (\mn@eprint {arXiv}
  {1707.05873})

\bibitem[\protect\citeauthoryear{Rudie et~al.,}{Rudie et~al.}{2012}]{rudie12}
Rudie G.~C.,  et~al., 2012, The Astrophysical Journal, 750, 67

\bibitem[\protect\citeauthoryear{{Scannapieco}}{{Scannapieco}}{2017}]{scannapieco17}
{Scannapieco} E.,  2017, \mn@doi [\apj] {10.3847/1538-4357/aa5d0d}, \href
  {http://adsabs.harvard.edu/abs/2017ApJ...837...28S} {837, 28}

\bibitem[\protect\citeauthoryear{{Schaal} et~al.,}{{Schaal}
  et~al.}{2016}]{schaal16}
{Schaal} K.,  et~al., 2016, \mn@doi [\mnras] {10.1093/mnras/stw1587}, \href
  {http://adsabs.harvard.edu/abs/2016MNRAS.461.4441S} {461, 4441}

\bibitem[\protect\citeauthoryear{{Schaye}, {Carswell}  \& {Kim}}{{Schaye}
  et~al.}{2007}]{schaye07}
{Schaye} J.,  {Carswell} R.~F.,   {Kim} T.-S.,  2007, \mn@doi [\mnras]
  {10.1111/j.1365-2966.2007.12005.x}, \href
  {http://adsabs.harvard.edu/abs/2007MNRAS.379.1169S} {379, 1169}

\bibitem[\protect\citeauthoryear{{Schneider} \& {Robertson}}{{Schneider} \&
  {Robertson}}{2017}]{schneider17}
{Schneider} E.~E.,  {Robertson} B.~E.,  2017, \mn@doi [\apj]
  {10.3847/1538-4357/834/2/144}, \href
  {http://adsabs.harvard.edu/abs/2017ApJ...834..144S} {834, 144}

\bibitem[\protect\citeauthoryear{{Schneider}, {Robertson}  \&
  {Thompson}}{{Schneider} et~al.}{2018}]{schneider18}
{Schneider} E.~E.,  {Robertson} B.~E.,   {Thompson} T.~A.,  2018, preprint,
  \href {http://adsabs.harvard.edu/abs/2018arXiv180301005S} {} (\mn@eprint
  {arXiv} {1803.01005})

\bibitem[\protect\citeauthoryear{{Shen}, {Madau}, {Guedes}, {Mayer},
  {Prochaska}  \& {Wadsley}}{{Shen} et~al.}{2013}]{shen13}
{Shen} S.,  {Madau} P.,  {Guedes} J.,  {Mayer} L.,  {Prochaska} J.~X.,
  {Wadsley} J.,  2013, \mn@doi [\apj] {10.1088/0004-637X/765/2/89}, \href
  {http://adsabs.harvard.edu/abs/2013ApJ...765...89S} {765, 89}

\bibitem[\protect\citeauthoryear{{Silk}}{{Silk}}{1977}]{silk77}
{Silk} J.,  1977, \mn@doi [\apj] {10.1086/154972}, \href
  {http://adsabs.harvard.edu/abs/1977ApJ...211..638S} {211, 638}

\bibitem[\protect\citeauthoryear{{Simcoe}, {Sargent}, {Rauch}  \&
  {Becker}}{{Simcoe} et~al.}{2006}]{simcoe06}
{Simcoe} R.~A.,  {Sargent} W.~L.~W.,  {Rauch} M.,   {Becker} G.,  2006, \mn@doi
  [\apj] {10.1086/498441}, \href
  {http://adsabs.harvard.edu/abs/2006ApJ...637..648S} {637, 648}

\bibitem[\protect\citeauthoryear{Springel}{Springel}{2010}]{springel10}
Springel V.,  2010, Monthly Notices of the Royal Astronomical Society, 401, 791

\bibitem[\protect\citeauthoryear{{Springel} \& {Hernquist}}{{Springel} \&
  {Hernquist}}{2003}]{spr03}
{Springel} V.,  {Hernquist} L.,  2003, \mn@doi [\mnras]
  {10.1046/j.1365-8711.2003.06206.x}, 339, 289

\bibitem[\protect\citeauthoryear{{Springel}, {White}, {Tormen}  \&
  {Kauffmann}}{{Springel} et~al.}{2001}]{springel01}
{Springel} V.,  {White} S.~D.~M.,  {Tormen} G.,   {Kauffmann} G.,  2001,
  \mn@doi [\mnras] {10.1046/j.1365-8711.2001.04912.x}, \href
  {http://adsabs.harvard.edu/abs/2001MNRAS.328..726S} {328, 726}

\bibitem[\protect\citeauthoryear{{Steidel}, {Erb}, {Shapley}, {Pettini},
  {Reddy}, {Bogosavljevi{\'c}}, {Rudie}  \& {Rakic}}{{Steidel}
  et~al.}{2010}]{steidel10}
{Steidel} C.~C.,  {Erb} D.~K.,  {Shapley} A.~E.,  {Pettini} M.,  {Reddy} N.,
  {Bogosavljevi{\'c}} M.,  {Rudie} G.~C.,   {Rakic} O.,  2010, \mn@doi [\apj]
  {10.1088/0004-637X/717/1/289}, \href
  {http://adsabs.harvard.edu/abs/2010ApJ...717..289S} {717, 289}

\bibitem[\protect\citeauthoryear{Suresh, Bird, Vogelsberger, Genel, Torrey,
  Sijacki, Springel  \& Hernquist}{Suresh et~al.}{2015}]{suresh15}
Suresh J.,  Bird S.,  Vogelsberger M.,  Genel S.,  Torrey P.,  Sijacki D.,
  Springel V.,   Hernquist L.,  2015, Monthly Notices of the Royal Astronomical
  Society, 448, 895

\bibitem[\protect\citeauthoryear{{Suresh}, {Rubin}, {Kannan}, {Werk},
  {Hernquist}  \& {Vogelsberger}}{{Suresh} et~al.}{2017}]{suresh17}
{Suresh} J.,  {Rubin} K.~H.~R.,  {Kannan} R.,  {Werk} J.~K.,  {Hernquist} L.,
  {Vogelsberger} M.,  2017, \mn@doi [\mnras] {10.1093/mnras/stw2499}, \href
  {http://adsabs.harvard.edu/abs/2017MNRAS.465.2966S} {465, 2966}

\bibitem[\protect\citeauthoryear{Thom et~al.,}{Thom et~al.}{2012}]{thom12}
Thom C.,  et~al., 2012, ApJ Letters, 758, L41

\bibitem[\protect\citeauthoryear{{Thompson}, {Quataert}, {Zhang}  \&
  {Weinberg}}{{Thompson} et~al.}{2016}]{thompson16}
{Thompson} T.~A.,  {Quataert} E.,  {Zhang} D.,   {Weinberg} D.~H.,  2016,
  \mn@doi [\mnras] {10.1093/mnras/stv2428}, \href
  {http://adsabs.harvard.edu/abs/2016MNRAS.455.1830T} {455, 1830}

\bibitem[\protect\citeauthoryear{Torrey, Vogelsberger, Sijacki, Springel  \&
  Hernquist}{Torrey et~al.}{2012}]{torrey12}
Torrey P.,  Vogelsberger M.,  Sijacki D.,  Springel V.,   Hernquist L.,  2012,
  MNRAS, 427, 2224

\bibitem[\protect\citeauthoryear{{Torrey}, {Vogelsberger}, {Genel}, {Sijacki},
  {Springel}  \& {Hernquist}}{{Torrey} et~al.}{2014}]{torrey14}
{Torrey} P.,  {Vogelsberger} M.,  {Genel} S.,  {Sijacki} D.,  {Springel} V.,
  {Hernquist} L.,  2014, \mn@doi [\mnras] {10.1093/mnras/stt2295}, \href
  {http://adsabs.harvard.edu/abs/2014MNRAS.438.1985T} {438, 1985}

\bibitem[\protect\citeauthoryear{{Turner}, {Schaye}, {Steidel}, {Rudie}  \&
  {Strom}}{{Turner} et~al.}{2015}]{turner15}
{Turner} M.~L.,  {Schaye} J.,  {Steidel} C.~C.,  {Rudie} G.~C.,   {Strom}
  A.~L.,  2015, \mn@doi [\mnras] {10.1093/mnras/stv750}, \href
  {http://adsabs.harvard.edu/abs/2015MNRAS.450.2067T} {450, 2067}

\bibitem[\protect\citeauthoryear{{Vogelsberger}, {Genel}, {Sijacki}, {Torrey},
  {Springel}  \& {Hernquist}}{{Vogelsberger} et~al.}{2013}]{vogelsberger13}
{Vogelsberger} M.,  {Genel} S.,  {Sijacki} D.,  {Torrey} P.,  {Springel} V.,
  {Hernquist} L.,  2013, \mn@doi [\mnras] {10.1093/mnras/stt1789}, \href
  {http://adsabs.harvard.edu/abs/2013MNRAS.436.3031V} {436, 3031}

\bibitem[\protect\citeauthoryear{{Weinberger} et~al.,}{{Weinberger}
  et~al.}{2017}]{weinberger17}
{Weinberger} R.,  et~al., 2017, \mn@doi [\mnras] {10.1093/mnras/stw2944}, \href
  {http://adsabs.harvard.edu/abs/2017MNRAS.465.3291W} {465, 3291}

\bibitem[\protect\citeauthoryear{Werk, Prochaska, Thom, Tumlinson, Tripp,
  O'Meara  \& Peeples}{Werk et~al.}{2013}]{werk13}
Werk J.~K.,  Prochaska J.~X.,  Thom C.,  Tumlinson J.,  Tripp T.~M.,  O'Meara
  J.~M.,   Peeples M.~S.,  2013, The Astrophysical Journal Supplement, 204, 17

\bibitem[\protect\citeauthoryear{{White} \& {Rees}}{{White} \&
  {Rees}}{1978}]{wr78}
{White} S.~D.~M.,  {Rees} M.~J.,  1978, \mnras, \href
  {http://adsabs.harvard.edu/abs/1978MNRAS.183..341W} {183, 341}

\bibitem[\protect\citeauthoryear{{Wotta}, {Lehner}, {Howk}, {O'Meara}  \&
  {Prochaska}}{{Wotta} et~al.}{2016}]{wotta16}
{Wotta} C.~B.,  {Lehner} N.,  {Howk} J.~C.,  {O'Meara} J.~M.,   {Prochaska}
  J.~X.,  2016, \mn@doi [\apj] {10.3847/0004-637X/831/1/95}, \href
  {http://adsabs.harvard.edu/abs/2016ApJ...831...95W} {831, 95}

\bibitem[\protect\citeauthoryear{{Zhang}, {Thompson}, {Quataert}  \&
  {Murray}}{{Zhang} et~al.}{2017}]{zhang17}
{Zhang} D.,  {Thompson} T.~A.,  {Quataert} E.,   {Murray} N.,  2017, \mn@doi
  [\mnras] {10.1093/mnras/stx822}, \href
  {http://adsabs.harvard.edu/abs/2017MNRAS.468.4801Z} {468, 4801}

\bibitem[\protect\citeauthoryear{{Zhu} et~al.,}{{Zhu} et~al.}{2014}]{zhu14}
{Zhu} G.,  et~al., 2014, \mn@doi [\mnras] {10.1093/mnras/stu186}, \href
  {http://adsabs.harvard.edu/abs/2014MNRAS.439.3139Z} {439, 3139}

\bibitem[\protect\citeauthoryear{{van de Voort} \& {Schaye}}{{van de Voort} \&
  {Schaye}}{2012}]{vdv12}
{van de Voort} F.,  {Schaye} J.,  2012, \mn@doi [\mnras]
  {10.1111/j.1365-2966.2012.20949.x}, \href
  {http://adsabs.harvard.edu/abs/2012MNRAS.423.2991V} {423, 2991}

\bibitem[\protect\citeauthoryear{{van de Voort}, {Springel}, {Mandelker}, {van
  den Bosch}  \& {Pakmor}}{{van de Voort} et~al.}{2018}]{vdv18}
{van de Voort} F.,  {Springel} V.,  {Mandelker} N.,  {van den Bosch} F.~C.,
  {Pakmor} R.,  2018, \mn@doi [\mnras] {10.1093/mnrasl/sly190}, \href
  {http://adsabs.harvard.edu/abs/2018MNRAS.tmpL.192V} {}

\makeatother
\end{thebibliography}

\end{document}